\documentclass{osa-article}
\journal{osajournal}
\articletype{Review Article}

\usepackage{braket}
\graphicspath{{./Figures/}}
\newcommand{\pare}[1]{\left( #1 \right)}
\newcommand{\cor}[1]{\left[ #1 \right]}

 \usepackage{pict2e,picture,graphicx}
 \usepackage{caption}
 \usepackage{subcaption}
 \usepackage{mathrsfs}
 \usepackage{mathtools}
 \usepackage[mathscr]{eucal}

\makeatletter
\DeclareRobustCommand{\Arrow}[1][]{%
\check@mathfonts
\if\relax\detokenize{#1}\relax
\settowidth{\dimen@}{$\m@th\rightarrow$}%
\else
\setlength{\dimen@}{#1}%
\fi
\sbox\z@{\usefont{U}{lasy}{m}{n}\symbol{41}}%
\begin{picture}(\dimen@,\ht\z@)
\roundcap
\put(\dimexpr\dimen@-.7\wd\z@,0){\usebox\z@}
\put(0,\fontdimen22\textfont2){\line(1,0){\dimen@}}
\end{picture}%
}
\makeatother
\newcommand{\veryshortrightarrow}{\hspace{.2mm}\scalebox{.8}{\Arrow[.1cm]}\hspace{.2mm}}

\begin{document}

\title{Entanglement-Based Quantum Information Technology}

\author{Zheshen Zhang,\authormark{1*} Chenglong You,\authormark{2} Omar S. Maga\~na-Loaiza,\authormark{2} Robert Fickler,\authormark{3} Roberto de J. Le\'{o}n-Montiel,\authormark{4} Juan P. Torres,\authormark{5,6} Travis Humble,\authormark{7} Shuai Liu,\authormark{1} Yi Xia,\authormark{8} Quntao Zhuang\authormark{9}}

\address{\authormark{1}Department of Electrical Engineering and Computer Science, University of Michigan, Ann Arbor, MI 48109, USA\\
\authormark{2}Quantum Photonics Laboratory, Department of Physics \& Astronomy, Louisiana State University, Baton Rouge, LA 70803, USA\\
\authormark{3}Tampere University, Photonics Laboratory, Physics Unit, Tampere, FI-33720, Finland\\
\authormark{4}Instituto de Ciencias Nucleares, Universidad Nacional Aut\'onoma de M\'exico, Apartado Postal 70-543, 04510 Cd. Mx., M\'exico\\
\authormark{5}ICFO - Institut de Ciencies Fotoniques, Mediterranean Technology Park, 08860 Castelldefels (Barcelona), Spain\\
\authormark{6}Department of Signal Theory and Communications, Campus Nord D3, Universitat Politecnica de Catalunya, 08034 Barcelona, Spain\\
\authormark{7}Oak Ridge National Laboratory, Oak Ridge, Tennessee USA\\
\authormark{8}Institute of Physics, Swiss Federal Institute of Technology Lausanne (EPFL), Lausanne, Switzerland\\
\authormark{9}Ming Hsieh Department of Electrical and Computer Engineering \& Department of Physics and Astronomy, University of Southern California, CA 90089, USA
}

\email{\authormark{*}zszh@umich.edu}



\begin{abstract}
Entanglement is a quintessential quantum mechanical phenomenon with no classical equivalent. First discussed by Einstein, Podolsky, and Rosen and formally introduced by Schrödinger in 1935, entanglement has grown from a scientific debate to a radically new resource that sparks a technological revolution. This review focuses on the fundamentals and recent advances in entanglement-based quantum information technology (QIT), specifically in photonic systems. Photons are unique quantum information carriers with several advantages, such as their ability to operate at room temperature, their compatibility with existing communication and sensing infrastructures, and the availability of readily accessible optical components. Photons also interface well with other solid-state quantum platforms. We will first provide an overview on entanglement, starting with an introduction to its development from a historical perspective followed by the theory for entanglement generation and the associated representative experiments. We will then dive into the applications of entanglement-based QIT for sensing, imaging, spectroscopy, data processing, and communication. Before closing, we will present an outlook for the architecture of the next-generation entanglement-based QIT and its prospective applications.
\end{abstract}

\tableofcontents

\section{Introduction}
Quantum information technology (QIT) is revolutionizing the way information is gathered, encoded, manipulated, transmitted, stored, and decoded. Among the various genres of quantum-information carriers, photons feature robustness at room temperature, compatibility with existing communication and sensing infrastructure, and the availability of optical devices. Over the past three decades, photonic QIT has led to a range of capabilities in communication, sensing, and computing that surpass classical information technology.

Entanglement, a unique quantum mechanical phenomenon, has become a valuable resource without any classical equivalent. The nonlocal and strong correlations present in entangled objects are the backbone of various QIT protocols. Photonic entanglement can now be routinely generated, processed, and measured in quantum optics platforms in laboratory settings. With the advent of integrated quantum photonics platforms compatible with the established semiconductor processes, the path is open for photonic entanglement to have a wide impact across various applications, including sensing, imaging, spectroscopy, communication, networking, and computing.

This review article is dedicated to introducing the foundations of entanglement-based QIT and reviewing its enabled applications. The article is structured as follows. Sec.~\ref{sec: foundation} will focus on the foundation for entanglement-based QIT. Specifically, Sec.~\ref{subsec: QIT_classes} will classify entanglement-based QIT protocols into three major categories and compare them to their classical counterparts from an architectural perspective. This classification sets the stage for a discussion of QIT applications, as the sensing, imaging, spectroscopy, data-processing, and communication protocols discussed in later sections fall into these three classes. Next, Sec.~\ref{subsec: entanglement_intro} will delve into the history, theoretical foundation, and experimental generation of entanglement. The following seven sections will explore the applications of entanglement-based QIT. Sec.~\ref{sec: quantum_metrology} will discuss entanglement-based quantum metrology, first introducing the Fisher information formalism (Sec.~\ref{subsec: QFI}) followed by quantum-metrology protocols based on N00N states (Sec.~\ref{subsec: N00N}) and squeezed states (Sec.~\ref{subsec: squeezed}), which are closely linked to entanglement (see Sec.~\ref{subsec: entanglement_intro}). The subsequent sections will cover various entanglement-based QIT applications for sensing and imaging, including quantum illumination (Sec.~\ref{sec: quantum_illumination}), quantum imaging (Sec.~\ref{sec: quantum_imaging}), light-matter interactions and spectroscopy (Sec.~\ref{sec: spectroscopy}), and distributed quantum sensing (Sec.~\ref{sec: distributed_quantum_sensing}). Sec.~\ref{sec: quantum_machine_learning} will focus on emergent entanglement-based quantum machine learning protocols and platforms, focusing on recent advances in photonic quantum machine learning built on noisy intermediate-scale quantum (NISQ) hardware platforms for classical (Sec.~\ref{subsec: quantum_machine_learning_classical_processing}) and quantum processing (Sec.~\ref{subsec: quantum_machine_learning_quantum_processing}). Sec.~\ref{sec: EACOMM} will examine two types of entanglement-assisted communication protocols based on discrete variables (Sec.~\ref{subsec: EACOMM_DV}) and continuous variables (Sec.~\ref{subsec: EACOMM_CV}). Before closing, we will look ahead to the future architecture of entanglement-based QIT, based on fast-advancing quantum modules such as quantum memories, quantum transducers, and quantum gates, and explore its potential capabilities and applications.

\section{Foundation for Entanglement-Based Quantum Information Technology}
\label{sec: foundation}
This section will first elucidate the theoretical and experimental foundation for entanglement-based QIT from an architectural perspective, describing three classes of QIT protocols and comparing them with their classical counterparts. Next, we will formally define entanglement, starting with a brief review on the history of entanglement, from its discovery to the verification experiments. We will then articulate the theory for the generation of two important genres of entanglement widely employed in QIT, namely the continuous-variable entanglement and the discrete-variable entanglement, followed by an introduction to representative experiments for the generation of the two genres of entangled states. Before closing this section, we will introduce recent advances in entanglement generation in integrated platforms as a route toward scalable entanglement-based QIT.

\subsection{Classical versus Entanglement-Based Protocols: an Architectural Perspective}
\label{subsec: QIT_classes}

\begin{figure}[!ht]
    \centering
    \includegraphics[width=1\textwidth]{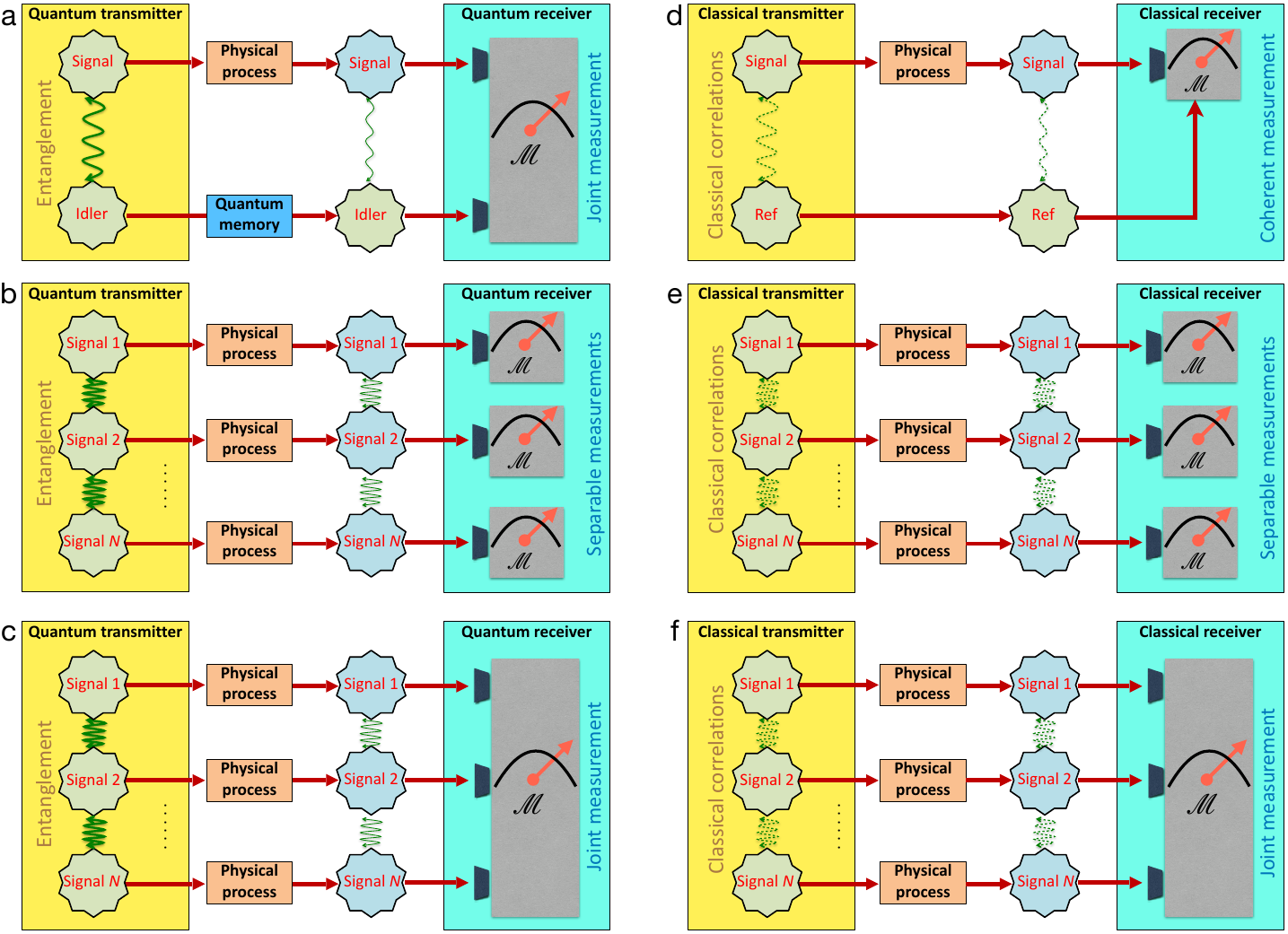}
    \caption{\label{fig:QIT_class}
    Three classes of entanglement-based QIT protocols and their classical counterparts. (a) Class 1: signal and idler are entangled at the transmitter. The signal undergoes a physical process while the idler is retained in a quantum memory. A quantum receiver takes a joint measurement on the signal and idler. For the classical counterpart in (d), the signal shares classical correlations with a reference at the transmitter, and they are measured by a classical receiver that conducts a coherent measurement. (b) Class 2: multiple signals are entangled at the transmitter. The signals undergo physical processes and are then measured by a quantum receiver that performs separable measurements on the signals. For the classical counterpart in (e), signals at the transmitter share classical correlations. (c) Class 3: multiple signals are entangled at the transmitter. The signals undergo physical processes and are then measured by a quantum receiver that performs a joint measurement on the signals. For the classical counterpart in (f), signals share classical correlations at the transmitter. In all figures, thick solid green curve: entanglement; thin solid green curve: entanglement or classical correlations; thin dashed green curve: classical correlations.}
\end{figure}

Entanglement is a quantum-mechanical property, with no classical description, shared by two or multiple objects. Entanglement gives rise to a stronger-than-classical nonlocal correlation, which can serve as a powerful resource for achieving capabilities that surpass the limitations of classical physics. The left panel of Fig.~\ref{fig:QIT_class} illustrates the architectures for three classes of entanglement-based QIT protocols with increasing technological complexity, categorized by how entanglement is utilized and measurements are performed. The classical counterparts for these QIT protocols are depicted in the right panel as a comparison.

In Class 1 protocols, depicted in Fig.~\ref{fig:QIT_class}a, a quantum transmitter prepares entangled signal and idler particles and sends the signal through a physical process such as a communication channel or a sensor module. The idler, sometimes referred to as the ancilla, is stored in a quantum memory without undergoing any alteration. A quantum receiver combines the sigal and idler to perform a joint measurement and capture the information carried on the signal. QIT protocols falling into this class include quantum illumination~\cite{lloyd2008enhanced, tan2008quantum, zhuang2017optimum, shapiro2020quantum, torrome2020introduction, zhuang2022ultimate}, entanglement-assisted communication~\cite{bennett2002entanglement, shi2020practical, zhuang2021quantum, hao2021entanglement}, and entanglement-assisted absorption spectroscopy~\cite{shi2020entanglement}.

In Class 2 protocols, summarized in Fig.~\ref{fig:QIT_class}b, a quantum transmitter generates multiple signals that are in an entangled state. Each signal undergoes a distinct physical process and is individually measured at a quantum receiver. This class of QIT protocols encompasses various applications, including quantum metrology based on multipartite entanglement~\cite{giovannetti2006quantum, giovannetti2011advances}, distributed quantum sensing~\cite{ge2018distributed, zhuang2018distributed, eldredge2018optimal, guo2020distributed, xia2020demonstration, zhang2021dqs}, quantum machine learning for distributed data processing~\cite{xia2021quantum}, and quantum secret sharing~\cite{tittel2001experimental, gottesman2000theory, hillery1999quantum, xiao2004efficient, zhang2005multiparty}.

Class 3 protocols, as illustrated in Fig.~\ref{fig:QIT_class}c, are similar to the second class, with the distinction that a joint measurement is performed on the signals at the quantum receiver. QIT protocols falling into this class include joint encoding in quantum communication to achieve superadditivity~\cite{smith2008quantum, hastings2009superadditivity, smith2011quantum, zhu2017superadditivity, zhu2018superadditivity}, quantum machine learning with joint processing~\cite{zhuang2019physical}, entangled two-photon absorption spectroscopy~\cite{schlawin2018}, and quantum error correction~\cite{noh2020encoding}. Table 1 lists the QIT protocols belonging to each of the three classes. 

\begin{table}[h!]
\begin{center}
\begin{tabular}{ |c||p{5cm}|p{5cm}|  }
 \hline
 \multicolumn{3}{|c|}{Table 1: List of entanglement-based quantum information technology protocols} \\
 \hline
 Class 1 &  
 \begin{itemize}
     \item Quantum metrology using entangled states (Fig.~\ref{Fig:metrology_intro}, Sec.~\ref{sec: quantum_metrology})
     \item Quantum illumination and covert sensing (Fig.~\ref{Fig:QI}, Sec.~\ref{sec: quantum_illumination})
 \end{itemize} & 
  \begin{itemize}
    \item Entanglement-assisted communication (Fig.~\ref{Fig:EACOMM_concept}, Sec.~\ref{sec: EACOMM})
    \item Quantum imaging (Fig.~\ref{fig:Corr_Imaging}a and Fig.~\ref{fig:NL_imaging}, Sec.~\ref{sec: quantum_imaging})
  \end{itemize}\\
 \hline
 Class 2 &
 \begin{itemize}
    \item Quantum key distribution~\cite{pirandola2020advances}
    \item Entanglement-enhanced light-matter interactions and spectroscopy (Fig.~\ref{Fig:ETPA_0}, Sec.~\ref{sec: spectroscopy})
 \end{itemize} & 
 \begin{itemize}
    \item Distributed quantum sensing (Fig.~\ref{fig:DQS_concept}, Sec.~\ref{sec: distributed_quantum_sensing})
    \item Quantum machine learning with distributed processing (Fig.~\ref{fig:QML_concept}, Sec.~\ref{sec: quantum_machine_learning})
 \end{itemize}\\
 \hline
 Class 3 &
 \begin{itemize}
    \item Quantum machine learning with joint processing (Fig.~\ref{fig:QML_concept}, Sec.~\ref{sec: quantum_machine_learning})
    \item Hybrid quantum classical communication~\cite{wilde2012quantum}
 \end{itemize} & 
 \begin{itemize}
    \item Quantum error correction~\cite{lidar2013quantum,terhal2015quantum}
 \end{itemize}\\
 \hline
\end{tabular}
\end{center}
\end{table}

\subsection{Entanglement in a Nutshell: a Brief Chronicle, Basic Theory, and Experiments}
\label{subsec: entanglement_intro}

The effect of entanglement was discussed in the seminal paper by Einstein, Podolsky, and Rosen (EPR) in 1935 with the aim of showing that the quantum mechanical description of correlated systems was seemingly incomplete~\cite{einstein1935can}, pending the introduction of hidden-variable models. Later, Schrödinger formally introduced the concept of entanglement in his response to EPR~\cite{schrodinger1935discussion} to describe systems whose correlation cannot be described by classical physics, i.e., it is a nonclassical correlation. Even back then, Schrödinger described entanglement as "not one but rather the characteristic trait of quantum mechanics", emphasizing the important difference from classically known correlations. Later, Bohm and Aharonov built upon the idea of the original EPR argument and reformulated it using analog considerations for spin-1/2 particles, which are a prime example of two-level systems, i.e., qubits~\cite{bohm1957discussion}. However, it was not until 1964 when Bell translated the mostly philosophical discussion into a setting where experiments could show a clear difference between classical correlations and their quantum counterparts~\cite{bell1964einstein}, that the testing of the counter-intuitive behavior of entangled systems incentivized experimentalists.

The subsequent seminal work by Clauser, Horne, Shimony, and Holt (CHSH) presented a bound, now called the CHSH inequality, for local hidden-variable theories that could be tested using experimentally tangible polarization-entangled photons. Pioneering experiments conducted between the 1970s and 1980s~\cite{freedman1972experimental, aspect1981experimental, aspect1982experimentalrealization, aspect1982experimental} suggested that local hidden-variable theories were incompatible with the observed correlations embedded in bipartite entanglement, subject to several loopholes that needed to be addressed~\cite{larsson2014loopholes}.

In the early 1990s, more intriguing properties of entanglement were unveiled as studies incorporated multipartite systems. Notably, Mermin showed that tripartite entanglement in the Greenberger-Horne-Zeilinger (GHZ) state exhibited nonclassical correlations that could be verified without invoking inequalities~\cite{mermin1990quantum}. Then, in the late 1990s, the GHZ state was experimentally prepared~\cite{bouwmeester1999observation} and employed to demonstrate the unique quantum correlations carried by the tripartite state~\cite{pan2000experimental}. Moving into the 21st century, with pivotal technological advances in entangled-photon sources~\cite{kim2006phase, evans2010bright, bernien2013heralded} and single-photon detectors~\cite{marsili2013detecting}, tests of Bell's inequalities culminated in a series of unambiguous loophole-free experiments~\cite{shalm2015strong, giustina2015significant, hensen2015loophole,bigbelltest}.

The current focus of research on entanglement has shifted from its scientific roots to the development of entanglement-based protocols, in tandem with the experimental toolbox, to harness it as a resource for QIT. In recognition of the remarkable trajectory of entanglement from fundamental studies to the onset of a revolutionary technological era, the 2022 Nobel Prize in Physics was awarded to John Clauser, Alain Aspect, and Anton Zeilinger for their pioneering research on entangled quantum states. To gain a more in-depth chronological review on the theoretical and experimental developments pertaining to entanglement, readers may refer to a recent comprehensive review article~\cite{paneru2020entanglement}.

\subsubsection{Theory for Entanglement Generation via Spontaneous Parametric Down-Conversion}
\label{sec:entanglement_generation_theory}
In general, entanglement is a type of correlation between two or more subsystems of a composite quantum system. The $n$ subsystems are denoted by indices $k=1, 2, \cdots, n$ ($n\ge2$). A state is considered entangled if it cannot be written as a probabilistic mixture of separable states~\cite{horodecki2009quantum}, i.e.,
\begin{align}
\label{eq: separable_states}
\rho\neq \sum_i p_i \rho_1^i \otimes \rho_2^i \otimes \cdots \otimes \rho_n^i,
\end{align}
where $\rho_k^i$ is the density matrix for subsystem $k$, and $p_i$ is the probability for $\rho_1^i \otimes \rho_2^i \otimes \cdots \otimes \rho_n^i$ to occur. Here, a subsystem can refer to a photon in a system of multiple photons, an optical mode in a multimode state, or sometimes different degrees of freedom for the same photon. When all states are pure, the global state is considered entangled if
\begin{align}
\label{eq: separable_states1}
|\psi\rangle \neq  |\psi\rangle_1 \otimes |\psi\rangle_2 \otimes \cdots \otimes |\psi\rangle_n.
\end{align}

Quantifiers of entanglement can be derived from the framework of the quantum resource theory~\cite{chitambar2019quantum}. For bipartite pure states $\psi_{12}$, the entanglement entropy is given by
\begin{align}
E(\psi_{12})=S(\rho_1)=S(\rho_2),
\end{align}
where $\rho_1$ and $\rho_2$ are the reduced states of subsystems 1 and 2 for the joint pure state $\psi_{12}$. The quantification of entanglement in mixed states or multipartite systems is more involved. Later in this section, we will discuss the entanglement criteria for a specific class of continuous-variable quantum states. Other entanglement quantifiers are beyond the scope of this review. For further details on technical aspects of entanglement, readers may refer to review articles in this area~\cite{horodecki2009quantum,paneru2020entanglement}.

We will now address the problem of entanglement generation. In the field of QIT, two main classes of entanglement have been extensively exploited: discrete-variable entanglement based on the properties of single photons, and continuous-variable entanglement built on the field properties of optical modes. We will first present a theoretical framework that unifies the generation of these two classes of entangled states then discuss the processes commonly used in QIT to produce bipartite entangled states, with the understanding that significant progress has been made in generating both continuous-variable~\cite{asavanant2019generation, chen2014experimental, larsen2019deterministic} and discrete-variable~\cite{schwartz2016deterministic, erhard2020advances} multipartite entangled states. These multipartite entangled states form the basis for one-way optical quantum computing~\cite{wu2020quantum, menicucci2008one, menicucci2006universal, alexander2018universal, alexander2016one, humphreys2014continuous}, which is beyond the scope of this review article.

One well-developed approach for generating optical entanglement is through the process of spontaneous parametric down-conversion (SPDC). In SPDC, a small portion of the higher-energy pump photons is converted into pairs of lower-energy photons, known as the signal and idler photons, due to the nonlinearity of an optical material such as lithium niobate (LiNbO$_3$), potassium titanyl phosphate (KTP), or beta barium borate (BBO), among others. We will first consider non-degenerate SPDC, where the signal and idler photons are distinguished by their physical properties such as their wavelengths, polarization, or spatial modes. For the sake of simplicity, we will adopt a single-mode interaction model that describes the signal and idler fields using annihilation operators $\hat{a}_S$ and $\hat{a}_I$, respectively. A complete theoretical treatment of multimode SPDC processes in continuous time-frequency space can be found in Refs.~\cite{walls2008quantum,shapiro20166}.

The Hamiltonian associated with the SPDC process is given by
\begin{align}
\label{eq:SPDC_Hamiltonian}
\hat{H} &= \hbar\omega_S \hat{a}^\dag_S\hat{a}_S + \hbar\omega_I \hat{a}^\dag_I \hat{a}_I + i\hbar\chi\left(\hat{a}_S^\dag \hat{a}_I^\dag e^{-i\omega_P t} - \hat{a}_S\hat{a}_I e^{i\omega_P t}\right)\notag\\
&= \hat H_0 + \hat H_I,
\end{align}
where $H_0 = \hbar\omega_S \hat{a}^\dag_S\hat{a}_S + \hbar\omega_I \hat{a}^\dag_I \hat{a}_I$, $\omega_S$ ($\omega_I$) is the angular frequency of the signal (idler) photon, $\chi$ represents the strength of the nonlinear interaction, which is determined by the nonlinearity of the material, the magnitude of the pump field, and the power density of the fields. Here, $\omega_P = \omega_S + \omega_I$ is the angular frequency of the pump photon, dictated by energy conservation. In writing the Hamiltonian, we assume that the pump is intense enough to be treated as a classical field. The term $\hat{a}_S^\dag \hat{a}_I^\dag$ describes the process of a pair of signal and idler photons being created by annihilating a pump photon, while the term $\hat{a}_S \hat{a}_I$ corresponds to the inverse process.

The dynamics of the system can be derived in the interaction picture using the following Heisenberg equations of motion~\cite{walls2008quantum}
\begin{align}
\frac{d \hat{a}_S}{dt} &= \frac{i}{\hbar}\left[\hat{H}_I, \hat{a}_S\right],\notag\\
\frac{d \hat{a}_I}{dt} &= \frac{i}{\hbar}\left[\hat{H}_I, \hat{a}_I\right].
\end{align}

Expanding the commutators above leads to the following two coupled operator equations:
\begin{align}
\frac{d \hat{a}_S}{dt} &= \chi \hat{a}^\dag_I,\notag\\
\frac{d \hat{a}_I}{dt} &= \chi \hat{a}^\dag_S.
\end{align}
Upon solving the equations, the dynamics of the annihilation operators are given by
\begin{align}
\label{eq:dynamics_nondegenSPDC}
\hat{a}_S &= \sqrt{g} \hat{a}_{S_0} + \sqrt{g-1} \hat{a}_{I_0}^\dag,\notag\\
\hat{a}_I &= \sqrt{g} \hat{a}_{I_0} + \sqrt{g-1} \hat{a}_{S_0}^\dag,
\end{align}
where $g \equiv \cosh^2 \chi t$. Here, $\hat{a}_{S_0}$ and $\hat{a}_{I_0}$ represent the initial annihilation operators prior to the interaction. In the context of SPDC, the initial modes are in the vacuum state. The parameter $g$ is commonly known as the gain of the parametric process and can be determined experimentally.

To further characterize the process, we define the mean photon number as $N_S \equiv \langle \hat a_S^\dag \hat a_S\rangle = g-1$. This quantity represents the average number of photons in the output mode $\hat{a}_S$ and is indicative of the degree of amplification achieved.

\begin{figure}
\centering
\includegraphics[width=\textwidth]{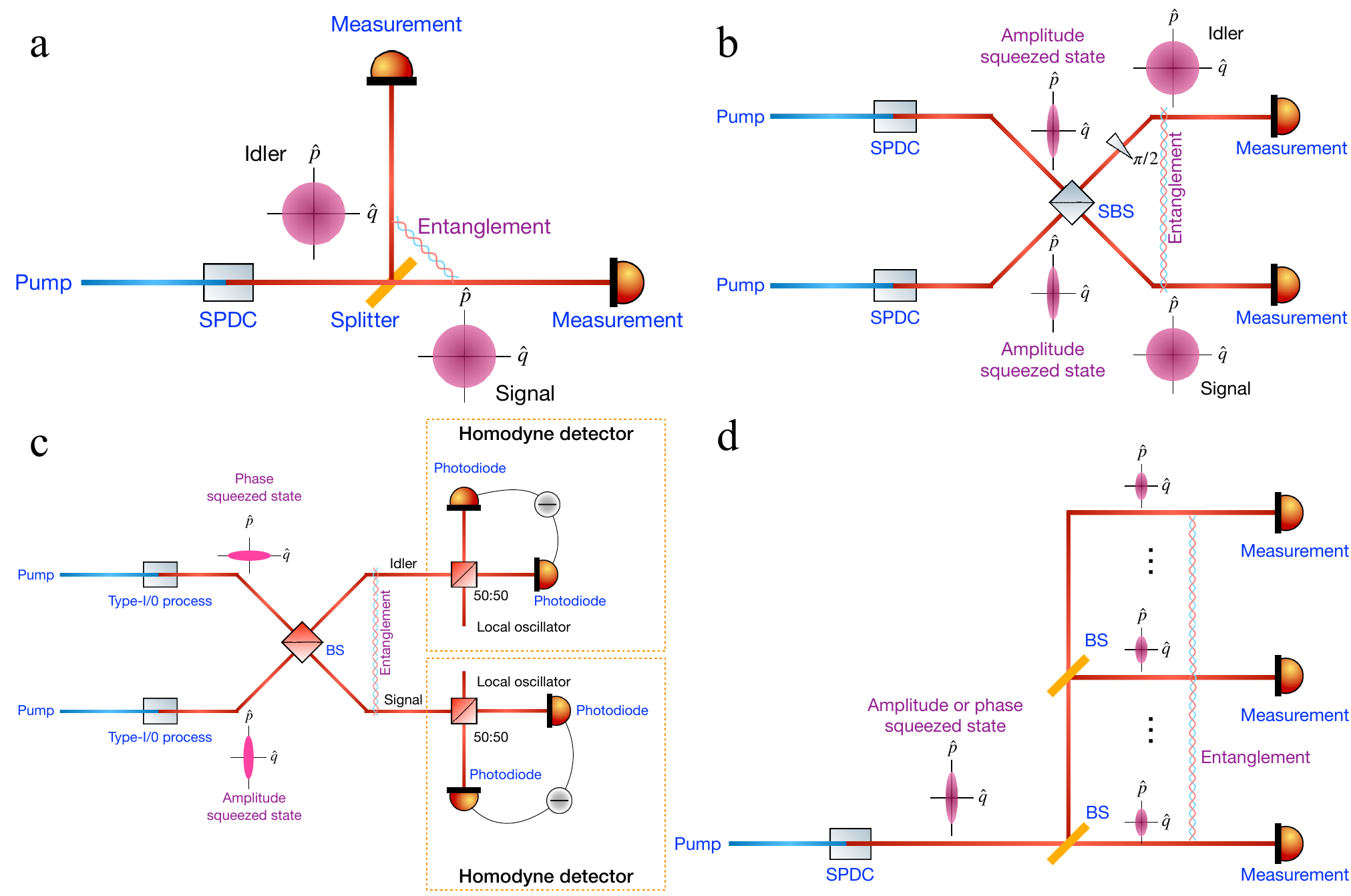}
\caption{Four types of sources for continuous-variable entanglement. The marginal states at the two arms are depicted in the phase space in purple. (a) Entanglement produced from a non-degenerate process. (b) Entanglement produced by combining two amplitude-squeezed states on a symmetric beam splitter with a transfer matrix given by Eq.~\eqref{eq:BS_Transformation_sym}. (c) Entanglement produced by combining an amplitude and a phase-squeezed state on a beam splitter with a transfer matrix given by Eq.~\eqref{eq:BS_transfer}. (d) Entanglement produced by splitting a single-mode squeezed state.}
\label{fig:CV_entanglement_sources}
\end{figure}

We proceed by defining the quadrature operators as follows:
\begin{align}
\hat{q} &\equiv \frac{1}{2}\left(\hat{a} + \hat{a}^\dag\right),\notag\\
\hat{p} &\equiv \frac{1}{2i}\left(\hat{a} - \hat{a}^\dag\right).
\end{align}
Here, $\hat{q}$ ($\hat{p}$) represents the real (imaginary) part of the annihilation operator and is often referred to as the amplitude (phase) quadrature. These operators analogously correspond to the sine and cosine components of a classical field relative to a phase reference. Experimentally, these quadratures can be measured through homodyne detection, where the output field is interfered with a strong reference field, commonly known as the local oscillator, as depicted in Fig.~\ref{fig:CV_entanglement_sources}a. For classical states, such as the initial vacuum states of the signal and idler modes, the variance of the quadrature measurements satisfies $\langle\Delta \hat{q}^2\rangle \geq 1/4$ and $\langle\Delta \hat{p}^2\rangle \geq 1/4$. A state is considered nonclassical if the variance of either of its quadratures falls below the classical limit of $1/4$.

Now, let us consider two modes, $\hat{a}_1$ and $\hat{a}_2$, in a classical state. It can be shown that $\langle \Delta (\hat{q}_1 - \hat{q}_2)^2 \rangle + \langle \Delta (\hat{p}_1 + \hat{p}_2)^2 \rangle \geq 1$. The Duan-Giedke-Cirac-Zoller inseparability criterion, established in Ref.~\cite{duan2000inseparability}, states that if $\langle \Delta (\hat{q}_1 - \hat{q}_2)^2 \rangle + \langle \Delta (\hat{p}_1 + \hat{p}_2)^2\rangle < 1$, the two modes are entangled. This criterion plays a crucial role in entanglement detection.

Applying this criterion to the two output modes, $\hat{a}_S$ and $\hat{a}_I$, from the SPDC process, it can be derived that $\langle \Delta (\hat{q}_S - \hat{q}_I)^2 \rangle + \langle \Delta (\hat{p}_S + \hat{p}_I)^2\rangle = (\sqrt{g}-\sqrt{g-1})^2 < 1$ for any gain $g\neq 1$. Hence, it confirms that the signal and idler modes obtained from SPDC are entangled. This entangled state is known as the two-mode squeezed vacuum (TMSV) state or the EPR state \cite{einstein1935can}. It serves as a fundamental resource for numerous continuous-variable QIT protocols.

We next consider degenerate SPDC in which the signal and idler modes are indistinguishable in all physical degrees of freedom. Mathematically, we let $\hat{a}_S = \hat{a}_I$ and $\hat{a}_{S_0} = \hat{a}_{I_0}$ and transform Eq.~(\ref{eq:dynamics_nondegenSPDC}) into
\begin{equation}
    \label{eq:dynamics_degenSPDC}
    \hat{a}_S = \sqrt{g}\hat{a}_{S_0} + \sqrt{g-1}\hat{a}_{S_0}^\dag.
\end{equation}
The quantum state associated with $\hat{a}_S$ is known as the single-mode squeezed vacuum state, with $\langle\Delta \hat{p}_S^2 \rangle = (\sqrt{g} - \sqrt{g-1})^2/4$ being the variance of the squeezed quadrature below the classical limit, and $\langle\Delta \hat{q}_S^2 \rangle = (\sqrt{g} + \sqrt{g-1})^2/4$ being the variance of the anti-squeezed quadrature. We next suppose that a second single-mode state associated with the annihilation operator $\hat{a}_S'$ is squeezed in its $\hat{p}_S'$ quadrature. As Fig.~\ref{fig:CV_entanglement_sources}b illustrates, mixing $\hat{a}_S$ and $\hat{a}_S'$ on a  symmetric beam splitter~\cite{loudon2000quantum} characterized by the transfer matrix
\begin{equation}
\label{eq:BS_Transformation_sym}
    \hat{U} = \frac{1}{\sqrt{2}}\begin{bmatrix}
1 & i \\
i & 1 
\end{bmatrix},
\end{equation}
followed by applying a $\pi/2$ phase shift on one of the output arms yields the two output modes
\begin{align}
    \hat{a} &= \frac{1}{\sqrt{2}}\left(-\hat{a}_S'+i\hat{a}_S\right),\notag\\
    \hat{b} &= \frac{1}{\sqrt{2}}\left(\hat{a}_S'+i\hat{a}_S\right).
\end{align}
One then obtains
\begin{align}
    \hat{p}_a &= \frac{1}{\sqrt{2}}\left(-\hat{p}_S' - \hat{q}_S\right)\notag\\
    \hat{q}_a &= \frac{1}{\sqrt{2}}\left(-\hat{q}_S' + \hat{p}_S\right)\notag\\
    \hat{p}_b &= \frac{1}{\sqrt{2}}\left(\hat{p}_S'- \hat{q}_S\right)\notag\\
    \hat{q}_b &= \frac{1}{\sqrt{2}}\left(\hat{q}_S'+ \hat{p}_S\right).
\end{align}
It would not be difficult to verify that $\langle \Delta (\hat{p}_a - \hat{p}_b)^2 \rangle + \langle \Delta (\hat{q}_a + \hat{q}_b)^2 \rangle < 1$, therefore proving $\hat{a}$ and $\hat{b}$ are entangled. The correlations embedded in the quadratures of $\hat{a}$ and $\hat{b}$ are very similar to those in $\hat{a}_S$ and $\hat{a}_I$, suggesting that $\hat{a}$ and $\hat{b}$ are also in a TMSV state.

The TMSV state can also be generated by interfering two modes squeezed, respectively, in the amplitude and phase quadratures as sketched in Fig.~\ref{fig:CV_entanglement_sources}c. To show this, let us consider $\hat{a}_S$ squeezed in the amplitude quadrature with $\langle \Delta \hat{q}_S^2 \rangle < 1/4$ and $\hat{a}_S'$ squeezed in the phase quadrature. Mixing them on a beam splitter with the transfer matrix of
\begin{equation}
    \label{eq:BS_transfer}
    \hat{U} = \frac{1}{\sqrt{2}}\begin{bmatrix}
1 & 1 \\
1 & -1 
\end{bmatrix},
\end{equation}
yields the following output modes:
\begin{align}
    \hat{a} &= \frac{1}{\sqrt{2}}\left(\hat{a}_S + \hat{a}_S'\right),\notag\\
    \hat{b} &= \frac{1}{\sqrt{2}}\left(\hat{a}_S-\hat{a}_S'\right).
\end{align}
A quick derivation gives
\begin{align}
    \hat{p}_a &= \frac{1}{\sqrt{2}}\left(\hat{p}_S + \hat{p}_S'\right),\notag\\
    \hat{q}_a &= \frac{1}{\sqrt{2}}\left(\hat{q}_S + \hat{p}_S'\right),\notag\\
    \hat{p}_b &= \frac{1}{\sqrt{2}}\left(\hat{p}_S - \hat{p}_S'\right),\notag\\
    \hat{q}_b &= \frac{1}{\sqrt{2}}\left(\hat{q}_S - \hat{p}_S'\right).
\end{align}
Hence, $\langle \Delta (\hat{p}_a - \hat{p}_b)^2 \rangle + \langle \Delta (\hat{q}_a + \hat{q}_b)^2 \rangle < 1$, proving the state is entangled.

A different type of continuous-variable entangled state, often referred to as the multimode squeezed state or the v-class entanglement, can be generated using a series of beam splitters to divert a single-mode squeezed vacuum state into multiple modes $\hat{a}_{S_1}, \hat{a}_{S_2}, ..., \hat{a}_{S_N}$, as sketched in Fig.~\ref{fig:CV_entanglement_sources}d. One can readily show that this state is entangled by using the Fisher information approach~\cite{qin2019characterizing}. 

The dynamics of non-degenerate SPDC can also be solved in the Schr\"{o}dinger picture. The initial quantum states are vacuum for both the signal and idler, denoted as $|0\rangle_S|0\rangle_I$. The evolution of the quantum state is given by~\cite{walls2008quantum}
\begin{equation}
    |\psi(t)\rangle = \exp\left[\chi t \left(\hat{a}_S^\dag \hat{a}_I^\dag -\hat{a}_S\hat{a}_I\right)\right]|0\rangle_S|0\rangle_I.
\end{equation}
The output state of SPDC, i.e., the TMSV state, expressed in the number basis reads
\begin{equation}
\label{eq:TMSV}
    |\psi\rangle_{SI} = \sum_{n = 0}^{\infty} \sqrt{\frac{(g-1)^n}{g^{n+1}}}|n\rangle_S|n\rangle_I.
\end{equation}
One can verify that the TMSV state described by Eq.~\eqref{eq:TMSV} is inseparable, i.e., cannot be written in the form of Eq.~\eqref{eq: separable_states}. In the limit of $g-1 \ll 1$, i.e., with a weak pump, the only dominant non-vacuum contribution is from the single-pair state $|1\rangle_S|1\rangle_I = \hat a_S^\dag \hat a_I^\dag |0\rangle_S|0\rangle_I$ while all the multi-pair contributions with $n\geq 2$ can be neglected. This is the regime for the generation of photon pairs. In tandem with auxiliary components, one can produce discrete-variable entanglement in the polarization, time-energy, time bin, or path degree of freedom, as we will articulate in the next sections.

\subsubsection{Continuous-Variable Entanglement Sources}
\label{sec:CV_entanglement_sources}
As discussed in Section~\ref{sec:entanglement_generation_theory}, TMSV states can be generated through a non-degenerate SPDC process or by interfering two single-mode squeezed states. Both approaches have been pursued in experimental studies. The generation of continuous-variable EPR entanglement traces back to a pioneering experiment conducted with a type-II KTP crystal in an optical cavity~\cite{ou1992realization}, which demonstrated the EPR paradox~\cite{einstein1935can}. However, the quality of the produced entangled state was affected by dispersion among the three interacting modes inside the optical cavity.

\begin{figure}[hbt!]
\centering
\includegraphics[width=1\textwidth]{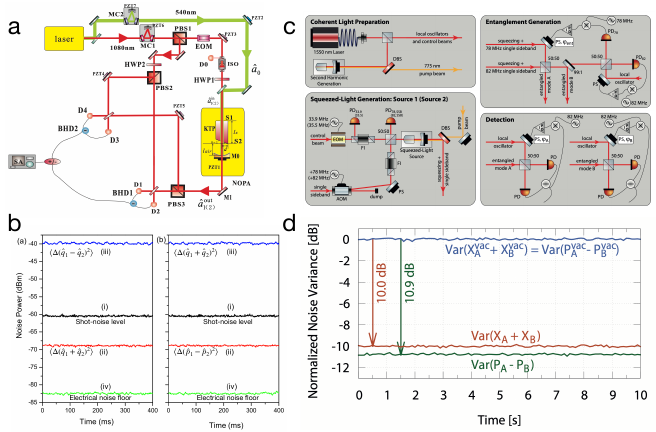}
\caption{\label{fig:TMSV}
Experiments to generate continuous-variable entanglement. (a) A source based on non-degenerate SPDC. The signal and idler modes are represented by two polarization modes at the output of the optical cavity, where a type-II PPKTP crystal resides. The signal and idler modes are separated by a PBS and then directed to two balanced homodyne detectors (BHDs) for characterization. MC: mode cleaner; PZT: piezo-electric transducer; EOM: electro-optic modulator; ISO: isolator; HWP: half-wave plate; BHD: balanced homodyne detectors; NOPA: non-degenerate optical parametric oscillator; M: mirror; PBS: polarizing beam splitter; D: detector; SA: spectrum analyzer. (b) Characterization data on the amplitude and phase quadratures of the entangled state from (a), showing that the variance of the sum of the amplitude quadrature $\langle \Delta (\hat{q}_1+\hat{q}_2)^2 \rangle$ and the variance of the difference of the phase quadrature $\langle \Delta (\hat{p}_1-\hat{p}_2)^2 \rangle$ both fall below the shot-noise level, thereby fulfilling the Duan-Giedke-Cirac-Zoller inseparability criterion~\cite{duan2000inseparability}. (c) A source based on interfering single-mode squeezed states produced from two independent degenerate SPDC processes. Phase-locked loops are implemented to stabilize the phase between the reference and the entangled modes. PD: photo diode; DBS: dichroic beam splitter; AOM: acousto-optic modulator; FI: Faraday Isolator; PS: phase shifter. (d) Characterization data showing that the variances for the sum of the amplitude quadrature $\hat{X}_A^{\rm vac} + \hat{X}_B^{\rm vac}$ and the difference of the phase quadrature $\hat{P}_A^{\rm vac} - \hat{P}_B^{\rm vac}$ fall below the corresponding vacuum noise levels by more than 10 dB. (a, b) reproduced from Ref.~\cite{zhou2015experimental}. (c, d) reprinted from Ref.~\cite{eberle2013stable}.}
\end{figure}

To overcome this challenge, more recent experiments have utilized wedged type-II periodically poled KTP (PPKTP) crystals to realize EPR sources, as shown in Fig.~\ref{fig:TMSV}a~\cite{zhou2015experimental,jinxia2017generation}. These experiments engineered the poling periods of the KTP crystals to tailor the wavelengths of the entangled beams at 1550 nm~\cite{zhou2015experimental} and 1080 nm~\cite{jinxia2017generation}. By adjusting the position of the PPKTP crystal inside the optical cavity, tri-resonance of the pump, signal, and idler modes was achieved, significantly enhancing the parametric interaction. As a result, correlations exceeding 8 dB below the classical limit were achieved for both the phase and amplitude quadratures in these experiments, as shown in Fig.~\ref{fig:TMSV}b.

The aforementioned experiments produced entangled states that exhibited quantum correlations in both the amplitude and phase quadratures. Another study by Ref.~\cite{laurat2005effects} explored the generation of entangled states with correlations in general quadratures using non-degenerate SPDC. The experiment introduced tunable coupling between the signal and idler modes in the optical cavity, enabling the generation of a more diverse set of continuous-variable entanglement.

Continuous-variable entanglement can also be generated by interfering two single-mode squeezed states from degenerate SPDC processes, as schematically sketched in Fig.~\ref{fig:CV_entanglement_sources}b and \ref{fig:CV_entanglement_sources}c. Such a configuration has been exploited to produce continuous-variable entanglement in both free-space~\cite{bowen2004experimental,eberle2013stable,furusawa1998unconditional} and optical fiber~\cite{silberhorn2001generation} platforms. In the experiment sketched in Fig.~\ref{fig:TMSV}c, the outputs from two independent single-mode squeezed-light sources interfered on a 50:50 beam splitter to produce a two-mode entangled state. The two entangled modes were characterized by two balanced homodyne detectors. A key ingredient in the experiment was the implementation of phase-locked loops, which involves creating and detecting the locking signals at a series of different RF frequencies. The measurement data presented in Fig.~\ref{fig:TMSV}d show that the variances of both the sum of the amplitude quadratures ($\hat{X}_A + \hat{X}_B$) and the difference of the phase quadratures ($\hat{P}_A - \hat{P}_B$) situate more than 10 dB below the vacuum noise levels ($\hat{X}_A^{\rm vac} + \hat{X}_B^{\rm vac}$ and $\hat{P}_A^{\rm vac} - \hat{P}_B^{\rm vac}$). This allowed for the validation of the Duan-Giedke-Cirac-Zoller inseparability criterion~\cite{duan2000inseparability} and the Reid EPR paradox criterion~\cite{reid1989demonstration}.

In optical fibers, the Kerr nonlinearity is leveraged, in lieu of SPDC, to generate continuous-variable entangled states. In an experiment reported in Ref.~\cite{silberhorn2001generation}, a pump was launched into a Sagnac loop in a polarization-maintaining fiber at approximately $45^{\circ}$ relative to the fiber axes, generating amplitude-squeezed states in the two orthogonal $p$ and $s$ polarizations. A 50:50 beam splitter with the transformation matrix presented in Eq.~\eqref{eq:BS_Transformation_sym} then interfered the two squeezed states to generate a two-mode entangled state.

\begin{figure}[!hbt]
\centering
\includegraphics[width=1\textwidth]{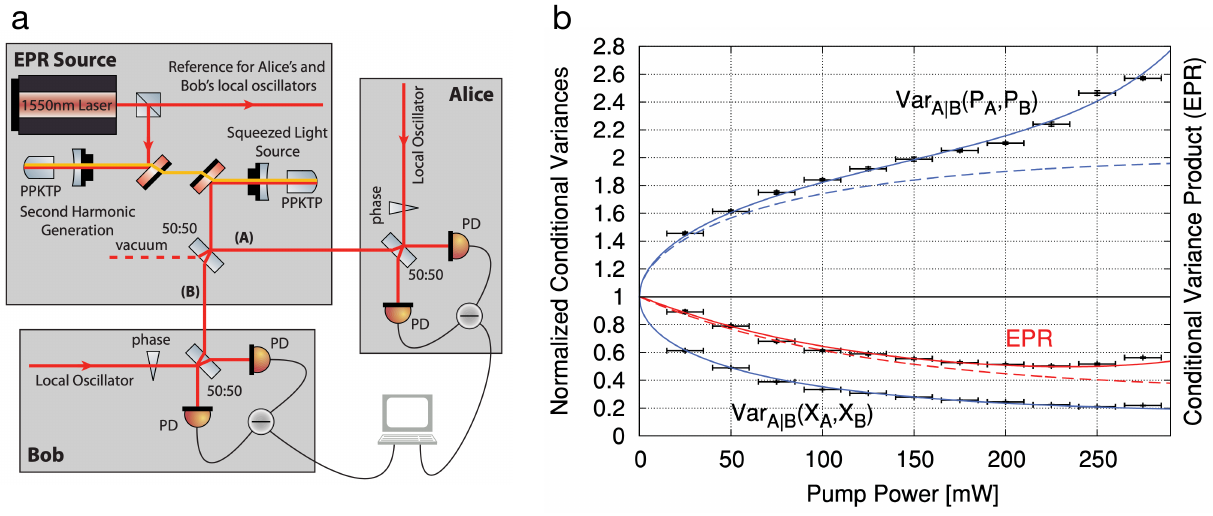}
\caption{\label{fig:distributed_squeezing}
Multimode squeezed-state source. (a) Experimental setup for the generation of a multimode squeezed state by splitting a single-mode squeezed state into two modes. PD: photodiode. (b) The conditional variances for the amplitude quadratures ${\rm Var_{A|B}}(X_A, X_B)$ and phase quadratures ${\rm Var_{A|B}}(P_A, P_B)$ are shown in blue. The red curves represent the Reid criterion for the EPR paradox, surpassing the classical limit of 1. Dashed curves: idealized theoretical model. Solid curves: theoretical model with additional excess noise factored in. Dots: experimental measurements. Figures reprinted from Ref.~\cite{eberle2011strong}.}
\end{figure}

Finally, as depicted in Fig.~\ref{fig:CV_entanglement_sources}d, multimode
squeezed state or v-class entanglement can be generated by splitting a single-mode squeezed state using a network of beam splitters. This type of entangled state has been experimentally verified~\cite{eberle2011strong} to satisfy the Reid criterion for the EPR paradox~\cite{reid1989demonstration}. In the experimental setup shown in Fig.~\ref{fig:distributed_squeezing}a, a single-mode squeezed state produced from a PPKTP crystal within an optical cavity is evenly split into two arms by a 50:50 beam splitter, resulting in the generation of two entangled modes. Balanced homodyne detectors, supplied with local oscillators as phase references, are used to measure the quadratures of the entangled modes. The measured conditional variances of the amplitude quadratures ${\rm Var_{A|B}}(X_A, X_B)$ and the phase quadratures ${\rm Var_{A|B}}(P_A, P_B)$ are plotted in Fig.~\ref{fig:distributed_squeezing}b. The calculated Reid criterion for the EPR paradox, given by ${\rm Var_{A|B}}(X_A, X_B)\times{\rm Var_{A|B}}(P_A, P_B)$, surpasses the classical limit of 1, thus confirming the entanglement between the two modes. More recently, the multimode squeezed state has found applications in entangled sensor networks, addressing various problems related to optical phase \cite{liu2021distributed,guo2020distributed,Zhao2021PRX,hong2021quantum}, radiofrequency \cite{xia2020demonstration}, optomechanical \cite{xia2022entanglement}, and data processing \cite{xia2021quantum}. For detailed information, we refer the readers to Sec.~\ref{sec: distributed_quantum_sensing} and Sec.~\ref{sec: quantum_machine_learning}.

\subsubsection{Polarization Entangled-photon Sources}
\label{sec: polarization_entanglement}
One of the most widely studied and utilized forms of discrete variable photonic entanglement is polarization entanglement. Photons entangled in their polarization degree of freedom have played a central role in numerous key experiments, ranging from the early stages of quantum information science \cite{freedman1972experimental,aspect1981experimental,aspect1982experimentalrealization,aspect1982experimental} to recent ground-satellite quantum communication testbeds \cite{yin2017satellite}. Polarization states can be easily manipulated, transmitted, and detected using readily available optical components with minimal disturbances, making them highly attractive for long-distance quantum communication.

The pioneering experiments to test Bell's inequalities \cite{freedman1972experimental,aspect1981experimental,aspect1982experimentalrealization,aspect1982experimental} utilized polarization entangled photons generated from atomic systems and measured using photomultiplier tubes. Subsequently, the use of SPDC in solid-state quantum optics platforms offered a promising approach to generate high-flux, wavelength-tailored, and high-fidelity polarization entangled photons. The non-collinear type-II SPDC process in a BBO crystal, for instance, enabled the construction of the first polarization entangled-photon source without the need for post-selection \cite{kwiat1995new}. Various geometries utilizing type-I SPDC processes in BBO crystals were later developed for polarization entanglement, leading to the generation of entangled states with more than 10 photons \cite{kwiat1999ultrabright,Wang2016PRL,zhong201812}. By employing quasi-phase-matching techniques, such as using LiNbO$_3$ or KTP crystals, polarization entangled photons spanning a wide spectral range can now be generated and then detected using state-of-the-art single-photon detectors.

\begin{figure}[hbt!]
\centering
\includegraphics[width=1\textwidth]{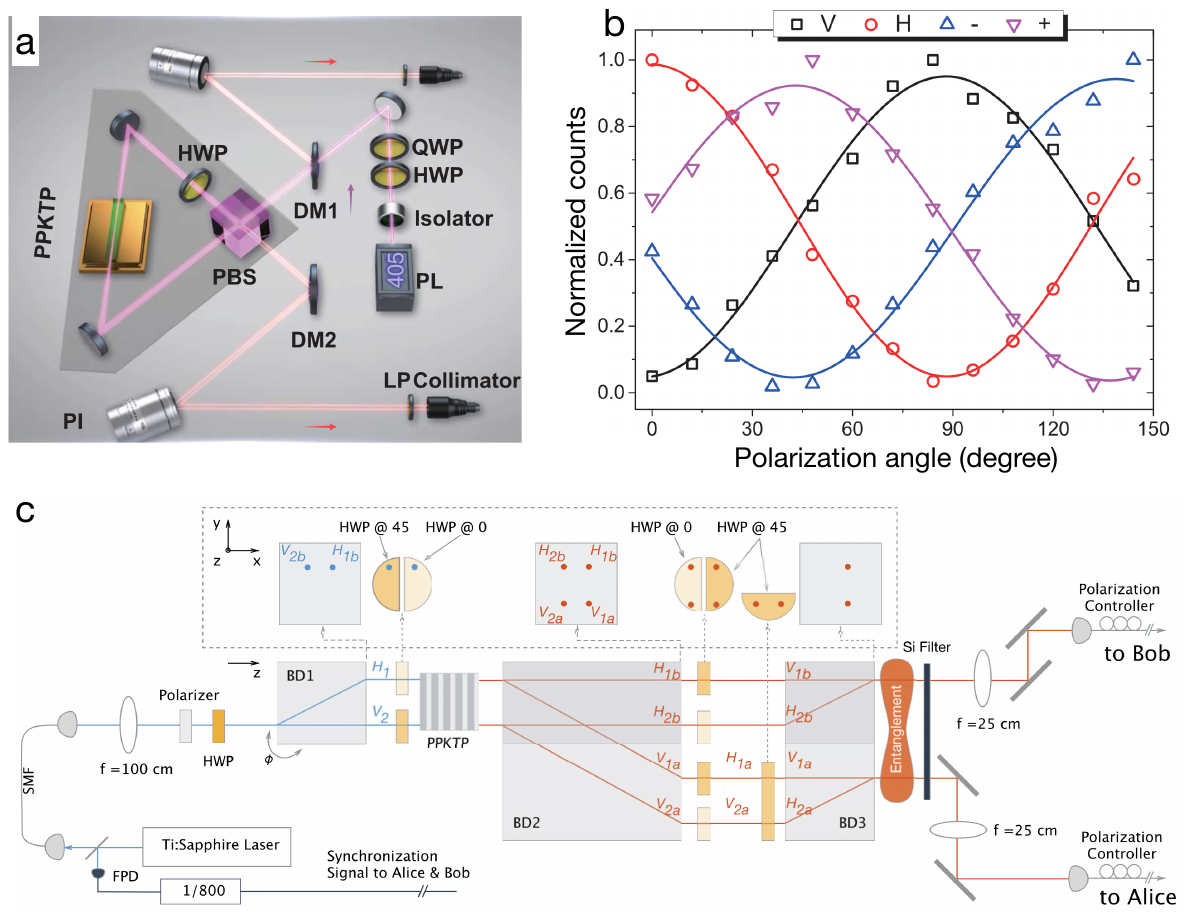}
\caption{\label{fig:polarization_entanglement}
Polarization entanglement sources. (a) Schematic of the polarization entanglement source in a Sagnac configuration. Such a source was deployed in the Micius quantum-communication satellite. HWP: half-wave plate; QWP: quarter-wave plate; PL: pump laser; PBS: polarizing beam splitter; DM: dichroic mirror; LP: long-pass filter. (b) Measurement data illustrating the correlation of polarization entangled photons on-satellite. (c) Schematic of the polarization entanglement source in a one-way configuration. (a, b) reprinted from Ref.~\cite{yin2017satellite}. (c) reprinted from Ref.~\cite{shalm2015strong}.}
\end{figure}

A particularly robust configuration for stable polarization entangled-photon sources is based on the Sagnac interferometer \cite{kim2006phase,fedrizzi2007wavelength}. Such sources have been extensively used in quantum optics and quantum information experiments. Notably, the polarization entangled-photon source deployed in the Micius quantum-communication satellite (Fig.~\ref{fig:polarization_entanglement}a) follows this scheme~\cite{yin2017satellite}: a type-II PPKTP crystal is pumped bidirectionally within a Sagnac loop, generating counter-propagating signal-idler photon pairs in the state $|H\rangle |V\rangle$. A half-wave plate rotates the polarization of the counterclockwise photon pairs by 90$^\circ$, transforming the state to $|V\rangle|H\rangle$. The counterclockwise and clockwise photons are then combined using a polarizing beam splitter. When a pair of photons is detected at the two output arms, it is impossible to distinguish whether they were generated by the clockwise or counterclockwise pump. As a result, the two quantum states must be coherently added, yielding
\begin{equation}
|\psi\rangle = \frac{1}{\sqrt{2}}\left(|H\rangle|V\rangle + e^{i\varphi}|V\rangle|H\rangle\right),
\end{equation}
where $\varphi$ represents the overall phase difference between the clockwise and counterclockwise paths, which remains stable over time. The photon correlations of the polarization entangled-photon source were characterized on-satellite, and the measurement data is presented in Fig.~\ref{fig:polarization_entanglement}b. To generate each curve, the polarization of the signal photons was fixed while sweeping the polarization of the idler photon during detection. The recorded normalized coincidence rates clearly exhibit a strong dependence on the polarizations of the signal and idler photons.

Apart from the Sagnac structure, an alternative configuration for generating polarization entangled photons is the one-way configuration first introduced in Ref.\cite{evans2010bright}. As shown in Fig.~\ref{fig:polarization_entanglement}c, a PPKTP crystal is pumped by two parallel beams from the same laser, which are then separated by a beam displacer (BD1). A second beam displacer (BD2) diverts the different polarization states from the upper and lower arms into four beams. A set of wave plates processes the polarizations, followed by a third beam displacer (BD3) that merges the four beams into the entangled signal and idler. By carefully designing the optical paths for the two combining beams in the upper or lower arm to be identical, the which-path information is effectively erased, resulting in high-quality polarization entangled photons. The one-way configuration offers robustness against environmental disturbances, leading to a measured visibility of coincidence detections exceeding 99.5\% for both the horizontal-vertical and diagonal-antidiagonal bases. The high quality of this source has recently been leveraged in the loophole-free test of Bell's inequalities~\cite{shalm2015strong}, along with another demonstration that relies on a Sagnac-like implementation \cite{giustina2015significant}.

\subsubsection{High-Dimensional Entangled Photons in Space and Time}
In addition to the polarization degree of freedom, SPDC processes have been utilized to generate photons that exhibit entanglement in the spatial and temporal domains \cite{erhard2020advances} that arises due to the conservation of energy and momentum during the SPDC process.

In the regime of low pump power and paraxial approximation, where the pump laser is treated as a classical light field and the SPDC crystal is thin, the resulting biphoton state can be described as \cite{walborn2010spatial,torres2011engineering,baghdasaryan2022generalized}
\begin{eqnarray}
   \ket{\psi}_{\rm SPDC}&=&\iint d\mathbf{q_S}d\mathbf{q_I}d\omega_S d\omega_I\Phi(\mathbf{q_S},\mathbf{q_I},\omega_S,\omega_I) \hat{a}_S^\dag(\mathbf{q_S},\omega_S)\hat{a}_I^\dag(\mathbf{q_I},\omega_I)\ket{0}_S\ket{0}_I\notag\\
   &=& \iint d\mathbf{q_S}d\mathbf{q_I}d\omega_S d\omega_I\Phi(\mathbf{q_S},\mathbf{q_I},\omega_S,\omega_I) \ket{\mathbf{q_S},\omega_S}\ket{\mathbf{q_I},\omega_I}.\label{eq:SPDC_high_dim}
\end{eqnarray}
Here, the signal and idler photons are characterized by their energies $\omega_{S,I}$ and transverse momenta $\mathbf{q}_{S,I}$. Unlike the Fock notation, which represents the state of the entire system, here we use separate ket-vectors to describe each photon in the generated pair. Each ket-vector specifies the spatio-temporal mode in which the photon can be measured or the mode that the single photon excites.

The joint bi-photon amplitude $\Phi(\mathbf{q_S},\mathbf{q_I},\omega_S,\omega_I)$ is determined by the phase matching of the SPDC process, which involves the interplay between the frequencies and wave vectors of the pump, signal, and idler photons. When the parameters of the SPDC process are appropriately chosen, the function $\Phi(\mathbf{q_S},\mathbf{q_I},\omega_S,\omega_I)$ becomes non-separable in momentum or frequency (or both), indicating the entangled nature of the generated photons.

This feature can also be intuitively understood in both domains. On one hand, the photon pair is naturally generated at the same time and in the same transverse spatial location, which is often referred to as the ``common birth zone'' \cite{schneeloch2016introduction}, resulting in perfect time and position correlations. On the other hand, due to energy and momentum conservation in the SPDC process governed by the phase matching conditions, one also observes frequency and momentum anti-correlations, given by $\omega_P = \omega_S + \omega_I$ and $\mathbf{q}_P = \mathbf{q}_S + \mathbf{q}_I$, respectively.

Perfect correlations in the complementary bases of time and energy, as well as position and momentum, indicate the generation of entanglement in either domain. Exemplary correlation measurements and the corresponding joint spectral, temporal, spatial, and momentum amplitudes are shown in Fig. \ref{fig:TimeEnergy_entanglement}a and \ref{fig:PosMom_entanglement}a. In both cases, the correlations were strong enough to verify the presence of entanglement.

However, it is important to note that in both domains, the coherence of the pump field plays a crucial role in entanglement generation. The coherence of the pump field is transferred to the generated bi-photon state and, consequently, to the entanglement between the two photons \cite{kulkarni2017transfer,zhang2006experimental}. Additionally, it is worth mentioning that in the ideal case of a single-frequency, plane-wave pump, along with an infinitely thin and infinitely extended crystal in the transverse direction, the down-converted photons would exhibit perfect correlations in frequencies and momenta. However, in realistic scenarios where the pump field has a non-zero frequency bandwidth and momentum spread, the correlations become less well-defined, leading to wider joint amplitudes. The limited frequency and momentum spread of the photon pair, determined by the phase matching conditions, defines the strength of the correlations and, consequently, the achievable dimensionality of the entanglement.

To generate discrete high-dimensional entangled states, the continuous spatio-temporal domain needs to be discretized into a specific number of orthogonal modes or discrete units, known as bins. The number of modes or bins that can be used depends on the resolution of the employed devices as well as the strength of the correlations discussed earlier.

These different modes or bins are commonly labeled by the letter $d$, representing a system that allows for $d$ orthogonal states. This terminology gives rise to the term ``qudits'', as compared qubits in two-level systems such as polarization. By utilizing the temporal or spatial domain to address photon pairs generated in the SPDC process, it becomes possible to generate high-dimensional entangled quantum states. These states offer many advantages in various aspects of quantum photonics, spanning from fundamental research to information technologies. One of the primary benefits is the ability to encode more than just a single bit of quantum information. In fact, high-dimensional entangled states can encode log$_2d$ qubits. Furthermore, they exhibit increased resilience to noise \cite{bechmann2000quantum,cerf2002security,sheridan2010security,ecker2019overcoming} and can simplify operations in quantum information processing \cite{lanyon2009simplifying}. From a fundamental perspective, such states are also valuable in tests of quantum versus classical theories \cite{kaszlikowski2000violations,collins2002bell}. For more detailed information, interested readers can refer to recent review articles \cite{cozzolino2019high,wang2020qudits,erhard2020advances}, which provide further insights into the field of high-dimensional entanglement and its applications in various research areas.

As discussed earlier, in an ideal scenario with perfect momentum and frequency anti-correlations in the SPDC process, the joint bi-photon amplitude $\Phi(\mathbf{q_S},\mathbf{q_I},\omega_S,\omega_I)$ can be approximated by delta functions, allowing for the discretization of continuous space into a large set of correlated spatio-temporal modes or bins $|n\rangle$. This discretization enables the generation of a state that closely resembles the maximally entangled state given by
\begin{equation}
    |\psi\rangle_{\rm qudit} = \frac{1}{\sqrt{d}} \sum_{n=0}^{d-1} |n\rangle|n\rangle. \label{eq:maxHD}
\end{equation}

In the laboratory, various sets of orthogonal light modes or non-overlapping bins in space and time can be utilized for this purpose. The exact discretization and the resulting dimensionality of entanglement depend on several factors, including the specific SPDC parameters, the chosen discretization scheme, and the resolution of the devices used in the experiment.

In the following sections, we will provide a brief overview on important recent advances in both the spatial and temporal domains, highlighting advancements in generating high-dimensional entanglement.

\subsubsection{Time-Energy Entanglement} 
For the case of time-energy entanglement, let us consider a simplified scenario with a single spatial mode for both the pump and the down-converted photon pair. This allows for a further simplification of Eq.~\eqref{eq:SPDC_high_dim} to
\begin{equation}
   \ket{\psi}_{\omega}=\int d\omega_S d\omega_I\Phi(\omega_S,\omega_I)\ket{\omega_S}\ket{\omega_I},\label{eq:SPDC_high_dim_time}
\end{equation}

\begin{figure}[!t]
    \centering
    \includegraphics[width=1\textwidth]{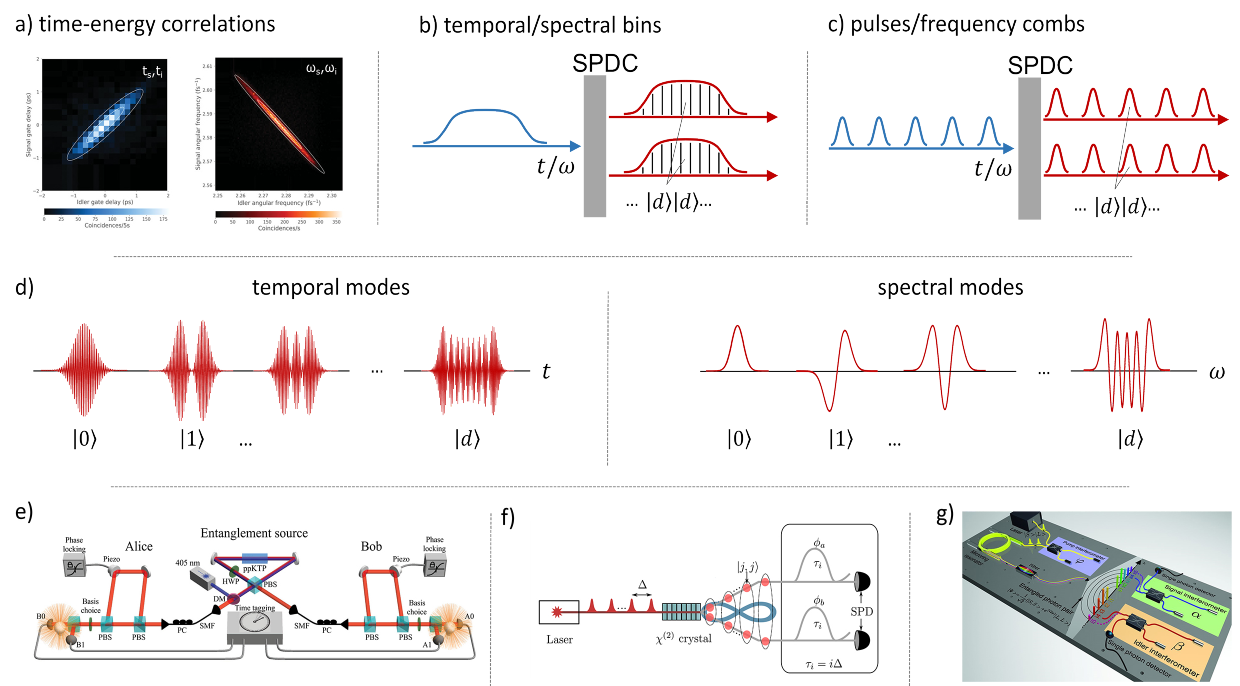}
    \caption{\label{fig:TimeEnergy_entanglement}
    Time-energy entanglement sources. (a) The SPDC process leads to correlation in time and anti-correlation in frequency, due to the conservation laws. To obtain high-dimensional entanglement, the correlations can be discretized into temporal/spectral bins. Figure reprinted from Ref.~\cite{maclean2018direct}. (b) Using a discretization scheme or (c) frequency comb as a pump can directly lead to high-dimensional entanglement in $d$ possible photon pulses. (d) A temporal quantum state can also be discretized into orthogonal temporal modes, which correspond to discrete orthogonal spectral modes in frequency. (e) A typical setup to generate and measure temporal correlations using a continuous-wave laser as the pump and Franson interferometers to verify the entanglement. Figure reprinted from Ref.~\cite{ecker2019overcoming}. (f) A setup generating high-dimensional entanglement using a pulse train. Figure reprinted from Ref.~\cite{martin2017quantifying}. (g) A setup to generate frequency bin entangled states using an integrated frequency comb. Figure reprinted from Ref.~\cite{reimer2016generation}.
    }
\end{figure}

In experiments, the correlations observed in frequency and their corresponding Fourier-related time domain are often expressed and measured using the joint spectral amplitude \cite{avenhaus2009experimental} and the joint temporal amplitude \cite{maclean2018direct} (see Fig. \ref{fig:TimeEnergy_entanglement}a). The discretization in either basis is set by the spectral or timing resolution of the detectors, and the resulting states are commonly referred to as spectral or time bins.

In Fig. \ref{fig:TimeEnergy_entanglement}b, a schematic of such a discretization scheme is depicted, illustrating how the continuous time and frequency domains are divided into discrete bins or modes, with each bin corresponding to a specific range of frequencies or time intervals. The correlations between these bins capture the entanglement between the photons.

In the time domain, achieving high-dimensional entanglement requires the timing resolution of the detectors to be much finer than the coherence time of the pump laser. Moreover, a high level of stability is necessary for the pump laser to enable the measurement of correlations not only in time but also in superpositions of time bins. To accomplish the latter, which is essential for entanglement certification, a commonly used measurement technique is an unbalanced Mach-Zehnder interferometer, also known as a Franson interferometer \cite{franson1989bell}. A schematic of an exemplary setup is depicted in Fig. \ref{fig:TimeEnergy_entanglement}e.

These approaches have facilitated the observation of high-dimensional entanglement, surpassing hundreds of time bins and even higher dimensionalities \cite{zhong2015photon, ecker2019overcoming}. Additionally, the utilization of pulse shaping techniques \cite{weiner2011ultrafast} in conjunction with nonlinear interactions has enabled the measurement of timing correlations at resolutions beyond the capabilities of conventional detectors \cite{donohue2013coherent, maclean2018direct}. This breakthrough overcomes the existing technological limitations of detectors and achieves significantly enhanced precision in temporal measurements.

Another strategy to achieve high-dimensional entanglement in the temporal domain involves employing a coherent mode-locked laser that emits a sequence of coherent pump pulses. Each pulse presents an opportunity for generating a photon pair, effectively defining distinct time bins as sketched in Fig. \ref{fig:TimeEnergy_entanglement}c. The utilization of coherent mode-locked lasers with a high repetition rate is essential for this purpose \cite{de2002creating}. The repetition rate plays a crucial role as it determines the time interval between two successive pulses and, consequently, the delay between the individual time bins. It is important to note that the high repetition rate poses challenges for the detection system due to the substantial imbalance between the two arms of the Franson interferometer.

Despite these challenges, experimental studies have successfully demonstrated entanglement dimensions of up to 18 using a setup illustrated in Fig. \ref{fig:TimeEnergy_entanglement}f \cite{martin2017quantifying}. This approach highlights the potential of coherent mode-locked lasers in generating high-dimensional temporal entanglement.

Another highly promising approach is the study of temporal modes, which constitute an orthogonal set of Hermite-Gauss shaped wave packets as depicted in Fig. \ref{fig:TimeEnergy_entanglement}d. These modes can be considered the most natural basis in the temporal domain \cite{brecht2015photon,raymer2020temporal}. In recent years, significant progress has been made in utilizing advanced detection and manipulation tools \cite{ansari2018tailoring} as well as direct entanglement generation through nonlinearity engineering \cite{graffitti2020direct}.

In addition to exploring the temporal domain, it is also possible to leverage the generated frequency correlations. In this context, frequency bins serve as an analogous discretization to time bins \cite{olislager2010frequency}, and techniques similar to pulse shaping methods \cite{bernhard2013shaping} can be employed to address them. Integrated frequency combs have emerged as a particularly promising approach, as evidenced by recent advancements \cite{reimer2016generation,kues2017chip,imany201850,kues2019quantum,chang2021648,borghi2023reconfigurable}. Integrated frequency combs can be seen as the analog counterpart to the aforementioned pulse trains, as depicted in Fig. \ref{fig:TimeEnergy_entanglement}b and c. Utilizing frequency combs to generate frequency bin entanglement offers a well-controlled and potentially scalable approach. Fig. \ref{fig:TimeEnergy_entanglement}g provides a sketch of such a setup. With entanglement dimensions already demonstrated in the tens to hundreds range, this approach is particularly attractive due to its integrated nature and compatibility with telecom equipment. Additionally, by employing a continuous pump laser and discretizing the biphoton state using a Fabry-Pérot cavity to form a biphoton frequency comb, it has been shown that high dimension can be achieved without post-selection \cite{xie2015harnessing,chang2021648}.

By combining the technology to shape single photons in time \cite{weiner2011ultrafast} and frequency \cite{kues2019quantum}, both approaches for generating high-dimensional entanglement in the energy-time domain hold great promise for expanding the size of the underlying Hilbert space. Moreover, their compatibility with integrated systems and optical fibers positions time-energy entanglement as a vital player in future high-dimensional QIT.

\subsubsection{Position-Momentum Entanglement}
\label{sec: position-momentum entanglement}
\begin{figure}[t!]
\centering
\includegraphics[width=1\textwidth]{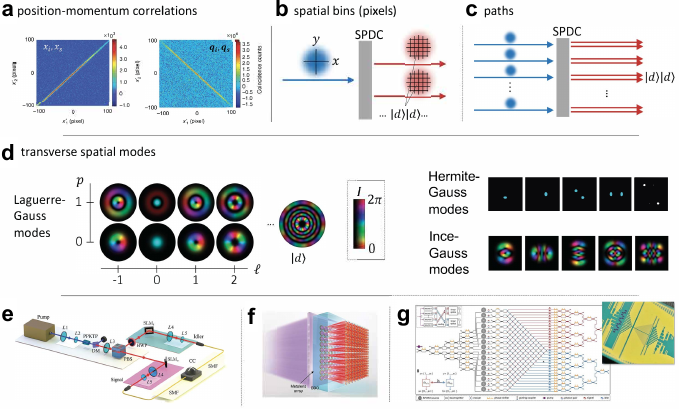}
\caption{\label{fig:PosMom_entanglement}
Position-momentum entanglement sources. (a) The SPDC process leads to correlation in space and anti-correlation in momentum, due to its phase matching conditions. Figure reprinted from Ref.~\cite{edgar2012imaging}. (b) To obtain high-dimensional entanglement, the correlations can be discretized into spatial bins, i.e., pixels. (c) Generating a superposition of multiple pump paths in one (or many) SPDC crystal generates high-dimensional entanglement entangled in $d$ different paths. (d) Another popular approach to discretize space into $d$-dimensional quantum states is by using different spatial mode families, three of which are shown here with the color depicting their phase structure and the brightness representing their intensity. (e) A sketch of a typical setup to study spatial correlations using spatial light modulators to project onto pixels or spatial modes. Figure reprinted from Ref.~\cite{srivastav2022characterizing}. (f) By using a metalens array, tens of parallel SPDC processes can be pumped, thereby generating high-dimensional path entanglement. Figure reprinted from \cite{li2020metalens}. (g) Integrated waveguide chips can be used to generate, manipulate, and analyze high-dimensional path-entangled photon pairs (inset shows a photo of the chip). Figure reprinted from Ref.~\cite{wang2018multidimensional}.
}
\end{figure}

Another widely exploited degree of freedom to generate high-dimensional entanglement through SPDC is the spatial domain, where non-classical correlations in position and momentum can be observed. Similar to the discussion in the time-energy domain, we can simplify the general SPDC state introduced in Eq. \eqref{eq:SPDC_high_dim} by considering correlations only in space. This simplification involves assuming a single pair of frequencies for the two photons, allowing us to write the state as
\begin{equation}
\ket{\psi}_{\mathbf{q}}=\int d\mathbf{q}_S d\mathbf{q}_I\Phi(\mathbf{q}_S,\mathbf{q}_I)\ket{\mathbf{q}_S}\ket{\mathbf{q}_I}.\label{eq:SPDC_high_dim_space}
\end{equation}
The conservation of transverse momentum in the SPDC process, i.e., $\mathbf{q}_P = \mathbf{q}_S + \mathbf{q}_I$, leads to highly correlated joint transverse momentum amplitudes. These anti-correlations have been leveraged to couple the photons into multiple fibers in a bundle. In doing so, the continuous space is discretized to generate high-dimensional path entanglement \cite{rossi2009multipath}. In addition to momentum anti-correlations, correlations in the complementary space, i.e., the transverse spatial locations, have been utilized to verify entanglement using the EPR criteria over the last few decades~\cite{howell2004realization,edgar2012imaging,ndagano2020imaging}. Fig. \ref{fig:PosMom_entanglement}a shows the measurements of the joint spatial and momentum amplitude on an electron-multiplied CCD camera.

As conceptually depicted in Fig. \ref{fig:PosMom_entanglement}b, the use of single-photon sensitive cameras, particularly in the pixel basis, has made it appealing to discretize the continuous transverse space, similar to time bins in the temporal domain. This approach to discrete high-dimensional entanglement has greatly benefited from the latest advancements in camera technologies, which have enabled the demonstrations of entanglement in more than 100 dimensions \cite{schneeloch2019quantifying}. In conjunction with advanced compressed sensing methods, a channel capacity of 8.4 bits per photon has been achieved, corresponding to over 300 dimensions \cite{howland2013efficient}.

In another approach, spatial light modulators (SLMs) are inserted into the beam path of each down-converted photon to introduce filtering masks that define macro-pixels of customized shapes and sizes. This configuration offers great flexibility in adjusting the macro-pixels to match the experimental setup that comprises crystals and imaging optics. Fig.~\ref{fig:PosMom_entanglement}e shows a schematic for such a setup in which SLMs enabled the generation of a nearly ideal maximally entangled state described in Eq.~\eqref{eq:maxHD} \cite{valencia2020high,nape2021measuring}. Additionally, the use of SLMs has allowed for measurements in superposition states of the macro-pixels to bolster in-depth characterization of the generated state, achieving entanglement dimensions of more than 50 \cite{valencia2020high,srivastav2022characterizing}.

The discretization of the transverse space into specific areas can also be achieved within the pump beam itself. In this configuration, a single pump is sent into the SPDC crystal with coherence across locations, as schematically sketched in Fig. \ref{fig:PosMom_entanglement}c. Recently, this approach was demonstrated using a lens array implemented through a nano-structured metasurface, enabling 100 parallel, coherently driven SPDC processes that generated up to six photons in a compact arrangement, as depicted in Fig. \ref{fig:PosMom_entanglement}f. The main challenge with such a scheme lies in the verification of high-dimensional entanglement, as it requires a method to project onto superposition states. To date, only three-dimensional entanglement in a subset of the whole array has been demonstrated \cite{li2020metalens}.

Following a similar concept, splitting the pump into multiple paths, each passing through an SPDC crystal, leads to what is known as path entanglement. This approach was recently extended to achieve an entanglement dimension of 32 in an efficient bulk-optics setup \cite{hu2020efficient}. Encoding the high-dimensional quantum state in different paths is particularly appealing not only because it can be directly implemented using single-mode fibers compatible with fiber networks \cite{schaeff2015experimental}, but also because it can be realized with integrated photonic waveguide chips (see Sec.~\ref{sec: quantum chips}). These chips hold great promise for scaling up quantum states to very large dimensions. An experiment using this technology demonstrated 14-dimensional entanglement, where the generation, manipulation, and analysis of the state were performed on the chip, as illustrated in Fig. \ref{fig:PosMom_entanglement}g \cite{wang2018multidimensional}.

Notably, the spatial domain of photons can be discretized into complex orthogonal transverse structures of light known as the transverse spatial modes. Fig. \ref{fig:PosMom_entanglement}d illustrates various mode families that exist subject to the underlying symmetry and coordinate system used to describe them. While entanglement has been demonstrated for all common mode families, such as Hermite-Gauss modes \cite{walborn2005conservation}, Laguerre-Gauss (LG) modes \cite{d2021full}, Ince-Gauss modes \cite{krenn2013entangled}, as well as modes with special properties like Bessel modes that are self-healing \cite{mclaren2012entangled} and Airy modes that self-accelerate \cite{lib2020spatially}, the most popular ones are the LG modes in cylindrical coordinates. LG modes possess a special property known as orbital angular momentum (OAM), which is based on a twisted phase structure. In fact, the first demonstration of spatial mode entanglement was achieved using OAM-carrying photons generated through the SPDC process, which conserves angular momentum \cite{mair2001entanglement}. Entanglement in OAM has been further verified by correlating OAM with its complementary variable, i.e., angle \cite{leach2010quantum}. The popularity of high-dimensional OAM entanglement is also driven by virtue of the technical simplicity. OAM-carrying photons can be relatively easily measured using phase modulations on SLMs and single-mode fibers \cite{Mirhosseini2016PRL, bhusal2021AQT}. By removing the helical phase structure of OAM modes through phase modulations, photons with flat phase profiles can be efficiently coupled into single-mode fibers, which is not the case for other mode types \cite{bouchard2018measuring}. Additionally, OAM quanta are conserved in cylindrically symmetric systems, making quantum states encoded in OAM-carrying spatial modes particularly robust when transmitted through cylindrical-shaped optical systems. By exploiting the full-field correlations of LG mode entanglement involving the azimuthal OAM degree of freedom and the radial degree of freedom, entanglement in dimensions up to 100 has been demonstrated \cite{krenn2014generation}. However, it is important to note that the physical size of the modes increases when higher-order modes are utilized, and the apertures of the optical system impose a practical limitation on high-dimensional entanglement. Nevertheless, LG modes are known to be the maximally dense solution in cylindrical systems in terms of the information capacity \cite{kahn2016twist}, with entanglement of OAM quanta up to 10,000 already achieved \cite{fickler2016quantum}. Generating high-dimensional entanglement using spatial modes of photon pairs from an SPDC process poses a challenge due to the decreasing likelihood of higher-order modes appearing in the down-converted pair, resulting in a non-maximally entangled state. Various methods have been developed to address this drawback. One approach is to carefully tune the phase-matching conditions \cite{romero2012increasing} to increase the probability of higher-order mode generation and, consequently, enhance the entanglement dimensionality. Another strategy is to shape the spatial mode of the pump light in the SPDC process to tailor the generated mode content and improve the quality of entanglement. However, this approach has thus far been limited to up to 5-dimensional entanglement \cite{kovlakov2018quantum, liu2020increasing}. More recently, there have been proposals to enhance the dimensionality by not only shaping the pump, but also engineering the nonlinear interaction through crystal design \cite{rozenberg2022inverse} to offer greater control over the entanglement properties.

It is worth noting that a novel approach building upon the previously discussed concepts has been introduced to generate high-dimensional entanglement through path identity, leading to successful generation of three-dimensional entangled states \cite{kysela2020path}. By exploiting the indistinguishability of multiple SPDC processes, each contributing additional modes to the entangled photon pair, this approach offers unprecedented flexibility and customization in realizing the desired entangled state. While the concept of path identity draws inspiration from well-known schemes used for generating polarization entanglement \cite{kwiat1999ultrabright}, it can also be extended to entangle more than two photons and can be linked to quantum imaging schemes \cite{hochrainer2022quantum}. This innovative approach opens up new possibilities for exploring and harnessing high-dimensional entanglement in various quantum information processing applications.

In the future, it is conceivable that transversely entangled photons in their spatial modes could serve as flying qudits, which are high-dimensional quantum information carriers, for long-distance communication in both fiber-based \cite{cao2020distribution} and free-space \cite{krenn2015twisted} scenarios, as they exhibit robust propagation characteristics. Following transmission, these qudits could be utilized directly for quantum operations \cite{brandt2020high} or interfaced with waveguide chips by converting them into path qudits through simple transformations \cite{fickler2014interface}. On quantum photonic chips, the entangled state can be further manipulated and converted into polarization encoding or multi-mode waveguide structures \cite{feng2016chip}. This integration of transverse mode entanglement with various platforms and encoding schemes holds great promise for advancing QIT.

\subsubsection{Entanglement in Multiple Degrees of Freedom}
Instead of generating high-dimensional entangled states for each degree of freedom (DOF) separately, significant progress has been made in utilizing hyper-entanglement, which involves simultaneous entanglement in multiple DOFs, such as polarization, space, and time. By treating each DOF as an additional resource for encoding quantum information, the overall Hilbert-space dimension is multiplied, resulting in a significantly enlarged state space. Various combinations of hybrid entanglement have been explored, including polarization and momentum \cite{barbieri2007polarization,gao2010experimental,ciampini2016path}, polarization and spatial modes \cite{d2016entangled}, polarization and time \cite{steinlechner2017distribution,chapman2022hyperentangled}, and hyper-entanglement involving all DOFs \cite{barreiro2005generation,graffitti2020hyperentanglement}.

The expanded state space provided by hyper-entanglement has enabled notable achievements, such as efficient entanglement distillation \cite{ecker2021experimental}, the generation of an 18-qudit entangled state using six photons and their polarization, path, and spatial mode DOFs \cite{wang201818}, and the demonstration of an entanglement-based quantum network \cite{wengerowsky2018entanglement}. These advancements highlight the potential of hyper-entanglement for various QIT applications.

\subsubsection{Integrated Entanglement Sources}
\label{sec: quantum chips}
The remarkable progress in understanding and implementing quantum information processing techniques has sparked great enthusiasm for the development of large-scale quantum systems. As these systems continue to increase in complexity and fulfill real-world requirements, there is an urgent need to transition from traditional, bulky optical entanglement sources to compact photonic integrated chips (PICs). This shift is made possible by the rapid advancements in classical PICs, which not only enable superior generation of entanglement states but also offer improved manipulation and detection capabilities. Moreover, PICs offer several advantages, including reduced footprint and cost, enhanced integration of functionalities, improved fidelity, and increased phase stability. Furthermore, the well-established fabrication techniques and process design kits (PDKs) associated with PICs allow for the integration of a wide range of optical functional components onto a single chip, including beam splitters, waveguides, delay lines, filters, cavities, detectors, among others. Additionally, PICs provide a diverse selection of material platforms, allowing researchers to choose the most suitable platform for their specific goals and targets.

\begin{figure}[hbt!]
\centering
\includegraphics[width=\textwidth]{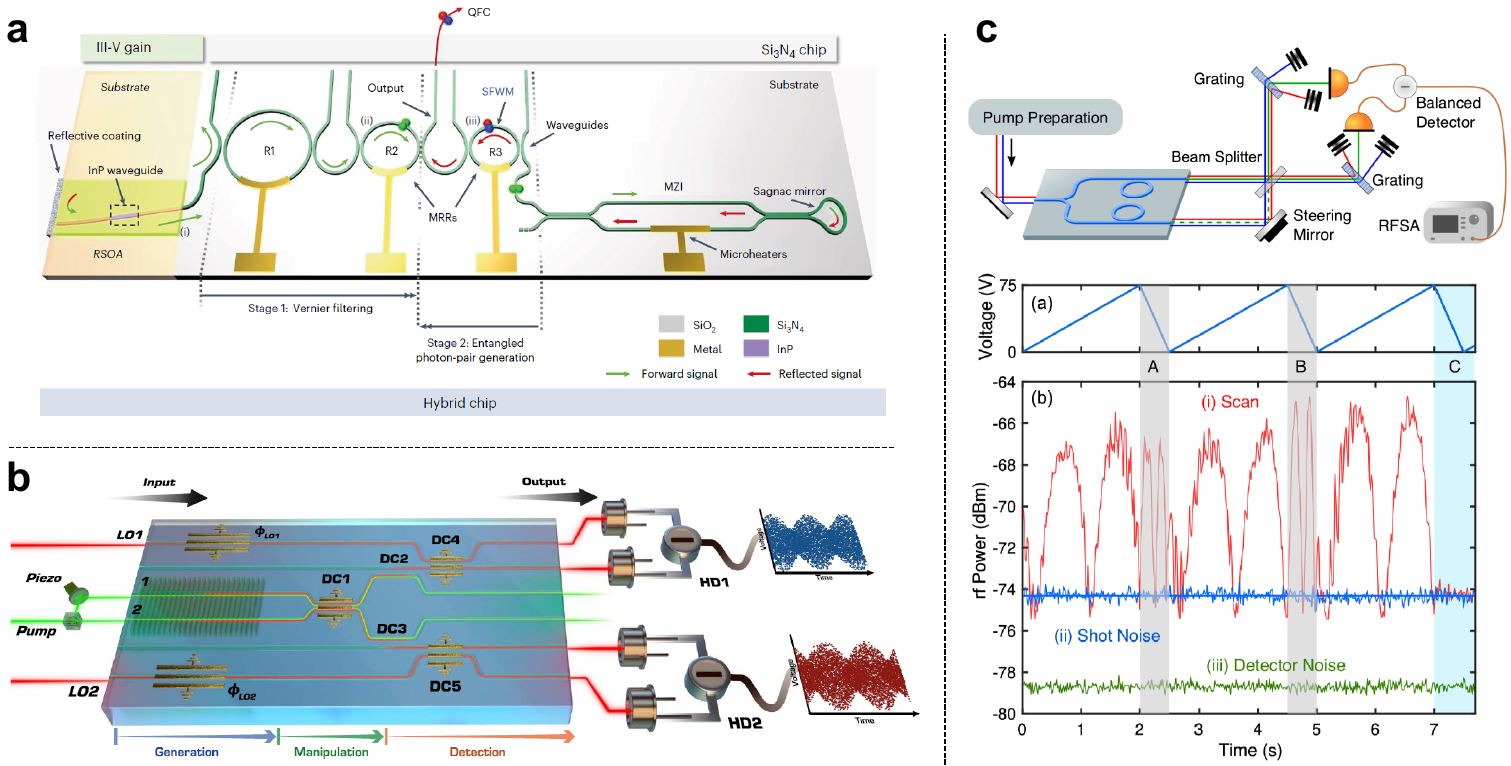}
\caption{\label{fig:integrated_entanglement_source}
Integrated entanglement sources. (a) Schematic of fully on-chip generation of frequency-bin entangled photons on an InP-SiNOI hybrid integration configuration. Figure reprinted from Ref.~\cite{Mahmudlu2023}. (b) Programmable integrated lithium niobate photonic circuits for the generation, manipulation, and detection of squeezed-vacuum and TMSV states through the SPDC process. Figure reprinted from Ref.~\cite{Lenzini2018}. (c) Sketch of dual-pump degenerate SFWM for squeezed-vacuum generation on the SiOI platform. Measurement data show the quadrature variance in relative to the shot noise level. Figure reprinted from Ref.~\cite{Zhao2020squeezing}.}
\end{figure}

In the discrete-variable regime, integrated entanglement sources can probabilistically generate entangled photons through SPDC in materials with second-order nonlinearity or spontaneous four-wave mixing (SFWM) through third-order nonlinearity. SPDC typically offers higher conversion efficiency, while SFWM platforms provide more mature and scalable low-cost fabrication possibilities.

The initial breakthrough in generating entangled states in PICs is traced back to 2008 when Alberto {\em et al.} achieved this milestone in a silica-on-silicon waveguide platform \cite{Politi2008}. They utilized single-mode waveguide interferometers to replace bulky beam splitters, enabling the realization of high-fidelity path-entangled states, two-photon quantum interference, and controlled-NOT gates. Since then, significant progress has been made in various PIC platforms over the past two decades, benefiting from the rapid development of micro- and nanofabrication technologies. These platforms include silicon-on-insulator (SOI) \cite{Silverstone2014SOI,Takesue2007SOI}, silicon nitride-on-insulator (SiNOI) \cite{Lu2019SiN}, thin film lithium niobate-on-insulator (TFLNOI) \cite{Zhao2020LN}, gallium arsenide-on-insulator (GaAsOI) \cite{Steiner2021GaAs}, and others.

Akin to the bulk-optics systems, PICs can generate entangled photons with multiple degrees of freedom (DOFs), such as time-energy \cite{Grassani2015}, polarization \cite{Sansoni2010Polarization}, optical path \cite{Silverstone2015path}, transverse mode \cite{feng2019chip}, time bin \cite{Xiong2015time}, and frequency bin \cite{reimer2016generation}. In particular, PICs provide a pathway for controlling and scaling up the dimensionality of entanglement using transverse modes supported by multi-mode waveguide structures \cite{feng2019chip}. Harnessing the capabilities of PICs, this method offers a compact and efficient solution to produce high-dimensional entanglement. The use of multi-mode waveguide structures also facilitates the manipulation and control on entangled states, paving the way for scalable entanglement-based QIT.

To integrate more quantum components in SOI PICs, foundry-compatible complementary metal-oxide-semiconductor (CMOS) fabrication processes have been leveraged to achieve the generation of multipartite multidimensional entanglement via monolithic integration of over 2,400 quantum components on a single chip \cite{bao2023very}. Recently, Hatam {\em et al.} realized fully on-chip turnkey entanglement sources on a SiNOI platform, as depicted in Fig.~\ref{fig:integrated_entanglement_source}a. They followed a hybrid integration approach, combining an electrically pumped indium phosphide (InP) reflective semiconductor optical amplifier (RSOA) with an efficient programmable SiN vernier filter as the low-noise pump source. As a result, frequency-bin entangled states were generated with up to 99\% fidelity without the need for an external bulky laser \cite{Mahmudlu2023}. Since SiNOI is a foundry-compatible PIC platform, their approach allows for seamless integration with other quantum processing circuits, paving the way for commercially available, cost-effective, scalable, and stable quantum PICs for QIT applications.

In contrast to the probabilistic SPDC process, semiconductor quantum dots (QDs) offer the capability of on-demand generation of single and entangled photons. While other platforms, such as color centers in crystals \cite{Bradac2019,Lohrmann2015}, rare earth atoms \cite{Dibos2018}, and two-dimensional materials \cite{Tran2016}, have been investigated for on-chip quantum-light generation, QDs exhibit advantages including high brightness and flexible operation wavelength through bandgap engineering. Efforts to surpass the fine structure splitting have led to several recent experiments that successfully demonstrated on-demand generation of high-fidelity entangled photons via biexciton-exciton cascade in an individual QD \cite{Huber2018QD,Wang2019QD,Liu2019QD}. Building upon this approach, Frejia {\em et al.} embedded QDs in a nanophotonic waveguide and transferred the intrinsic polarization entanglement into photon path entanglement \cite{stfeldt2022QD} to enable interfacing with other quantum components in PICs. However, integrating QDs with waveguides or cavities requires precise QD positioning and alignment lithography. This is particularly important for epitaxial QDs that are typically randomly distributed on the substrate. Despite the development of the wafer bonding and ``pick-and-place'' techniques, which often come with tradeoffs in fidelity and uniformity \cite{Davanco2017,Chanana2022}, achieving large-scale integration of QDs in PICs remains challenging. For more comprehensive discussions on QDs-based integrated quantum platforms, readers should refer to the review paper by Hepp {\em et al.} \cite{Hepp2019QD}.

In contrast to the discrete-variable counterpart, continuous-variable entanglement offers advantages of deterministic generation and efficient detection at room temperature, while its fidelity is more susceptible to loss. In the pioneering experiments, researchers employed single-pass quasi-phase-matched waveguides in lieu of bulk crystals to generate on-chip squeezed light. Squeezing levels of 12\% and 14\% were measured in the KTP \cite{Anderson1995} and lithium niobate (LiNbO$_3$)\cite{Serkland1995} platforms respectively. In these experiments, however, the experimental components apart from the SPDC processes remained off-chip and required precise beam shaping, alignment, and phase locking. More recently, Genta {\em et al.} demonstrated the on-chip generation and characterization of EPR entanglements by employing optical waveguide interferometers to interfere two off-chip single-mode squeezing states \cite{Masada2015}, an approach discussed in Sec.~\ref{sec:entanglement_generation_theory} and Sec.~\ref{sec:CV_entanglement_sources}. The use of optical waveguides inherently ensures near-perfect spatial mode matching between the squeezed light and the local oscillator (LO), significantly reducing the complexity associated with homodyne measurements. Later, Lenzini {\em et al.} integrated two individual single-pass periodically poled LiNbO$_3$ (PPLN) squeezed-light sources with waveguides, phase shifters, and interferometers on the same PIC to enable the reconfigurable generation, manipulation, and characterization of squeezed vacuum states and EPR entangled states \cite{Lenzini2018}, as shown in Fig.~\ref{fig:integrated_entanglement_source}b. For QIT based on continuous variables, high levels of squeezing are crucial. For example, entanglement swapping typically requires 3 dB of squeezing \cite{Loock1999} to surpass the classical limit; 2D cluster state generation needs 4.5 dB of squeezing \cite{asavanant2019generation}; and fault-tolerant quantum computing requires 10 dB  of squeezing \cite{Fukui2018}. To date, the highest detected on-chip squeezed vacuum is 6 dB over a 2.5 THz bandwidth, demonstrated by Takahiro {\em et al.} using a single-mode ZnO-doped PPLN waveguide\cite{Kashiwazaki2020}, showing great potential for large-scale QIT on PIC platforms.

In addition to the above low-index-contrast platforms, recent research has focused on generating squeezed states in more compact nanophotonic platforms. In 2015, Avik {\em et al.} made significant strides in this direction by demonstrating the use of a single silicon nitride (SiN) microring resonator for squeezed-light generation. Utilizing the non-degenerate SFWM process above the parametric oscillation threshold, they measured amplitude squeezing of 1.7 dB between the generated signal and idler beams \cite{Dutt2015}. Subsequently, Zhao {\em et al.} further advanced this field by demonstrating the generation of quadrature squeezed-vacuum states using a degenerate SFWM scheme, as illustrated in Fig.~\ref{fig:integrated_entanglement_source}c. In their approach, both the squeezed light and LO were simultaneously generated on the same chip \cite{Zhao2020squeezing}. They utilized two identical ring resonators operating in the over-coupling and critical coupling conditions for squeezing generation and LO generation. The large separation of the two pump signals helped minimize the noise caused by thermal-refractive effects. However, the individual parametric processes of each pump introduced unwanted noise fluorescence to the squeezing states. As a result, a 1.34 dB quadrature squeezed vacuum was detected, with an inferred on-chip squeezing level of 3.09 dB accounting for coupling losses. To address this issue, Zhao {\em et al.} proposed a novel approach by leveraging a photonic molecule design, which involved an auxiliary resonator strongly coupled with the squeezing resonator \cite{Zhang2021squeezing}. By effectively detuning the resonances supporting parametric processes, the noise from the pump were significantly suppressed, while the squeezing resonator remained unaffected. The directly measured squeezing level was 1.65 dB, corresponding to 8 dB on-chip squeezing, with the measured squeezing limited by the out-chip collection and detection efficiencies. Although non-degenerate SFWM suffers from higher thermal-refractive noise in squeezed-light generation, it inherently supports a large number of spectral modes, making it appealing for large-scale QIT applications. Based on non-degenerate SFWM, Vaidya {\em et al.} demonstrated 1 dB quadrature squeezed vacuum in a SiN ring resonator\cite{Vaidya2020}. Later, Yang {\em et al.} achieved a 1.6 dB squeezing in an ultra-high-Q silica microdisk with 20 operational pairs of TMSV states \cite{yang2021squeezed}. In their experiment, the LOs were generated by means of an electro-optic modulator to circumvent the need for complex phase locking of multiple lasers. Despite the recent tremendous advances, the generation of squeezed states in PICs remains nascent. Assuming optimal performance for each individual component in PIC platforms, it would become feasible to achieve squeezing levels in excess of 10 dB and high-quality 3D cluster states for one-way quantum computing \cite{wu2020quantum}. Remarkably, recent advancements in programmable photonic processors based on continuous-variable entanglement have shown quantum computational advantages at room temperature \cite{arrazola2021quantum,madsen2022quantum}.

\section{Quantum Metrology Using Entangled States}
\label{sec: quantum_metrology}
\begin{figure}[hbpt!]
    \centering
\includegraphics[width=0.8\textwidth]{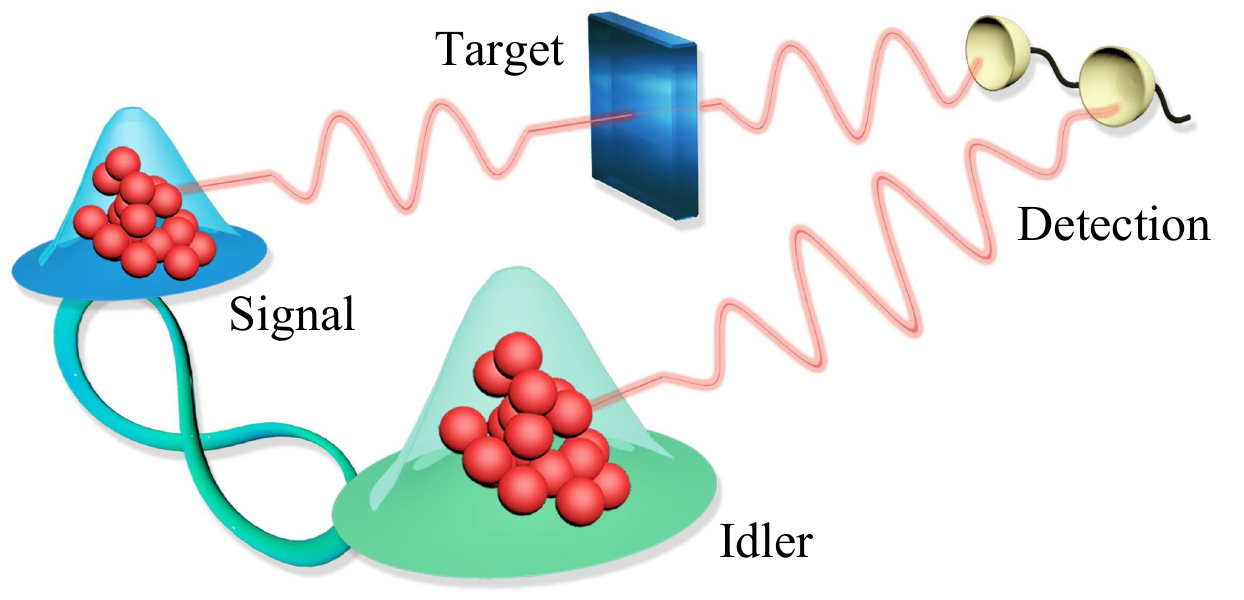}
    \caption{Schematic of quantum metrology using entangled states, which consists of three steps. The first step prepares an entangled state of signal and idler. The second step involves the interaction between the signal and the target. In the third step, the signal and infer are measured to infer the properties of the target.}
    \label{Fig:metrology_intro} 
\end{figure}
Metrology is one of the important topics of optical sciences. The use of photons to measure small quantities is one of the most important applications of light \cite{Giovannetti:04, giovannetti2011advances, demkowicz2015, Sciarrino20review,ShinOE13, MaganaPRL2014,MaganaPS2016}. Unfortunately, with classical states of light produced by, e.g., the laser, one cannot surpass the fundamental shot-noise limit \cite{Jarzyna:12, Lang:13, Takeoka:17}. Quantum metrology builds on nonclassical states of light such as entanglement and squeezing to beat the conventional limit. With the recent development in experimental capabilities, quantum metrology has become tangible in laboratory settings and is moving quickly toward real-world applications \cite{Giovannetti:04, giovannetti2011advances, demkowicz2015, caves:1981, Sciarrino20review}. Here, we will introduce the theory of quantum metrology followed by a review on its realizations. 

\subsection{Theoretical Description of Quantum Metrology}
\label{subsec: QFI}
A metrology protocol aims to measure a parameter $\varphi$ of a physical system. As shown in Fig. \ref{Fig:metrology_intro}, quantum metrology consists of three steps: Step 1 generates the probe state. In Step 2, the probe state is sent to interact with the system. In the last step, the signal is measured to estimate the parameter of interest for the system. In classical metrology, one has a data set consisting of $N$ copies of independent identically distributed random variables drawn from a probability density function (PDF), $p_{\varphi}(X)$, where the PDF is solely depending on an unknown parameter $\varphi$ being estimated. The goal of the estimation is to construct an estimator $\tilde{\varphi}_{N}(\mathbf{x})$ to minimize the uncertainty. To this end, one can use the Fisher information (FI) to derive an optimal estimator \cite{kay1993fundamentals}. In this case, one could construct the classical \emph{Cram\'er-Rao bound} (CRB) that lower bounds the mean square error (MSE) of any unbiased estimator $\tilde{\varphi}_{N}$:
\begin{equation}
\Delta^{2}{\tilde{\varphi}}_{N} \ge\frac{1}{N\, F\!\left[p_{\varphi}\right]}\,,
\label{eq:CRB}
\end{equation}
where $F\!\left[p_{\varphi}\right]$ is the FI defined as follows:
	\begin{equation}
		F\left[p_{\varphi}\right]=  \int\!\!\textrm{d}x\, \left(\frac{\partial }{\partial \varphi}\log p_{\varphi}(x)\right)^2 p_{\varphi}(x) =  \int\!\!\textrm{d}x\,\frac{1}{p_{\varphi}(x)}\left[\frac{\partial\, p_{\varphi}(x)}{\partial\varphi}\right]^{2}.
	\end{equation}
It should be noted that the classical CRB is dependent on the measurement performed on the signal. As such, one would ask: what is the best achievable performance for estimating the parameter $\varphi$ irrespective of the measurement? The \emph{quantum Cram\'{e}r-Rao bound} (QCRB) provides the answer. QCRB is a generalization of the classical CRB, setting the lower bound for the estimation variance over all locally unbiased estimators subject to measurements allowed by quantum mechanics \cite{Braunstein1994, helstrom1969quantum, holevo2011probabilistic}. The QCRB is derived from the quantum FI (QFI) defined as
\begin{equation}
    F_{\mathrm{Q}}\left(\rho_\varphi\right)=\max _{\left\{p_{\varphi}\right\}} F(p_{\varphi}).
\end{equation}
Like the classical CRB and FI, the QCRB is linked to the QFI by
\begin{equation}
    \Delta^2 \varphi_N \geq \frac{1}{N F_{\mathrm{Q}}\left[\rho_{\varphi}\right]}\geq \frac{1}{N F\left[p_{\varphi}\right]}.
\end{equation}
The QFI $F_{\mathrm{Q}}$ can be calculated by 
\begin{equation}
F_{\mathrm{Q}}\left[\rho_{\varphi}\right]=\operatorname{Tr}\left\{\rho_{\varphi} L\left[\rho_{\varphi}\right]^2\right\},
\end{equation}
where $L\left[\cdot\right]$ is called the symmetric logarithmic derivative operator. In simple scenario of a pure state $\rho_{\varphi}\!=\!\left|\psi_{\varphi}\right\rangle \!\left\langle \psi_{\varphi}\right|$,
the QFI is derived as
\begin{equation}
F_{Q} = 4(\langle\dot{\psi}_{\varphi}|\dot{\psi}_{\varphi}\rangle - |\langle\dot{\psi}_{\varphi}|\psi_{\varphi}\rangle|^2), \quad |\dot{\psi}_{\varphi}\rangle=\frac{d |\psi_{\varphi}\rangle}{d \varphi} .
\end{equation}
It is interesting to note that, in the context of optical interferometry, when the estimated parameter is encoded on the state by a unitary operation described as $\rho_{\varphi}\!=\! U_{\varphi}\rho U_\varphi^{\dagger}$ with $U_{\varphi}\!=\!\textrm{e}^{-\textrm{i}\hat{H}\varphi}$, QFI on a pure state $\rho = \ket{\psi}\bra{\psi}$ is proportional to the variance
of $\hat{H}$ \cite{giovannetti2011advances}:
\begin{equation}
F_Q(\ket{\psi_\varphi}) = 4 \Delta^2 H = 4 (\bra{\psi} \hat{H}^2 \ket{\psi} - \bra{\psi} \hat{H}  \ket{\psi}).
\end{equation}
We should note that care needs to be taken in calculating the QFI \cite{demkowicz2015, MaganaPS2016, Takeoka:17, You:19, Sciarrino20review}. Specifically, the QFI is deriving by optimizing over all possible measurement schemes, which may include cases where one already has prior knowledge of the parameter being estimated. To ensure a valid QFI calculation, it is necessary to identify a measurement that can attain the QCRB. For a more comprehensive review on the quantum estimation theory, the readers should refer to Refs. \cite{toth2014quantum, demkowicz2015, Sciarrino20review}. Finally, performance metrics need to be defined to benchmark metrology protocols. For optical-metrology protocols using, on average, $N$ photons to perform the estimation, $1 / \sqrt{N}$ is defined as the shot-noise limit (SNL) while $1 / N$ is referred to as the Heisenberg limit (HL) \cite{giovannetti2006quantum}. The SNL is considered the fundamental limit for a classical metrology protocol, dictating its maximal attainable sensitivity. However, quantum metrology protocols based on nonclassical states are not subject to the SNL. In the following sections, we will discuss experiments that utilized entanglement to surpass the SNL.

\subsection{Quantum Metrology Using N00N States}
\label{subsec: N00N}
The N00N state is maximally entangled with multiple particles distributed over two modes, presenting one of the most significant and paradigmatic classes of quantum states with a fixed number of particles \cite{Bollinger1996, Dowling:08}. Formally, the N00N state is described as 
\begin{equation}
\left|\psi_{\mathrm{N00N}}\right\rangle=\frac{|N0\rangle+e^{i N \varphi}|0N\rangle}{\sqrt{2}},
\end{equation}
where $N$ is the number of particles in the system, and $\varphi$ is the relative phase between the two modes. Using the N00N state in phase estimation, the ultimate sensitivity without loss is given by 
\begin{equation}
    \Delta \varphi_{\mathrm{N00N}} \geq \frac{1}{N},
\end{equation}
showing performance at the HL. 

This distinct feature of the N00N state makes it an attractive resource to address phase-estimation problems \cite{Dowling:08}. Nevertheless, one should note that creating a N00N state is daunting and typically probabilistic, rendering N00N-state-based sensing impractical in real-world scenarios. To date, only the N00N state with $N=2$ can be deterministically generated via the Hong-Ou-Mandel effect \cite{HOM1987}. The generation of N00N states with $N>2$ has to resort to probabilistic pre-selection or post-selection with the success rate decay exponentially with respect to $N$. Moreover, N00N state is susceptible to loss, causing the QFI to decrease exponentially with the photon number $N$. 

\begin{figure}[hbpt!]
    \centering
    \includegraphics[width=0.95\textwidth]{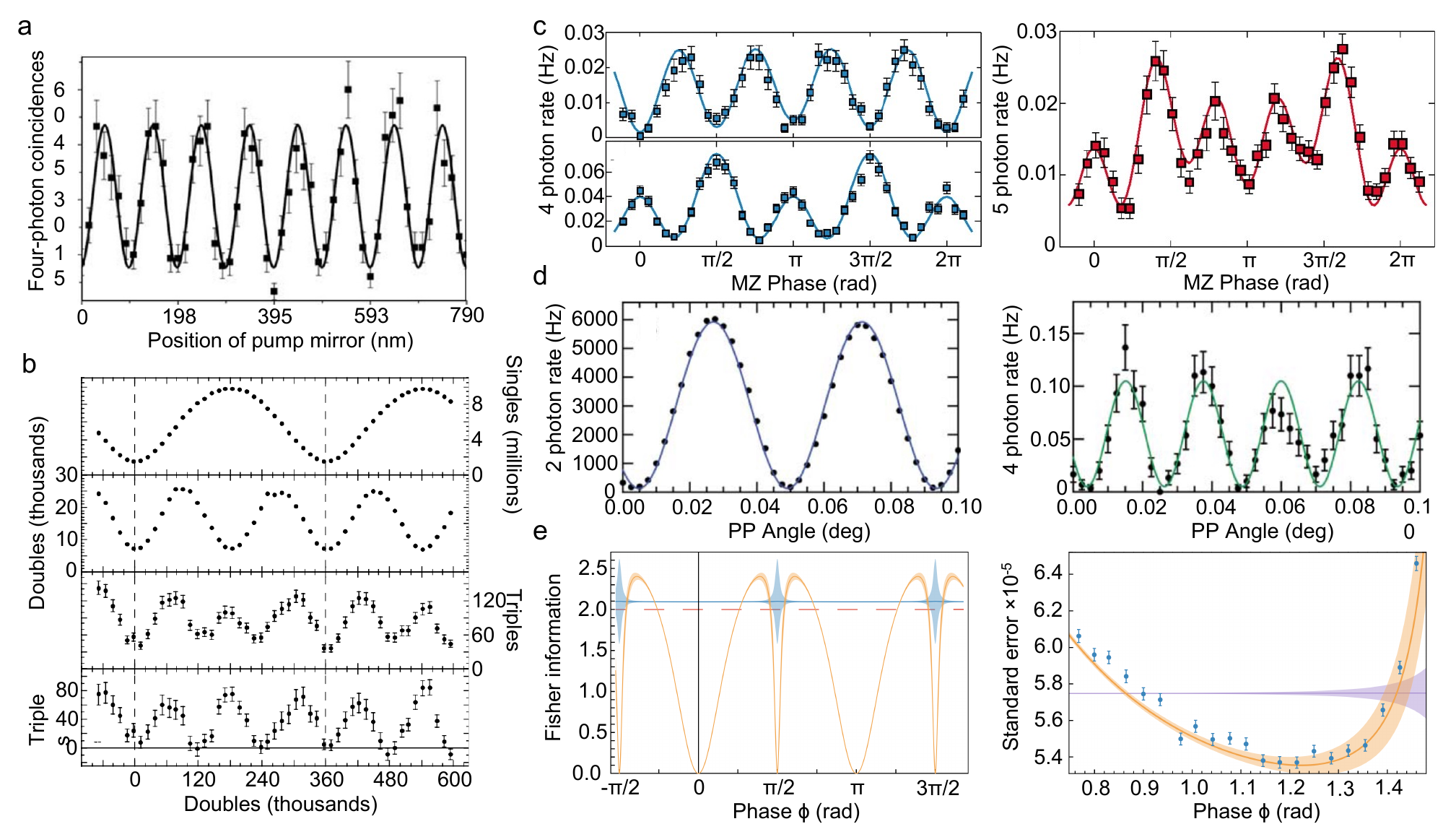}
    \caption{Experimental quantum metrology using N00N states. (a) Pure four-photon interference due to path-entangled four-photon state. Figure reprinted from Ref.~\cite{Walther:04}. (b) Super-resolution phase measurement with two and three photons. Figure reprinted from Ref.~\cite{mitchell2004super}. (c) Experimental results of coincidence measurements, demonstrating 4- and 5-fold super-resolution. Figure reprinted from Ref.~\cite{Afek879}. (d) Two-photon and four-photon count rate with a high-visibility fringe that beats the SNL. Figure reprinted from Ref.~\cite{Nagata:07}. (e)Experimental data for Fisher information and phase uncertainty, with unconditional violation of SNL achieved. Figure reprinted from Ref.~\cite{slusssarenko:17}.}
    \label{Fig:metrologyN00N} 
\end{figure}

The very first few experiments with N00N states were indeed based on post-selection on the detection events. For example, Walther {\em et al.} demonstrated a interferometer using four-photon N00N states generated by an SPDC source \cite{Walther:04}. As shown in Fig. \ref{Fig:metrologyN00N}a, the reduction of the oscillation wavelength derived from post-selected four-photon coincidence events confirmed the super-resolution enabled by the N00N state. At the same time, Mitchell {\em et al.} experimentally demonstrated a technique to produce three-photon N00N states \cite{mitchell2004super} and demonstrated super-resolving phase measurements with post-selected multiphoton events, as shown Fig. \ref{Fig:metrologyN00N}b. These two experiments proved the experimental plausibility to generate N00N states using linear optics in conjunction with post-selection. 
Another important experiment by Afek {\em et al.} generated N00N-like states by mixing a coherent state with the state from SPDC \cite{Afek879}. As illustrated in Fig. \ref{Fig:metrologyN00N}c, $N$-fold super-resolution with up to 5 photons was experimentally observed. These prominent experiments demonstrated the feasibility of producing N00N states and their use in quantum metrology protocols, sparking a series of follow-on experiments that demonstrated super-resolution using N00N states \cite{Okamoto_2008, Guo2008PRA, ChenPRL2010, Wang2016PRL ,hiekkamaki2021photonic, defienne2022pixel}. 
Despite these encouraging results, it is important to note that these experiments only showed super-resolution in estimating an optical phase inside the interferometer, manifesting as the interference fringes oscillating faster than the these obtained with classical states. In this regard, achieving super-resolution does not require surpassing the SNL \cite{demkowicz2015, Sciarrino20review}. In fact, one can achieve super-resolution using carefully prepared classical states without beating the SNL \cite{XiongPRL2005, MaganaPS2016,HashemiRafsanjaniOptica17,Fatemeh2022Nano}.

More recently, researchers explored using N00N states to surpass the SNL \cite{Nagata:07, gao2010experimental, Walmsley:10, israel2014supersensitive, slusssarenko:17, hong2021quantum, hong2022practical}. Fig. \ref{Fig:metrologyN00N}d shows one of the earliest attempts in this direction made by Nagata {\em et al.}, demonstrating that entangled four-photon interference can outperform the SNL \cite{Nagata:07}. However, it should be emphasized that Nagata {\em et al.}'s study employed conditional measurements based on post-selection and therefore did not constitute an unconditional violation of the SNL. The probabilistic nature of the protocol, as well as the loss incurred in the experiment and the low detection efficiency made the task of unconditional surpassing of the SNL arduous. Recently, Slussarenko {\em et al.} demonstrated the first unconditional violation of SNL in quantum metrology protocol based on a high-fidelity N00N state with $N=2$ photons \cite{slusssarenko:17}, with experimental data depicted in Fig. \ref{Fig:metrologyN00N}e. The high generation efficiency of the N00N state and the low overall loss in the experiment paved the way for such a breakthrough. 

\subsection{Quantum Metrology Using Squeezed States}
\label{subsec: squeezed}
As discussed in Sec. \ref{subsec: entanglement_intro}, the squeezed state has reduced noise in one field quadrature at the expense of increased noise in the other quadrature. The squeezed state can be leveraged to improve the sensing precision if the information of interest resides in the less noisy quadrature and is captured by homodyne measurement on that quadrature. In 1987, Xiao {\em et al.} utilized squeezed states to estimate the optical phase at a measurement precision exceeding the SNL \cite{Kimble:87}. Almost simultaneously, Grangier {\em et al.} experimentally demonstrated an squeezed-light-based interferometer beating the SNL \cite{Grangier1987PRL}. The two experiments spurred many subsequent quantum-metrology studies using squeezed states. At present, the most well-known application of squeezed-state quantum metrology is its application to gravitational wave detection \cite{abbott2009ligo, harry2010advanced, aasi2015advanced}. The original Laser Interferometer Gravitational-Wave Observatory (LIGO) was designed to measure gravitational waves that induce distance changes on the order of $10^{-18}$ m \cite{abbott2009ligo, harry2010advanced, aasi2015advanced}. However, the photodetection shot noise fundamentally limited the original LIGO's frequency of gravitational observations. To mitigate this impediment, the advanced LIGO enhanced by squeezed light was developed and employed, now operating below the SNL \cite{McKenzie2002PRL, goda2008quantum, GroteLIGOPRL2013, Virgo2019PRL, LIGOPRL2019}. Readers should refer to Refs. \cite{abbott2009ligo, schnabel2010quantum, harry2010advanced, aasi2015advanced, Barsotti_2019} for more comprehensive reviews on for gravitational-wave detection enhanced by squeezed light.

\begin{figure}[htbp!]
    \centering
    \includegraphics[width=\textwidth]{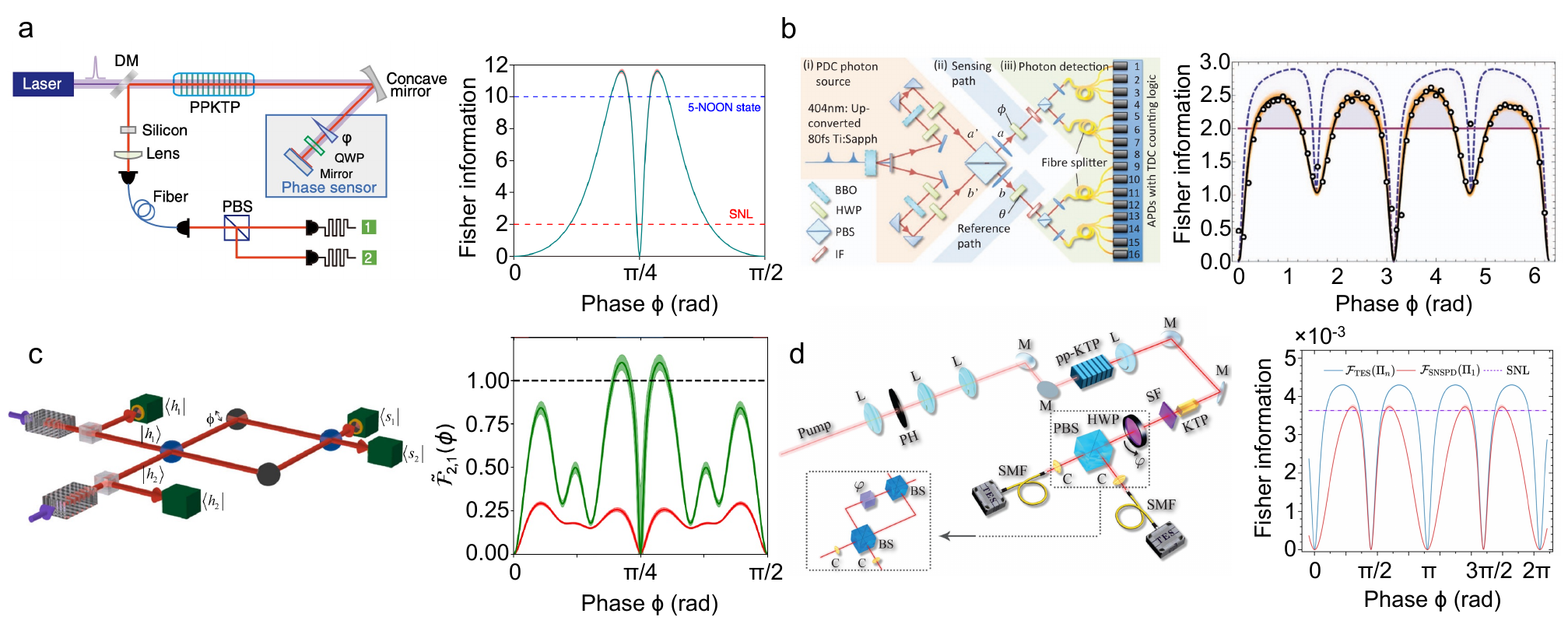}
    \caption{Experimental quantum metrology using squeezed states with photon counting. (a) The experimental setup of an unconventional nonlinear interferometer that achieves an unconditional and robust quantum metrological advantage. The Fisher information per photon exceeds the predicted value using an ideal five-photon N00N state and perfect detectors. Figure reprinted from Ref.~\cite{Qin2023PRL}. (b) Schematic of the experimental setup for quantum metrology with photon counting. The Fisher information extracted from multiphoton interference fringes shows advantage over the SNL. Figure reprinted from Ref.~\cite{Obrien:16}. (c) An interferometric scheme using TMSV states and photon-number-resolving detectors. The calculated Fisher information of the post-selected events beats the SNL. Figure  reprinted from Ref.~\cite{Thekkadath20} (d) Experimental setup to demonstrate multiphoton quantum-enhanced phase estimation. The Fisher information for detection scheme with photon-number resolution outperforms the SNL and beats the detection scheme without photon-number resolution. Figure reprinted from Ref.~\cite{You2021APR}.}
    \label{Fig:metrologysqz}
\end{figure}

Apart from detecting the field quadratures, one can also measure the photon statistics of squeezed states to achieve quantum enhancements in metrology. A common approach is to exploit the TMSV state generated through the SPDC process, as previously discussed in Sec.~\ref{subsec: entanglement_intro}, in phase sensing. In fact, as shown in Sec.~\ref{subsec: N00N}, many quantum metrology protocols create the desired N00N state from a TMSV-state source \cite{Walther:04, mitchell2004super}. As discussed in Sec.~\ref{sec:entanglement_generation_theory}, at weak squeezing levels, i.e., low gain of the parametric process, the generated TMSV state is effectively a photon-pair source with the multi-pair probability negligible. In this case, photon pairs from the source in tandem with photon statistics measurements are leveraged to achieve measurement precision superior to the SNL \cite{slusssarenko:17, Zhao2021PRX}.

Very recently, Qin {\em et al.} demonstrated a 5.8-fold enhancement above the SNL in terms of the FI extracted per photon, without discarding photons due to loss and imperfections \cite{Qin2023PRL}. This was achieved using a stimulated squeezing nonlinear interferometer consisting of a pair of two-mode squeezing operations together with two superconducting nanowire single-photon detectors (SNSPDs). The FI per photon as a function of the phase clearly surpassed the SNL and outperformed an ideal 5-N00N state in some regions of the phase space, as shown in Fig. \ref{Fig:metrologysqz}a.

Although an idealized model ignores the multi-pair contributions in a TMSV state, in practice the multi-pair events in SPDC, albeit occurring with small probabilities, induce errors in protocols without accounting for these contributions using photon-number-resolving detectors. To address this issue, Matthews {\em et al.} first used photon-number-resolving detectors based on multiplexing 16 avalanche photodiodes (APDs) \cite{Obrien:16} to measure a four-mode squeezed state (see Fig. \ref{Fig:metrologysqz}b). They were able to detect four-photon coincidences and beat the SNL with post-selection, despite the presence of significant loss. Indeed, the ability to detect multi-photon events opens the door to achieving further quantum advantages with squeezed states \cite{Omar:19}. In particular, the fast development of highly efficient photon-number-resolving detectors such as the superconducting transition edge sensors (TES) has greatly facilitated surpassing the SNL by retaining the photon-number information \cite{Gerrits:12,gerrits2016superconducting,Levine:14}. In this regard, Thekkadath {\em et al.} utilized high-gain SPDC sources and photon-number-resolving detectors to perform interferometry with heralded quantum probes of up to $N = 8$ \cite{Thekkadath20}, as illustrated Fig. \ref{Fig:metrologysqz}c. They employed two type-II parametric down-conversion sources to herald photon pairs and injected the heralded state into an interferometer. At the output, photon-number-resolving measurements were taken to estimate an unknown phase difference. Despite the protocol being intrinsically probabilistic, quantum advantages in phase estimation were observed. In 2021, You {\em et al.} experimentally demonstrated, for the first time, a scalable protocol for quantum-enhanced optical phase estimation across a broad range of phases, without any pre- or post-selected measurements \cite{You2021APR}. The efficient design of an SPDC source combined with photon-number-resolving detection enabled the utilization of all detected photons in an unconditional quantum-enhanced phase estimation. As shown in Fig. \ref{Fig:metrologysqz}d, the ability to probe all possible multi-photon interference events facilitated the identification of complex multipartite interactions occurring in the interferometer. Consequently, this scheme enabled the estimation of a broad range of phases with sensitivities unconditionally surpassing the SNL. Finally, we note that these experiments represent only a subset of the extensive body of research conducted on quantum metrology using squeezed states. Interested readers are encouraged to refer to Refs. \cite{xiang2013quantum, lawrie2019quantum, Sciarrino20review} for greater detail.

\section{Quantum Illumination}

\label{sec: quantum_illumination}

\begin{figure}[t!]
    \centering
    \includegraphics[width = 0.8\textwidth]{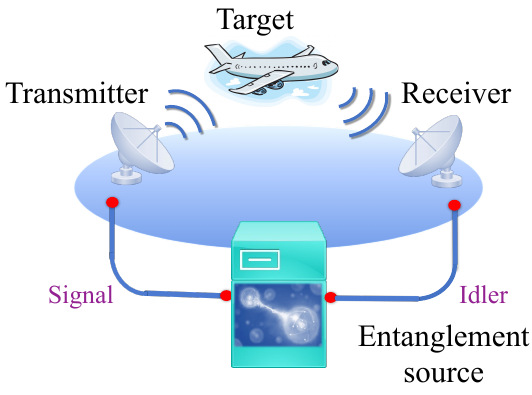}
    \caption{Schematic of quantum illumination for target detection. An entanglement source delivers signal and idler photons to the transmitter and receiver. The transmitter then sends the signal to interrogate a target residing in a lossy and noisy environment. The receiver takes a joint measurement on the signal received from the environment and the idler to infer the presence or absence of the target.}
    \label{Fig:QI}
\end{figure}

Quantum illumination (QI) is an entanglement-enhanced sensing scheme that aims to improve the precision and sensitivity of target detection in the microwave frequency region, as illustrated in Fig.~\ref{Fig:QI}. By entangling the signal probe with locally stored idler reference, QI seeks to enhance the performance of target detection compared to classical approaches~\cite{lloyd2008enhanced,tan2008quantum,shapiro2020quantum}.

In the original QI proposal, which focused on determining the absence or presence of a target, the error-probability exponent exhibited a 6 dB advantage resulting from the utilization of entanglement~\cite{tan2008quantum}. This advantage was achieved while consuming the same probe power as the classical target detection scheme. Recently, QI has been extended to enhance target ranging and angle detection~\cite{zhuang2021quantum,zhuang2022ultimate}. These extensions show the potential for even greater advantages over classical methods in the intermediate signal-to-noise ratio (SNR) region, thanks to the threshold phenomena of nonlinear parameter estimation~\cite{zhuang2022ultimate}. Furthermore, the same mechanism that enables the quantum advantage in target detection has also been applied to phase-sensing scenarios~\cite{hao2022demonstration}. This demonstrates the versatility and potential applicability of QI beyond its original context.

\subsection{Theory for Quantum Illumination}

In the QI protocol, a total of $M$ signal-idler pairs are utilized for the sensing task. Each pair consists of a signal mode $\hat a_S$ and an entangled idler mode $\hat a_I$, generated through the SPDC process described in Eq.~\eqref{eq:dynamics_nondegenSPDC}. The signal mode is sent to probe a potential target, while the returned mode $\hat a_R=\sqrt{\kappa} \hat a_S+\sqrt{1-\kappa} \hat e$ is collected at the quantum receiver. Here, $\kappa$ represents the transmissivity, and $\braket{\hat e^\dagger \hat e}=N_B/(1-\kappa)$ denotes the mean photon number of the background noise. The idler mode is stored locally, awaiting a joint measurement with the returned signal at the quantum receiver.

In the original QI target-detection protocol proposed by Tan \textit{et al.}~\cite{tan2008quantum}, where the target's presence or absence is equally probable, the asymptotic error probability derived from the quantum Chernoff bound~\cite{audenaert2007,pirandola2008} is given by
\begin{align}
P_{E}^{\rm QCB}\sim \frac{1}{2}\exp\left(-\frac{M\kappa N_S}{N_B}\right),
\end{align}
where $N_S \equiv \braket{\hat a_S^\dagger \hat a_S}$ represents the signal brightness. QI operates in the regime of weak signal and strong background noise, characterized by $N_S\ll1$ and $N_B\gg1$. On the other hand, with the same set of parameters, the optimal classical illumination (CI) scheme, employing a coherent-state transmitter and a homodyne-detection receiver, exhibits an error probability given by
\begin{align}
P_{C}\sim \frac{1}{2}\exp\left(-\frac{M\kappa N_S}{4N_B}\right),
\label{PE_QCB}
\end{align}
where the exponent is four times inferior to that of QI. Notably, the advantage of QI over CI persists even in the presence of entanglement-breaking noise and loss, which render the correlation between the returned signal and retained idler fully classical. However, the quantum Chernoff bound alone does not provide guidance on how to design a structured quantum receiver to fully exploit QI's advantage.

Before delving into the design problem of the quantum receiver, it is important to highlight that the original QI proposal utilizes a quantum transmitter that emits a TMSV state, which has been demonstrated to be optimal for target-detection tasks in various scenarios~\cite{de2018minimum,nair2020fundamental,bradshaw2021optimal}. Reference~\cite{de2018minimum} showed that the TMSV state is optimal in the asymmetric scenario, where the objective is to minimize the decay rate (maximize the error-probability exponent) of the probability of a false positive given a certain probability of a false negative. Conversely, the coherent state is the optimal input in the absence of an ancilla, such as the idler mode. Reference~\cite{nair2020fundamental} demonstrated that in the symmetric scenario of minimizing the overall error probability, the TMSV state is optimal in the regime of low signal brightness and bright thermal noise. Furthermore, Ref.~\cite{bradshaw2021optimal} showed that in the limit of zero reflectivity, the TMSV state minimizes the error probability, while the coherent state is optimal among unentangled sources. However, for scenarios with non-zero reflectivity and finite signal and noise brightness, other quantum states for the probe may slightly outperform the TMSV state.

The real challenge in QI lies in the design of the quantum receiver, which aims to implement a joint measurement on the returned signal and idler modes while harnessing the advantage of QI over CI. The optimal measurement for a specific quantum-state discrimination or parameter estimation problem can generally be solved mathematically by formulating a set of positive operator-valued measure (POVM) elements. However, the practical implementation of such a measurement requires a structured design comprising experimental modules that can realize the desired POVM elements while considering practical technological constraints. For example, in the microwave domain, quantum-limited detection requires extensive cooling due to ambient noise backgrounds, and basic measurement apparatuses such as photon-counting modules are less developed compared to the optical domain~\cite{dixit2021searching,assouly2022demonstration}.

The first generation of quantum receivers for QI was based on parametric amplification~\cite{Guha2009}. Two specific receiver designs emerged from this approach: the optical parametric amplifier receiver (OPAR) and the phase-conjugate receiver (PCR), both achieving half of the six decibel error-probability exponent advantage offered by QI. 

The OPAR operates by applying a two-mode squeezing operation, as described by Eq.~\eqref{eq:dynamics_nondegenSPDC}, to create the mode $\hat c=\sqrt{g} \hat a_{I}+\sqrt{g-1}\hat a_{R}^\dagger$ from each pair of signal and idler modes, $\hat a_{I}$ and $\hat a_{R}$. Photodetection is then performed on all the $\hat c$ modes to determine the presence or absence of the target. The optimal gain for the two-mode squeezing operation is derived as $G\sim 1+ \sqrt{N_S}/N_B$. Similarly, the PCR follows a similar principle by first generating a phase-conjugate mode using $\hat c=\sqrt{g} \hat a_{v}+\sqrt{g-1}\hat a_{R}^\dagger$, where $\hat a_v$ represents a vacuum mode. The phase-conjugate mode is then interfered with the idler mode, followed by balanced photodetection. The PCR offers similar performance to the OPAR but provides practical benefits in high-noise regimes~\cite{hao2021entanglement}. In the regime of weak signal and strong noise, where $N_S\ll1$ and $N_B\gg1$, the error probability for both the OPAR and PCR can be approximated as:
\begin{align} 
P_{E}^{\rm OPAR}\sim P_{E}^{\rm PCR}\sim \frac{1}{2}\exp\left(-\frac{M\kappa N_S}{2N_B}\right).
\end{align}
Although the OPAR and PCR are sub-optimal compared to the ultimate quantum limit, they offer the advantage of using off-the-shelf components and have been successfully implemented.

\subsection{Quantum Illumination Experiments}
\begin{figure}[t!]
    \centering
    \includegraphics[width = 0.8\textwidth]{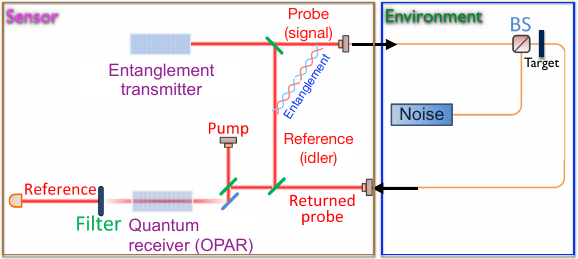}
    \caption{Diagram of the proof-of-concept quantum illumination target detection experiment reported in Ref.~\cite{zhang2015}. The setup entails an non-degenerate entanglement transmitter and a quantum receiver based on OPAR. The environmental noise is emulated by injecting optical thermal noise.}
    \label{Fig:QI_exp}
\end{figure}

A proof-of-concept demonstration of QI for target detection was carried out in the optical domain~\cite{zhang2015}. As schematically depicted in Fig.~\ref{Fig:QI_exp}, the QI sensor comprised an entanglement transmitter and a quantum receiver. The entanglement transmitter entailed a periodically poled lithium niobate (PPLN) crystal pumped by a laser to generate entangled signal and idler, serving respectively as the probe and reference, through a non-degenerate SPDC process as illustrated in Fig.~\ref{fig:CV_entanglement_sources}a. The probe was transmitted to a lossy and noisy environment where a target may reside. The environmental noise was emulated by injecting optical thermal noise to mimic the microwave scenario. The probe returned from the environment was first combined with the reference retained in a quantum memory composed of a fiber spool and then with the pump prior to being processed at an OPAR composed of a second PPLN crystal and other optics for filtering and collection. The OPAR converted the phase information carried on the returned signal into amplitude modulation on the reference. By detecting the reference at the OPAR output, one can infer the properties of interest for the target such as its presence or absence. In this experiment, the entanglement transmitter and OPAR enabled a $\sim 20\%$ SNR advantage despite a 14 dB overall environmental loss and a noise background 75 dB more intense than the signal.

The proof-of-concept QI experiment in the optical domain has verified the benefit of entanglement in an entanglement-breaking sensing environment. However, to demonstrate the entanglement-enabled advantage, the optical QI experiment had to artificially mixed thermal noise with the probe signal as the ambient noise at the optical wavelengths is negligible. In contrast, blackbody radiation noise at the microwave frequencies is appreciable, rendering QI a useful tool to address microwave sensing problems. To implement QI in the microwave domain, a proposal based on optical-microwave transduction was conceived~\cite{Barzanjeh_2015}. The proposed QI protocol exploits entangled microwave signal and optical reference. The storage of the optical reference does not require cooling as one can simply store it in an optical fiber loop with insignificant loss. The returned signal is up-converted from the microwave to the optical domain to enable quantum-limited joint measurements with the reference. Notably, the state-of-the-art optical-microwave transduction efficiencies~\cite{lauk2020perspectives,awschalom2021development,fan2018superconducting,han2021microwave} still fall short of what is needed by the proposed microwave QI protocol. 

\begin{figure}[t!]
    \centering
    \includegraphics[width = \textwidth]{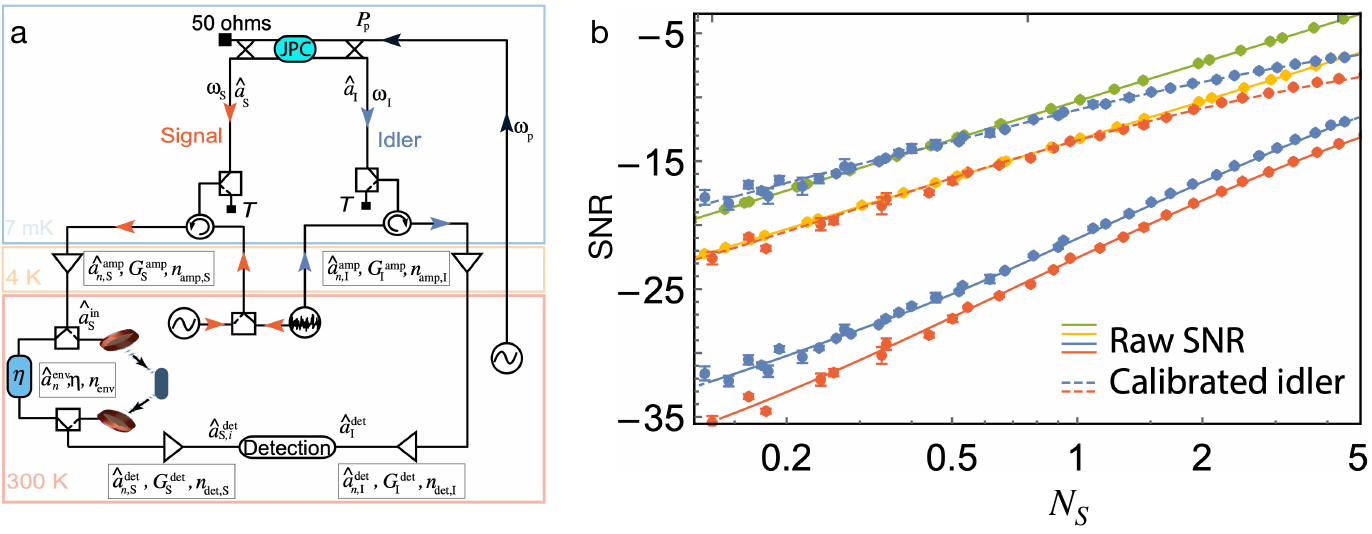}
    \caption{Microwave quantum illumination experiment. (a) Schematic for microwave QI based on a Josephson parametric converter (JPC) to generate entangled microwave photons and a digital PCR. Both the probe (signal) and reference (idler) photons are first amplified and subsequently detected at room temperature. (b) Measured SNR at different source brightness $N_S$. Solid blue: microwave QI based on a JPC entanglement transmitter and a digital PCR; solid orange: CI based on correlated probe and reference; solid green: CI based on a coherent-state transmitter and homodyne receiver; solid yellow: CI based on a coherent-state transmitter and heterodyne receiver; dashed blue: calibrated microwave QI. Figures reprinted from Ref.~\cite{Barzanjeh_2020}. }
    \label{Fig:microwave_QI}
\end{figure}

On the experimental front of microwave QI, Ref.~\cite{Barzanjeh_2020} reported a setup based on Josephson parametric converter (JPC) as the entanglement transmitter and a digital PCR as the quantum receiver, as shown in Fig.~\ref{Fig:microwave_QI}a. JPC generated entangled microwave signal photons serving as the probe and idler photons serving as the reference. The physical process of entanglement generation from a JPC resembles that of the SPDC in the optical domain as they can both be described mathematically by the Hamiltonian in Eq.~\eqref{eq:SPDC_Hamiltonian}. Unlike the OPAR and PCR that perform joint measurements on the returned probe and the reference at the receiver, the digital PCR took a measurement on the reference microwave photons at the transmitter and recorded the digitized measurement outcomes. A second set of measurements were taken on the returned probe to construct the covariance matrix of the probe and the reference. Shown in Fig.~\ref{Fig:microwave_QI}b, the SNR of the microwave QI plotted alongside these of the CI based on correlated probe and reference as well as with the coherent-state transmitter in tandem with a homodyne receiver. Due to the limitation of the digital PCR, its associated SNR could not surpass that of the optimal CI scheme based on a coherent-state transmitter and a homodyne receiver. Nonetheless, the experiment did show that the calibrated SNR for the microwave QI, assuming the availability of a physical PCR and perfect detectors, would beat that of the optimal CI, thereby constituting a significant step in experimental microwave QI. In a more recent microwave QI experiment, two microwave resonators coupled by a Josephson ring modulator were employed to produce entangled microwave photons and to serve as the OPAR~\cite{assouly2022demonstration}. The microwave OPAR exploited a superconducting qubit dispersively coupled to a microwave resonator, in lieu of a photodetector as used in the optical QI experiment, to read out the state of the microwave resonator that stored the reference photons. This microwave QI experiment achieved a $\sim 20\%$ advantage over the optimal CI scheme in the error-probability exponent, on par with the demonstrated quantum advantage in the optical QI experiment~\cite{zhang2015}.

The above experiments based on OPAR and PCR are limited in their quantum advantage, due to the experimental constraints as well as the receivers being sub-optimal. To design an optimal receiver, one needs to construct a measurement to perform the optimal quantum state discrimination between multiple copies of identical quantum states $\hat \rho_{RI, 0}^{\otimes M}$ versus $\hat \rho_{RI, 1}^{\otimes M}$, where the indices $0,1$ represent the target being present or absent. In general, for mixed states, local operations on each copy and collectively processing the data from $M$ copies cannot solve such an optimal state discrimination problem~\cite{calsamiglia2010local}. A joint measurement on $M$ copies is in general needed by an optimal QI receiver. 

The first optimal receiver design was proposed in Ref.~\cite{zhuang2017optimum}. The design utilizes a sum-frequency generation process to jointly interact the $M$ return-idler pairs and produce a sum mode in a coherent state embedded in weak thermal noise. The optimal receiver for QI adopts a feedforward mechanism, in analogy to the Dolinar receiver~\cite{dolinar_processing_1973} designed for optimal coherent-state discrimination. However, the optimal receiver requires unit-efficiency sum-frequency generation at the single photon level, therefore challenging experimental realizations. 

A recent proposal of the optimal receiver design based on correlation-to-displacement (`$\rm C\veryshortrightarrow D$') conversion surprisingly shows that an optimal receiver design can be achieved by separately heterodyne-detecting the return mode, and consequently processing the idler field~\cite{shi2022fulfilling}. Conditioned on the results of the heterodyne detection of the return modes, the associated idler modes are in coherent states embedded in weakly thermal noise. A beamsplitter array will then be combining the idlers---coherently add up the displacements while maintaining the same level of noise. By well-established coherent-state discrimination protocols~\cite{dolinar_processing_1973}, one can achieve the optimal error probability of quantum illumination in Eq.~\eqref{PE_QCB}. In fact, it achieves a tighter lower bound of the error probability of QI with an arbitrary entanglement source~\cite{nair2020fundamental}.

\begin{figure}[t!]
    \centering
    \includegraphics[width = 0.75\textwidth]{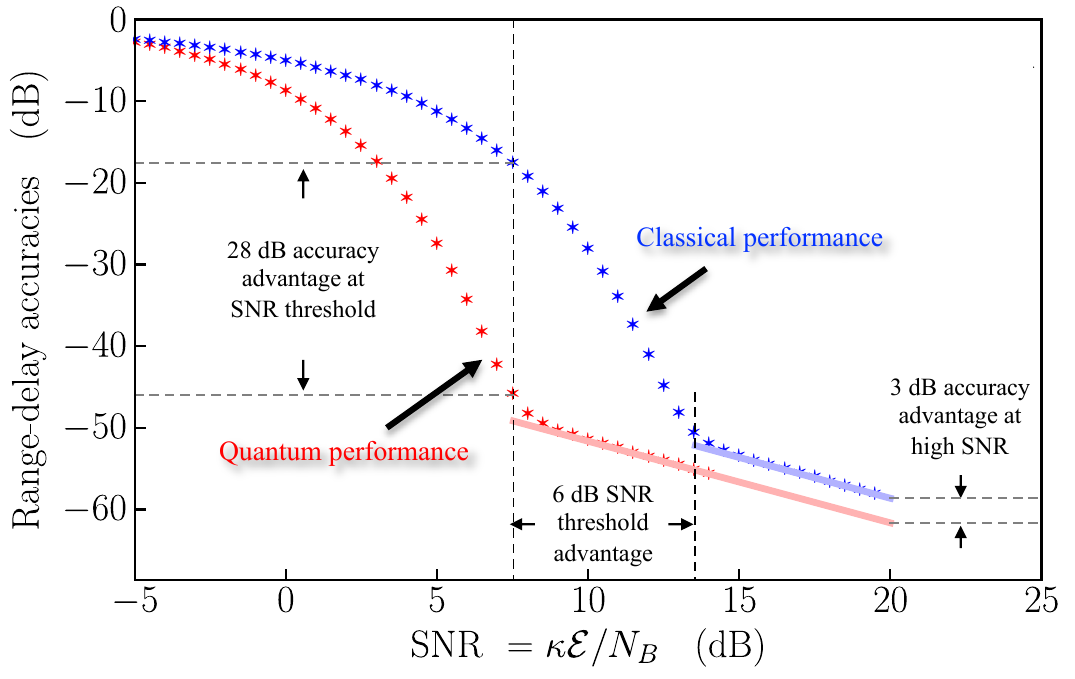}
    \caption{Quantum advantage in ranging at the long integration time limit. The plot assumes a bandwidth of $10^6$ HZ and an initial range uncertainty of $5$ kilometers. Figure reprinted from Ref.~\cite{zhuang2022ultimate}.
    }
    \label{Fig:QI_ranging}
\end{figure}

Before we close on the topics of QI for target detection, we would like to introduce extensions of QI to more complex target detection scenarios. While the original QI protocol considers detecting the presence or absence of a single target at a specific spatial spot and at a certain time, real world target detection scenarios are more complex, involving detecting target's range, angle, and speed~\cite{van2001detection3}. Therefore, it is important to extend the quantum advantages in the original QI to general scenarios. Refs.~\cite{zhuang2021quantum,zhuang2022ultimate} showed that quantum advantages in ranging---the estimation of the distance between the observer and the target---is possible. As shown in Fig.~\ref{Fig:QI_ranging}, due to the nonlinear nature of the parameter estimation problem, the quantum range accuracy is much better than the classical one at the intermediate signal-to-noise region. The caveat in such advantage is that the advantage still assumes the weak brightness of the signal field. Therefore, with practical modeling of target ranging, it is found that quantum advantages are limited to detecting targets at hundreds of meters away such that a reasonable integration time of seconds is adequate to provide useful information about target. 

\subsection{Covert Sensing}
\label{subsec: covert}

\begin{figure}[t!]
    \centering
    \includegraphics[width = 0.8\textwidth]{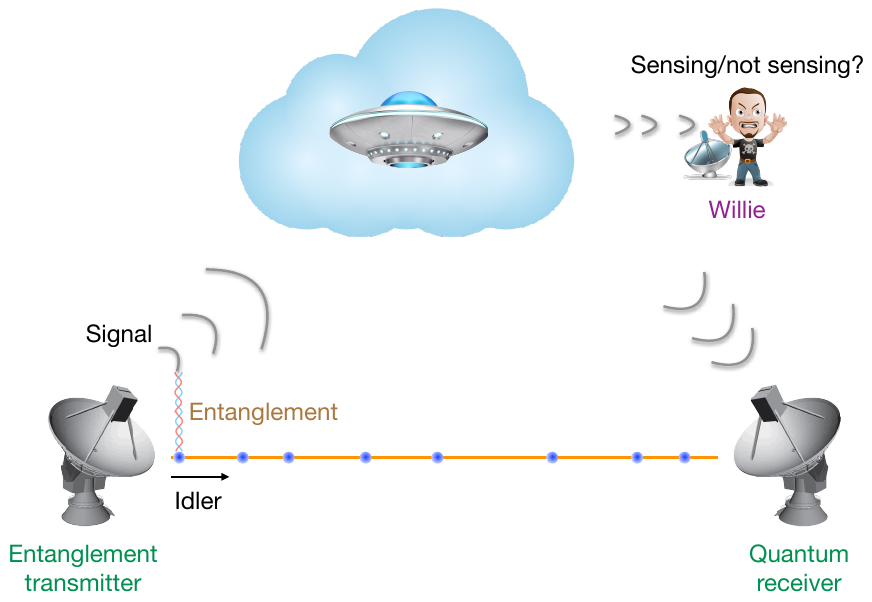}
    \caption{Schematic for entanglement-enhanced covert sensing. The entanglement transmitter generates signal and idler photons to execute the sensing task. The signal photons are used to probe a target residing in a very noisy environment while the idler photons are delivered to the quantum receiver. The opponent, Willie, attempts to detect the sensing operation by measuring the intensity of his captured photons. The quantum receiver performs a joint measurement on the signal photons received from the environment and the retained idler photons to estimate the probed physical properties of the target. By concealing the signal in a large noise background, one can ensure that Willie is unable to detect the sensing operation due to the statistical fluctuation in his measurement. Entanglement enhances the estimation precision and the covertness of the protocol. Figure reprinted from Ref.~\cite{hao2022demonstration}.}
    \label{Fig:covert_concept}
\end{figure}

A variant of the QI protocols is entanglement-enhanced covert sensing. In covert sensing, the probe signal is concealed in a bright noise background such that the opponent, Willie, who is granted access to the probe power scattered in the environment, is unable to detect the sensing attempt. Entanglement can be leveraged to enhance the performance and security of covert sensing, as schematically sketched in Fig.~\ref{Fig:covert_concept}. Akin to QI, in covert sensing the transmitter generates entangled signal and idler modes, transmits the signal to probe an object, and retains the idler for the quantum receiver. Willie aims to detect the signal embedded in a bright noise background but the statistical fluctuation of the noise impedes his capability to observe the sensing attempt. Willie's error probability $\mathbb{P}_e^{\rm (w)} \geq 1/2-\epsilon$, where $\epsilon$ is referred to as the covertness parameter. Within the relevant parameter region, one can show that
\begin{equation}
\label{eq:covertness}
    \epsilon \propto \frac{\sqrt{M}N_S}{N_B},
\end{equation}
where $M$ is the number of probe modes proportional to the integration time, $N_S$ is the per-mode mean photon number for the probe, and $N_B \gg 1$ is the per-mode mean photon number for the noise background. Eq.~\eqref{eq:covertness} suggests that $\sqrt{M}N_S = {\rm constant}$ is needed to fix the covertness parameter at a given background noise level, leading to ${\rm SNR} \propto \sqrt{M}$, known as the square-root law of covert communication and sensing~\cite{bash2015quantum}. In analogy to QI, the quantum receiver takes a joint measurement on the received signal and the idler to infer the physical properties of the object under investigation. The estimation precision is derived by the quantum Cram\'{e}r-Rao bound:
\begin{equation}
    \delta \theta^2_Q \geq \frac{1}{M F},
\end{equation}
where $F$ is the quantum Fisher information~\cite{marian2016quantum}.

\begin{figure}[t!]
    \centering
    \includegraphics[width = 1\textwidth]{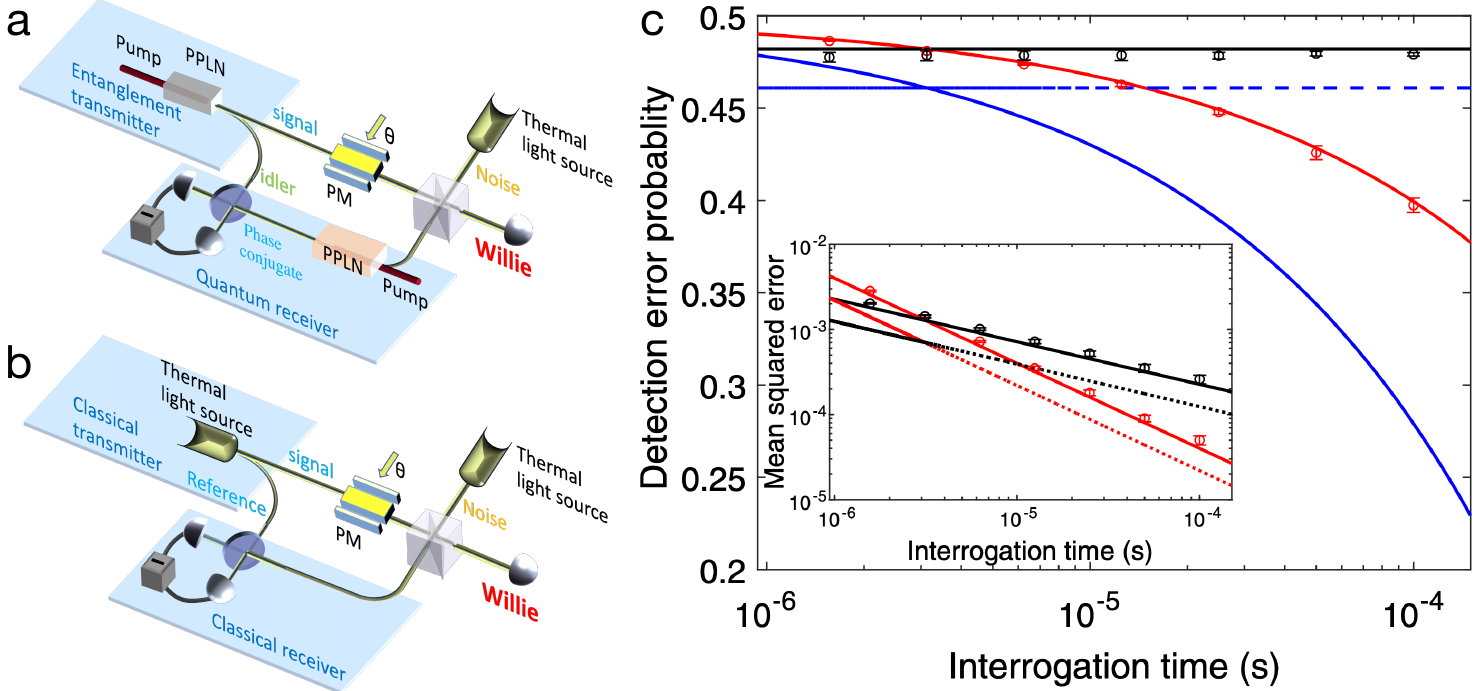}
    \caption{Entanglement-enhanced covert sensing experiment. (a) Diagram of entanglement-enhanced covert sensing experiment based on a non-degenerate entanglement source and a PCR. (b) Diagram of classical covert sensing experiment based on a thermal light source and a balanced homodyne receiver. (c) Experimental data of Willie's detection error probability when the square-root law is obeyed (black) or violated (red). Dots: experimental data. Red and black curves: theory. Blue curves: lower bound for Willie’s detection error probability. Inset: corresponding mean squared errors for classical (solid) and entanglement-enhanced (dotted) covert sensing, highlighting different slopes when the square-root law is obeyed and violated. Figures reprinted from Ref.~\cite{hao2022demonstration}.}
    \label{Fig:covert_data}
\end{figure}

An entanglement-enhanced covert sensing experiment built on a tabletop platform was recently reported~\cite{hao2022demonstration}, as depicted in Fig.~\ref{Fig:covert_data}a. The performance of entanglement-enhanced covert sensing was benchmarked against that of covert sensing based on classical light sketched in Fig.~\ref{Fig:covert_data}b. In the entanglement-based experiment, the transmitter consisted of a PPLN crystal that produced non-degenerate signal and idler modes. The signal served as the probe to measure the phase shift imparted by an object. The environmental noise was emulated by mixing thermal light with the signal. Willie captured a portion of the signal and noise power that did not reach the quantum receiver and aimed to measure its power to infer the sensing attempt. Different from the optical QI target detection experiment with an OPAR~\cite{zhang2015}, a PCR was implemented in the covert sensing experiment. The PCR comprised a second PPLN that generatd a phase-conjugate beam of the signal returned from the environment. The phase-conjugate beam interfered with the retained idler on a 50:50 beam splitter followed by two photodetectors in a balanced configuration. The classical covert sensing scheme employed a thermal light source and a balanced homodyne receiver while Willie's setup remained unchanged. Figure~\ref{Fig:covert_data}c plots Willie's detection error probability as a function of the interrogation time when the square-root law is obeyed (black) or violated (red), with the corresponding estimation mean squared error shown in the inset. One can observe that the mean squared error drops at a slower rate when the square-root law is obeyed, unveiling a tradeoff between performance and security. 

\section{Quantum Imaging with Entangled Photons}
\label{sec: quantum_imaging}

Another fast-growing field of entanglement-based QIT is quantum imaging, which promises enhanced resolutions at ultra-low light intensities and measurements in spectral regions where cameras do not exist. With the development of efficient entanglement sources and advanced measurement devices, a multitude of encouraging demonstrations of quantum imaging have been conducted in recent years. In the following, we will provide a brief introduction to the most important features required for a quantum-light source to be used in nearly all quantum-imaging protocols. We will then highlight a few significant recent developments in the field of quantum imaging. For more extensive background information, we encourage interested readers to refer to recent review articles on quantum imaging, such as Refs. \cite{magana2019quantum, moreau2019imaging, gilaberte2019perspectives}.

\subsection{Entangled-Photon Sources for Quantum Imaging}
Nearly all entanglement-based quantum-imaging protocols hinge on the position and momentum degrees of freedom, as introduced in Sec. \ref{sec: position-momentum entanglement}. The most common method to generate a two-photon position-momentum entangled state is through the SPDC process, in which a pump photon is down-converted into an entangled pair of photons.

A vital property for imaging is the spatial correlation in the position-momentum entanglement. As described in Sec. \ref{sec: position-momentum entanglement}, a paraxial Gaussian pump beam and a specific combination of pump, signal, and idler frequencies (i.e., $\omega_P = \omega_S + \omega_I$) generate entangled states expressed as
\begin{eqnarray}
\ket{\psi}_{\mathbf{q}} = \int d\mathbf{q}_Sd\mathbf{q}_I\Phi(\mathbf{q}_S,\mathbf{q}_I)\ket{\mathbf{q}_S}\ket{\mathbf{q}_I}. \label{eq:SPDC_Imaging}
\end{eqnarray}
Under ideal conditions that lead to perfect anti-correlations in transverse momenta, i.e., $\Phi(\mathbf{q}_S,\mathbf{q}_I)\approx\delta(\mathbf{q}_S+\mathbf{q}_I)$, one can detect, in the far-field of the crystal, a photon in precisely the opposite transverse position of its entangled partner.

In the near-field, which corresponds mathematically to the Fourier transform and experimentally to imaging the crystal onto the observation plane, the joint spectral amplitude can be approximated as $\Phi(\mathbf{x}_S,\mathbf{x}_I)\approx\delta(\mathbf{x}_S-\mathbf{x}_I)$, where $\mathbf{x}_S$ and $\mathbf{x}_I$ are the transverse position vectors. This approximation implies that detecting one photon at a particular transverse position requires the entangled partner photon to be at the same position, which can be utilized for imaging purposes.

For many quantum imaging tasks, the achievable resolution is directly linked to the strength of correlations, making it critical to obtain a state that closely resembles this ideal situation. To achieve strong correlations, the pump should be collimated as much as possible, indicating that larger apertures are advantageous. Additionally, the crystal should be as thin as the required detection rate allows. A highly collimated pump field minimizes the transverse momentum spread, while a thin crystal allows for a larger bandwidth of transverse momenta in the down-converted photons. When both conditions are met, the correlations are maximized, resulting in enhanced imaging resolution. Optimizing quantum imaging parameters shares many similarities with preparing high-dimensional entangled states, and research findings in either field often synergistically advance the other.

In a broad sense, entanglement-based quantum imaging protocols can be divided into three categories: correlation-based imaging, nonlinear interference-based imaging, and imaging using photon-number entangled states. The following sections will delve into each category.

\subsection{Correlation-Based Quantum Imaging}

In correlation-based quantum imaging protocols, the image is obtained by analyzing correlation between measurements, specifically coincidences between two or more involved photons. While these protocols impose high technological requirements on the imaging devices, they offer the advantage of improved signal-to-noise ratios, as discussed below.

One of the most popular imaging methods in the field of quantum imaging is known as ghost imaging falling into the category of correlation-based imaging protocols. In a ghost imaging protocol, the information on the imaged object is not obtained from the detection of individual photons but rather from the correlations between different measurements.

\begin{figure}[hbt!]
    \centering
    \includegraphics[width=1\textwidth]{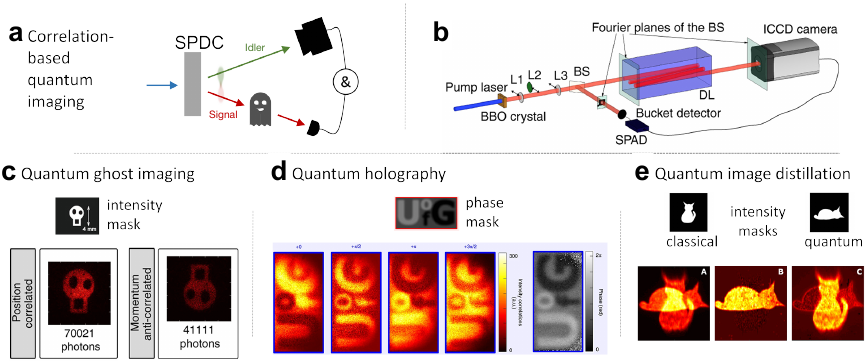}
    \caption{\label{fig:Corr_Imaging}
    Correlation-based quantum imaging. (a) Schematic of a correlation-based imaging protocol, where the object is placed in the path of the signal photon (red), which are not directly imaged. The image is obtained by recording the idler photon (green) and correlating the camera exposures to the detections of its entangled partner. (b) Sketch of an experimental setup for quantum ghost imaging with a triggered ICCD camera. Results obtained with this setup are displayed in (c), showing that by the virtue of the entanglement images can be obtained in the image plane (left) and momentum plane (right) depending on the location of the object and camera after the crystal. (d) Quantum holography using four phase-shifted correlation images (left), from which the phase pattern can be retrieved (right). (e) An image encoded into entangled photons can be distilled through correlation measurements, even when superposed with an image encoded in a strong classical light field. Results show the mixed images (left, A), the distilled quantum image (middle, B), and the classical image obtained through subtracting the two (right, C). (b, c) reprinted from Ref.~ \cite{aspden2013epr}. (d) reprinted from Ref.~\cite{defienne2021polarization}. (e) reprinted from Ref.~\cite{defienne2019quantum}.
    }
\end{figure}

In ghost imaging, a spatially entangled photon pair is split into two beams of single photons: a signal beam and an idler beam. The signal photon is used to illuminate the object to be imaged, and after transmission or reflection, it is detected by a single-pixel detector, known as a bucket detector, that does not provide any spatial information. The idler photons, on the other hand, do not interact with the object and are recorded by a spatially resolving measurement, such as raster scanning with a single-pixel detector or using a single-photon sensitive camera. Fig. \ref{fig:Corr_Imaging}a depicts a schematic of a ghost imaging protocol.

The camera recordings alone do not directly produce an image but rather show the incoherent mixture of all possible detection data, representing the cross-section of the SPDC emission. However, when the camera recordings of the idler photons are correlated with the bucket detector events from the signal photons (i.e., the camera only records the idler photons when their corresponding signal photons are measured), the image is reconstructed based on the spatial correlations.

An intuitive way to visualize such a ghost imaging protocol, and many other protocols based on photon pairs, is through the Klyshko picture \cite{pittman1996two}, in which the SPDC crystal generating entangled photons is replaced by a mirror, and the bucket detector in the signal-photon arm is replaced by a light source. Using this arrangement, the same image recorded by the camera in the idler arm can be observed \cite{aspden2014experimental}.

The first demonstration of ghost imaging in 1995 \cite{pittman1995optical} utilized a scanning approach to record the idler photons due to the limited camera technology at that time. Thanks to the advancement of modern cameras capable of nanosecond triggering and single-photon sensitivity, subsequent demonstrations have showcased the power of spatial correlations in ghost imaging \cite{aspden2013epr, morris2015imaging, moreau2018ghost, wolley2022quantum}. Figure \ref{fig:Corr_Imaging}b illustrates a sketch of a ghost imaging experiment. 

One intriguing feature of entanglement-based ghost imaging is the ability to obtain either the image or its inverted counterpart depending on whether the object and camera are placed in the image or Fourier plane of the crystal, which corresponds to utilizing position correlations or momentum anti-correlations, respectively, as shown in Figure \ref{fig:Corr_Imaging}c. Another significant advantage of ghost imaging is the flexibility to employ two different wavelengths for the signal and idler photons, known as two-color ghost imaging. This allows for wavelength adjustment based on detector and camera efficiencies or noise characteristics \cite{aspden2015photon}. 

While entanglement was initially considered a critical ingredient for ghost imaging, subsequent theoretical and experimental studies raised questions about its necessity. It was shown that thermal light sources and properly utilized coherent light can also be employed in correlation imaging protocols, exhibiting similar features (for further details on the debate and classical ghost imaging results, interested readers can refer to \cite{erkmen2010ghost, shapiro2012physics, padgett2017introduction, moreau2018ghost} and references therein). These debates helped to identify specific quantum advantages. One evident quantum feature is the possibility of achieving perfectly correlated images in the near and far fields, resulting in inverted images \cite{aspden2013epr}. Moreover, entanglement-based ghost imaging enables signal-to-noise ratios beyond what is attainable with classical correlation imaging, thanks to the perfect correlations of SPDC and its quantum statistics \cite{MaganaAPL2013, meda2017photon}. In fact, by employing a similar protocol to the aforementioned setup but imaging both entangled photons with the same camera, sub-SNL imaging has been demonstrated \cite{brida2010experimental}. Overall, the advantage of entanglement-based ghost imaging is particularly significant at very low light levels, as quantum statistics offer a means to surpass the limitations of classical Poissonian statistics, which exhibit increasing uncertainty in the few-photon regime.

An interesting question regarding the spatial resolution limits arises in ghost imaging protocols: which of the three wavelengths for pump, signal, and idler defines the diffraction limit? It has been found that the so-called effective de Broglie wavelength, which is half the wavelength of the photon pair, sets a fundamental limit \cite{santos2005generation}. Additionally, it has been demonstrated that the resolution is not only limited by the point spread function of the camera system but also by the quality of the spatial correlations \cite{moreau2018resolution}.

In recent decades, the concept of utilizing quantum correlations for imaging applications has been extended to various regimes and applications. Here are a few noteworthy examples:

1. {\em Quantum Secure Imaging}: Quantum correlations, in combination with polarization and the BB84 quantum cryptography protocol, have been employed to develop a quantum secure imaging protocol. This protocol enables the detection of any unwanted changes to the image, such as those caused by eavesdroppers \cite{malik2012quantum}.

2. {\em Combination with Interaction-Free Imaging}: Correlation-based quantum imaging has been successfully combined with interaction-free imaging, another popular quantum imaging technique. By incorporating spatially-resolved interference effects and correlation-based ghost imaging, it becomes possible to obtain both intensity and phase images of an object. This counter-intuitive approach allows imaging through the detection of photons that cannot have interacted with the object \cite{zhang2019interaction}.

3. {\em Image Recognition}: Recent research has shown that by structuring the pump beam and employing special pattern projection techniques during the detection process, an image recognition protocol based on ghost imaging principles can be realized. This approach offers promising possibilities for image recognition tasks \cite{qiu2019structured}.

4. {\em Four-Photon Entanglement}: Entanglement-based imaging has been extended to four-photon states using entanglement swapping. By projecting two photons onto an anti-symmetric state, it has been demonstrated that the image placed into one arm of a photon pair can be swapped to the other pair while being contrast-inversed, showcasing the complexity of achieved quantum imaging operations \cite{bornman2019ghost}.

5. {\em Holographic Tasks}: Quantum correlation-based imaging concepts have been adapted for holographic tasks. Fig. \ref{fig:Corr_Imaging}d presents some of the results. In such a scheme, spatial correlations derived from SPDC are utilized to perform holographic reconstructions of birefringent images encoded in polarization entangled states. This approach offers insensitivity to spatial phase disturbances, enhanced spatial resolution due to reduced de Broglie wavelength of the bi-photon state, and strong resistance to classical noise \cite{defienne2021polarization}.

6. {\em Quantum Image Distillation}: Quantum correlation-based imaging techniques have been applied to quantum image distillation. This refers to the retrieval of images based on quantum correlations (not necessarily entanglement) from a strong uncorrelated noisy background or when overlaid with a strong classical light field carrying its own imaging information. Quantum image distillation experiments have demonstrated the capability to extract quantum images from such challenging scenarios \cite{defienne2019quantum} (see Fig. \ref{fig:Corr_Imaging}e for experimental results).

These examples illustrate the diverse range of applications and advancements in quantum imaging based on correlation and entanglement principles.

\subsection{Quantum Imaging Using Nonlinear Interference}

While correlation-based quantum-imaging protocols offer numerous advantages, they also pose challenges, particularly associated with the camera technology, as many of these protocols require cameras to be single-photon sensitive with high timing resolution or necessitate special post-processing for correlation extraction. To combat these challenges, a new quantum-imaging approach based on nonlinear interferometry has emerged that leverages entanglement without the need for correlation measurements. In nonlinear interferometry, nonlinear crystals are used in lieu of beamsplitters in an interferometer \cite{yurke19862}, yielding nonlinear interference that has found various applications in quantum optics, most notably in quantum-enhanced phase sensing \cite{chekhova2016nonlinear}.

\begin{figure}[hbt!]
    \centering
    \includegraphics[width=1\textwidth]{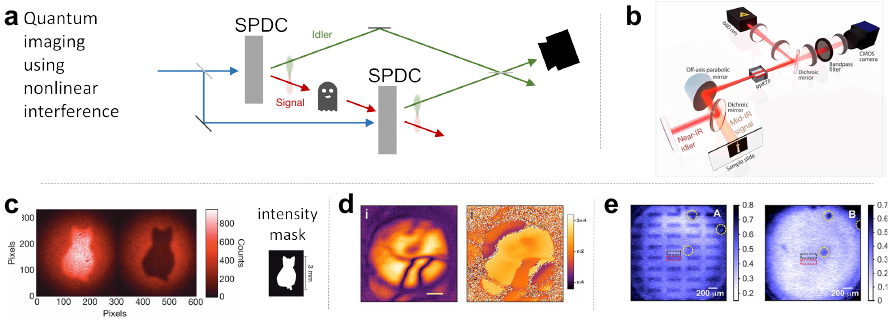}
    \caption{\label{fig:NL_imaging}
    Quantum imaging based on nonlinear interference. (a) Schematic of a quantum imaging protocol based on nonlinear interference, where the object is placed in the path of the signal photon (red) between the two crystals. The image is obtained by recording the interference of the idler photon (green) without the need for correlation measurements. (b) Experimental setup for quantum imaging using nonlinear interference implemented in the commonly used Michelson geometry to perform mid-infrared microscopy. Figure reprinted from Ref.~\cite{kviatkovsky2020microscopy}. (c) Captured interference images at both output ports of the beamsplitter in the idler arm when an intensity mask is placed into the arm of the signal photon. Figure reprinted from Ref.~\cite{lemos2014quantum}. (d) Mid-infrared absorption (left) and phase (right) microscopy images recorded with the idler at around 800 nm. Scale bar, 200 $\mu$m. Figure reprinted from Ref.~\cite{kviatkovsky2020microscopy}. (e) Hyperspectral absorption microscopy images in the mid-infrared for signal (idler) wavelengths of 3.18 $\mu$m (653.2 nm) on the left and 3.32 $\mu$m (638.8 nm) on the right showing the sensitivity of the sample to different probe wavelengths. Figure reprinted from  Ref.~\cite{paterova2020hyperspectral}.
    }
\end{figure}

Quantum imaging based on nonlinear interference builds on the demonstration by Zou, Wang, and Mandel to showcase the counterintuitive effect of induced coherence \cite{zou1991induced, lemos2014quantum}. The method involves coherently pumping two SPDC crystals in a way that only one pair is generated. The signal photon from the first crystal is then directed into the second crystal, aligned to perfectly overlap with the path of the signal photon from the second crystal, as sketched in Fig. \ref{fig:NL_imaging}a. In measuring the signal photon, it is impossible to distinguish which SPDC process has occurred, thereby also inducing coherence in the idler photon. As a result, the idler photons' paths from both crystals can be brought to interference using a beamsplitter. Importantly, the interference occurs regardless of whether the signal photon is detected, eliminating the need for correlations. Remarkably, an object placed in the signal path between the two crystals manifest as a reduction in the interference visibility in the detection of the idler photons. Therefore, the image of the object can be directly retrieved without correlation measurements, a technique known as ``quantum imaging with undetected photons'' \cite{lemos2014quantum}. Since the setup acts as an interferometer, phase images, in addition to amplitude images, can be captured by modifying the interference pattern (Fig. \ref{fig:NL_imaging}c-e). Similar to correlation-based imaging, the photon interacting with the object does not need to have the same wavelength as the photon recorded by the camera. Additionally, it does not require detecting the photons directly interacting with the object, offering the potential to image objects at wavelengths where cameras and detectors may not exist.

Since the introduction of quantum imaging with undetected photons in 2014, the concept has undergone significant developments and expansions in various settings. These advancements include achieving video rate recording speed \cite{gilaberte2021video}, exploring classical analogs \cite{cardoso2018classical}, extending its application to sensing in the THz frequency regime \cite{kutas2020terahertz}, utilizing it for mid-infrared microscopy \cite{kviatkovsky2020microscopy, paterova2020hyperspectral} (see Fig.~\ref{fig:NL_imaging}b), and employing it in holography tasks \cite{topfer2022quantum}. Furthermore, it has been applied to spectroscopy \cite{kalashnikov2016infrared, kutas2021quantum} and integrated into frequency-domain optical coherence tomography \cite{vanselow2020frequency}. In the imaging domain, recent advancements have combined it with single pixel imaging and interaction-free measurement techniques, leading to remarkable outcomes \cite{yang2023interaction}. Leveraging nonlinear imaging enables spatially resolved imaging without the need for cameras or raster scanning, without direct interaction with the object, and in wavelength regions where good detectors do not exist.

A current area of extensive study in quantum imaging with undetected photons revolves around understanding the resolution limits of the protocol. Research has shown that resolution strongly depends on the quality of momentum correlations, specifically the position-momentum entanglement of the SPDC process \cite{fuenzalida2022resolution}. In the paraxial regime, resolution is determined by the wavelength of the undetected photon. However, in tightly focused SPDC processes with significant non-paraxial effects, quantum imaging with undetected photons becomes fundamentally diffraction-limited by the longer wavelength between the signal and idler \cite{vega2022fundamental}. Nonetheless, proposals for achieving subdiffraction imaging with undetected photons have been put forward \cite{santos2022subdiffraction}.

Finally, it is worth mentioning that while the concept of nonlinear interference can be connected to different protocols for entanglement generation \cite{hochrainer2022quantum}, and the setup has been employed to quantify momentum correlations \cite{hochrainer2017quantifying}, entanglement may not be indispensable for such protocols as position correlations alone appear sufficient for image retrieval \cite{viswanathan2021position}. Consequently, the role of entanglement and its advantages in the domain of quantum imaging with undetected photons requires further investigation \cite{shapiro2015classical}.

\subsection{Imaging with Entangled Photon-Number States}

In the realm of quantum imaging, it is worth exploring the imaging-related advantages offered by entangled photon-number states, commonly referred to as N00N states (see also Sec. \ref{sec: quantum_metrology}). These states have been shown to surpass the standard diffraction limit of light wavelength $\lambda$ by a factor of $N$.

\begin{figure}[hbt!]
    \centering
    \includegraphics[width=1\textwidth]{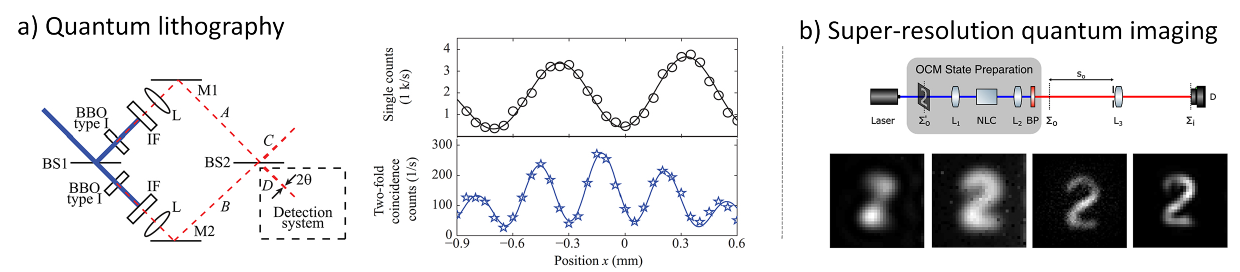}
    \caption{\label{fig:PhotNum_imaging}
    Quantum imaging using entangled photon-number states. (a) Schematic for quantum lithography based on 2-photon path-entangled N00N states generated through parallel pumping of two crystals (left). When brought to interference on a beam splitter (BS2), two-photon correlations together with optical centroid measurements result in a high-visibility fringe pattern with double the frequency (bottom right) compared to single-photon interference fringes (top right). Figure reprinted from \cite{shin2011quantum}. (b) Schematic for super-resolution imaging of an object illuminated with 405 nm light and imaged using 810 nm photons with the optical centroid measurement technique (top). The bottom images show recordings using spatially coherent light at 810 nm, spatially incoherent light at 810 nm, two-photon states at 810 nm, and coherent light at 405 nm, respectively (from left to right). Figures reprinted from \cite{unternahrer2018super}.
    }
\end{figure}

One early example of utilizing N00N states in imaging applications is quantum lithography sketched in Fig.~\ref{fig:PhotNum_imaging}a, where an interference pattern formed by $N$ photons is generated through the creation of N00N states in different paths, resulting in $N$ photons in a superposition between the two paths \cite{boto2000quantum}. By overlapping these paths with a slight tilt and measuring two-photon correlations in space, the resulting interference fringe pattern becomes $N$ times denser than the classical interference pattern produced by light of the same wavelength \cite{d2001two}. This phenomenon is the spatial equivalent of the phase super-sensitivity exhibited by N00N states \cite{mitchell2004super}. Notably, this phase super-sensitivity has been combined with polarization microscopy to image birefringent materials with enhanced sensitivity \cite{israel2014supersensitive}.

More recently, the concept of N00N states has been extended to transverse spatial modes instead of paths, enabling super-resolution measurements shown in Fig.~\ref{fig:PhotNum_imaging}b \cite{unternahrer2018super}. This extension has led to improvements in angular resolution for orbital angular momentum modes \cite{hiekkamaki2021photonic} and enhancements in longitudinal resolution for radial modes \cite{hiekkamaki2022observation}. Interestingly, the latter study also revealed that the commonly used intuitive explanation for the $N$-fold enhancement using the photonic de Broglie wavelength of the $N$-photon states, i.e., $\lambda/N$ \cite{edamatsu2002measurement}, was only valid for plane waves. When N00N states implemented through spatial modes are considered, the simple notion of the photonic de Broglie wavelength needs to be reevaluated \cite{hiekkamaki2022observation}.

In general, quantum-light sources with spatial correlations are known to enable resolution enhancement in imaging protocols, achieving the Heisenberg limit and providing an $N$-fold resolution improvement over the diffraction limit \cite{giovannetti2009sub}.
One can intuitively consider the spatially correlated state obtained from ideal SPDC, as described in Eq.~\eqref{eq:SPDC_Imaging}, as a multidimensional bi-photon N00N state where the two photons are in a superposition of all possible locations. This concept holds the potential for super-resolution imaging that surpasses the classical limit by a factor of two.
However, realizing this enhancement is challenging due to the limited availability of photon-number-resolving detectors and the exponential decrease in efficiency in an attempt to localize $N$ photons within a sub-diffraction-limited area.
To overcome this limitation, an optical centroid measurement was proposed \cite{tsang2009quantum} and subsequently implemented for photon numbers up to 4 \cite{shin2011quantum,rozema2014scalable}.
In this protocol, the detection does not rely on registering $N$-photon coincidence events at a specific location, but rather evaluates all events, particularly their centroid, which represents the mean position of all detected photons. Through appropriate post-processing, sub-wavelength fringes and quantum-enhanced resolution can be achieved.
Similar to other quantum-imaging techniques, recent advancements in the camera technology and data processing approaches have enabled successful super-resolution measurements \cite{unternahrer2018super,toninelli2019resolution,defienne2022pixel,Bhusal2022npj}.

The aforementioned results represent promising initial strides toward the development of quantum imaging protocols that benefit from entanglement. However, for these techniques to be applicable in real-world scenarios, various technological challenges must be overcome, including improved spatio-temporal resolution and enhanced single-photon measurements.

\section{Entanglement-Enhanced Light-Matter Interactions and Spectroscopy}
\label{sec: spectroscopy}

Spectroscopy techniques play a pivotal role in extracting valuable information about the energy dynamics and chemical structure of various substances, ranging from small molecules to large photosensitive complexes \cite{ernst_book,mukamel_book,hamm_book,cho_book}. Accurate identification of unknown molecular samples has been fundamental in advancing technologies used in modern society, including water and air pollution monitoring, homeland security, and healthcare applications \cite{workman_book}. Traditionally, these techniques rely on the use of laser light in the optical regime. However, recent investigations have revealed the potential of nonclassical light, such as entangled photon pairs, to open new and exciting avenues in experimental nonlinear spectroscopy \cite{dorfman2016,schlawin2017,shi2020entanglement,munoz2021,cutipa2022,Chen2022-1,Chen2022-2}.

\begin{figure}[h!]
    \centering
    \includegraphics[width=\textwidth]{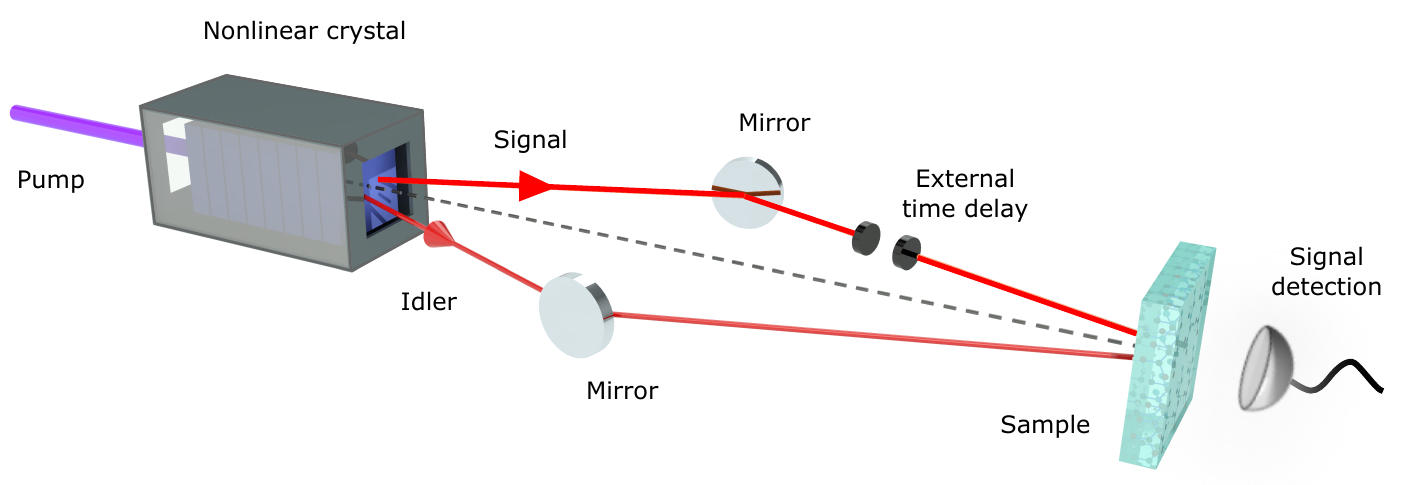}
    \caption{Schematic of an ETPA experiment. Entangled photons are produced by non-degenerate SPDC. In this process, a nonlinear crystal is pumped with light of central frequency $\omega_{P}$ to produce frequency-correlated photons with central frequencies $\omega_{S}$ and $\omega_{I}$. A tunable delay $\tau$ is introduced into the path of one of the photons. Finally, both photons impinge on the sample under investigation, producing a ETPA signal either as a ration between the generated and the transmitted photon pairs (transmission-based configuration) or as the fluorescence produced by the sample after the two-photon excitation (fluorescence-based configuration). Figure reprinted from Ref.~\cite{Mertenskotter:21}.}
    \label{Fig:ETPA_0}
\end{figure}

Remarkably, the time- and frequency-correlations exhibited by entangled photon pairs have facilitated the observation of intriguing phenomena known as entangled two-photon absorption (ETPA), as schematically sketched in Fig.~\ref{Fig:ETPA_0}. These correlations have played an important role in predicting and exploring various fascinating effects, including the linear dependence of the two-photon absorption rate on the photon flux \cite{javanainen1990,dayan2005,lee2006}. Such observations have led to the discovery of phenomena such as two-photon-induced transparency \cite{fei1997,guzman2010}, virtual-state spectroscopy \cite{saleh,KOJIMA,nphoton,roberto_spectral_shape,oka2010,villabona_calderon_2017,Varnavski2017,oka2018-1,oka2018-2,svozilik2018-1,svozilik2018-2,burdick2018,RobertoTemperatureControlled,Mukamel2020roadmap,Mertenskotter:21}, the induction of usually forbidden atomic transitions \cite{ashok2004}, the manipulation of matter's quantum pathways \cite{roslyak2009,roslyak2009-1,raymer2013,schlawin2016,schlawin2017-1,schlawin2017-2}, the analysis of many-body states \cite{kira2011quantum}, and the control of molecular processes \cite{shapiro2011,shapiro_book}. Notably, one of the most appealing features of ETPA is the linear relationship between two-photon absorption and the incident photon flux. This implies that multi-photon processes can be effectively excited using low-power, continuous-wave, single-frequency laser sources \cite{schlawin2018}. Such findings hold significant promise for the development of novel compact and cost-effective quantum-enhanced spectrometers.

While the theoretical foundation for the linearization of ETPA has been established \cite{javanainen1990,dayan2005}, there is an ongoing debate regarding the actual quantum enhancement that such a process can offer to spectroscopy \cite{raymer2021entangled,landes2021quantifying,raymer2021}. This discussion has sparked significant experimental work to investigate ETPA and its predicted linear behavior \cite{villabona2020,parzuchowski2021,tabakaev2021,landes2021,samuel2022,Tabakaev2022}. Interestingly, some of these studies have confirmed the predictions made in theory papers from the 1980s-1990s, while others suggest that previous results may have overestimated the potential enhancement that ETPA could provide for future quantum-enhanced spectroscopy. In essence, the question of whether ETPA can be observed using standard entangled-photon sources, such as those based on SPDC (see Sec.~\ref{subsec: entanglement_intro} for an introduction), and common fluorophores like Rhodamine and Tetraporphyrins, remains open. There is a need to establish a series of tests to probe genuine ETPA phenomena \cite{Tapia2023}. This is particularly important considering recent arguments that single-photon-loss mechanisms, such as scattering \cite{cushing2022} and hot-band absorption \cite{mikhaylov2022}, may mimic the expected linear absorption behavior of entangled photons.

In this section, we aim to equip the reader with fundamental tools to delve into the research field of ETPA. We begin by presenting a simple theoretical model that describes the interaction between entangled light and matter. This model serves as a basis for designing practical schemes for experimental implementation of entangled-photon absorption spectroscopy. Subsequently, we delve into current endeavors focused on the development of innovative applications enabled by ETPA. By providing this foundation, we hope to facilitate further exploration and advancement in the field of ETPA research.

\subsection{Theoretical Description of Entangled Two-Photon Absorption}

We begin by examining the interaction between an absorbing medium and a two-photon optical field, denoted as $\ket{\Psi}$. The interaction is described by the interaction Hamiltonian $\hat{H}_{\text{int}}(t) = \hat{d}(t) \hat{E}^{(+)}(t)$, where $\hat{d}$ is the dipole moment operator and $\hat{E}^{(+)}(t)$ represents the positive-frequency part of the electric field operator. Notably, the total field comprises both the signal ($S$) and idler ($I$) fields. Each electric field operator is given by:
\begin{equation}
\hat{E}^{(+)}_{S,I}(t) = \int d\omega_{S,I} \sqrt{\frac{\hbar \omega_{S,I}}{4\pi \epsilon_{0} c A}} \hat{a}_{S,I}(\omega_{S,I}) e^{-i \omega_{S,I} t},
\end{equation}
where $c$ represents the speed of light, $\epsilon_{0}$ is the vacuum permittivity, $A$ is the effective area of the field interacting with the sample, and $\hat{a}_{S,I}(\omega_{S,I})$ corresponds to the annihilation operator of a photonic mode with frequency $\omega_{S,I}$.

We consider the initial state of the absorbing medium to be its ground state $\ket{g}$, with energy $\varepsilon_{g}$. Using time-dependent second-order perturbation theory, we can calculate the probability of the medium being excited to a doubly-excited final state $\ket{f}$, with energy $\varepsilon_{f}$, through a two-photon absorption process. This probability is given by:

\begin{equation}
P_{g \rightarrow f} = \left\vert \frac{1}{\hbar^{2}} \int_{-\infty}^{\infty} dt_{2} \int_{-\infty}^{t_{2}} dt_{1} \mathcal{D}(t_1, t_2) \mathcal{E}(t_1, t_2) \right\vert^{2},
\label{Eq:prob}
\end{equation}
where
\begin{eqnarray}
\mathcal{D}(t_1, t_2) &=& \sum_{j=1} d_{fj} d_{jg} e^{-i(\varepsilon_{j}-\varepsilon_{f})t_2} e^{-i(\varepsilon_{g}-\varepsilon_{j})t_1}, \label{Eq:dipole}\\
\mathcal{E}(t_1, t_2) &=& \bra{0} \hat{E}_{S}^{+}(t_2) \hat{E}_{I}^{+}(t_1) \ket{\Psi} + \bra{0} \hat{E}_{S}^{+}(t_2) \hat{E}_{I}^{+}(t_1) \ket{\Psi},
\label{Eq:field}
\end{eqnarray}
Here, $d_{fj} = \bra{f} \hat{d} \ket{j}$ and $d_{jg} = \bra{j} \hat{d} \ket{g}$ are the transition matrix elements of the dipole moment operator, representing the coupling between the states $\ket{j}$ and $\ket{f}$, as well as $\ket{j}$ and $\ket{g}$. The sum over $j$ in Eq. (\ref{Eq:dipole}) indicates that the two-photon excitation of the medium occurs through intermediate states $\ket{j}$ with energy eigenvalues $\varepsilon_{j}$. Furthermore, since we are interested in the absorption of the two-photon states, the final state of the field is assumed to be the vacuum state.

We extend the model by assuming that the two-photon state is generated through type-II SPDC \cite{TORRES2011227}. In this process (see Fig. \ref{Fig:ETPA_0}), a second-order nonlinear crystal of length $L$ is pumped by a Gaussian pulse with temporal duration $T_{p}$, producing two photons with orthogonal polarizations: the signal photon and the idler photon. To ensure time indistinguishability of the paired photons, the signal and idler photons undergo polarization interchange and pass through a similar crystal of length $L/2$. Finally, an external time-delay is introduced between the photons. The resulting two-photon state is given by \cite{roberto_spectral_shape}:

\begin{eqnarray}
\ket{\Psi} &=& \left(\frac{T_{p}T_{e}}{2\pi\sqrt{\pi}}\right)^{1/2} \int_{-\infty}^{\infty} \int_{-\infty}^{\infty} d\omega_S d\omega_I \exp\left[-T_{P}^{2}(\omega_P-\omega_S-\omega_I)^2\right] \nonumber \\
& & \times \text{sinc}\left[T_{e}(\omega_{I}-\omega_{S})\right] e^{i\omega_{I}\tau}\hat{a}^{\dagger}_{S}(\omega_{S})\hat{a}^{\dagger}_{I}(\omega_{I})\ket{0},
\label{Eq:two-photon_1}
\end{eqnarray}
where $\omega_{j}$ ($j=P,S,I$) represents the frequencies of the pump, signal, and idler fields, respectively. The term $\text{sinc}[T_{e}(\omega_{I}-\omega_{S})]$ ensures the temporal overlap between the signal and idler photons. The entanglement or correlation time between the photon pairs is given by:
\begin{equation}
T_{e} = \frac{(N_S - N_I)L}{4},
\label{Eq:entanglement_time}
\end{equation}
where $N_{S,I}$ are the inverse group velocities of the signal and idler photons, respectively.

By substituting Eqs. (\ref{Eq:dipole})--(\ref{Eq:two-photon_1}) into Eq. (\ref{Eq:prob}), we can derive the expression for the two-photon absorption probability as follows:
\begin{eqnarray}
P_{g\rightarrow f}(T_{e},T_{p},\tau) &=& \frac{4T_{p}}{T_{e}}\frac{\sqrt{2\pi}\omega_{0}^{2}}{\hbar^2\epsilon_{0}^2c^2A^2}\exp\left[-2T_{p}^{2}(\omega_{p}-\varepsilon_{f})\right] \nonumber \\
& \times & \left\vert \sum_{j}d_{fj}d_{jg}\left( \frac{1 - e^{-i(\varepsilon_{j} - \omega_{0})(2T_{e}-\tau)}}{\varepsilon_{j}-\omega_{0}} + \frac{1 - e^{-i(\varepsilon_{j} - \omega_{0})(2T_{e}+\tau)}}{\varepsilon_{j}-\omega_{0}} \right) \right\vert^{2},
\label{Eq:prob_1}
\end{eqnarray}
where we have assumed that the photons are degenerate with central wave-packet frequencies $\omega_{S}^{0}=\omega_{I}^{0}=\omega_{0}=\omega_{P}/2$. In this expression, $\omega_{P}$ represents the central frequency of the pump field. We have also set $\varepsilon_{g}=0$ by displacing the energy levels for simplicity. Notably, Eq. (\ref{Eq:prob_1}) reveals two important features: (1) The probability of entangled two-photon absorption is determined by the coherent superposition (interference) of all possible pathways involving intermediate states $j$ for two-photon excitation, and (2) the ETPA signal can be controlled by adjusting the external delay between the paired photons, denoted by $\tau$.

\subsection{Entangled-Photon Absorption Spectroscopy: First Approach}

\begin{figure}[t!]
    \centering
    \includegraphics[width=\textwidth]{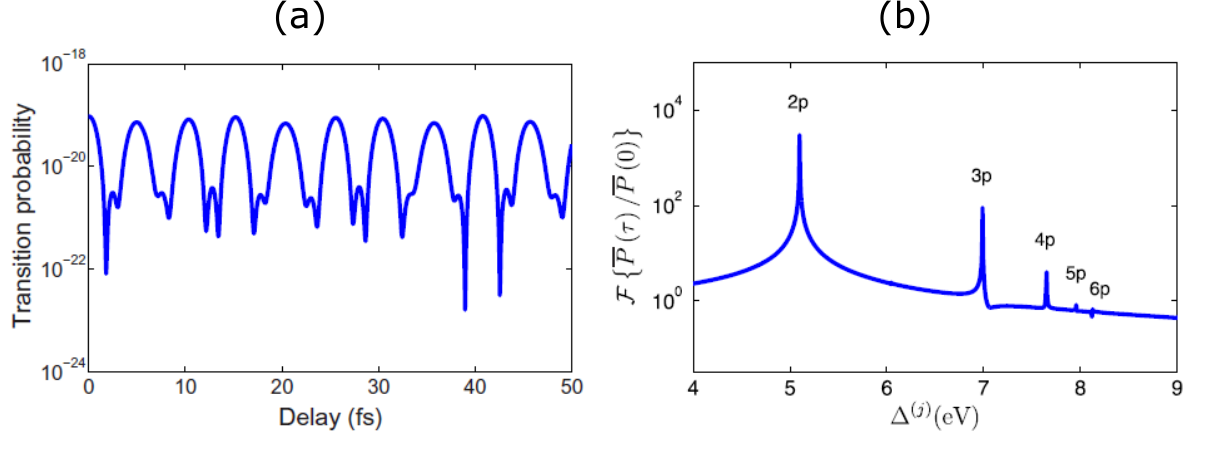}
    \caption{Entangled two-photon absorption spectroscopy. (a) Calculated non-monotonic absorption probability as a function of the delay between paired photons. (b) Fourier transform of the normalized weighted-and-averaged ETPA transition probability as a function of the energy mismatch $\Delta^{(j)} = \varepsilon_{j}-\omega_{0}$. The $y$-axis is shown on a logarithmic scale. Figures reprinted from Ref. \cite{roberto_spectral_shape}.}
    \label{Fig:ETPA_1}
\end{figure}

In 1997, Saleh \emph{et al.} introduced a technique called ``entangled-photon virtual-state spectroscopy'' based on the ETPA probability described by Eq. \eqref{Eq:prob_1} \cite{saleh}. This technique utilizes the interference pattern observed in the ETPA signal as a function of the external delay $\tau$ to extract spectroscopic information about the sample. Figure \ref{Fig:ETPA_1}a illustrates an example of the predicted ETPA signal for the 1s $\rightarrow$ 2s transition of atomic hydrogen, showcasing its non-monotonic behavior that encodes the sample's spectral information.

To extract this spectroscopic information, Saleh and colleagues proposed performing an average of Eq. (\ref{Eq:prob_1}) over a range of values of $T_{e}$, yielding the weighted-and-averaged ETPA transition probability \cite{saleh}:
\begin{equation}\label{Eq:average}
\overline{P}(\tau) = \frac{1}{T}\int_{T_{e}^{\rm min}}^{T_{e}^{\rm max}} P_{g\rightarrow f}(T_{e},T_{p},\tau)T_{e}dT_{e},
\end{equation}
where $T = T_{e}^{\rm max} - T_{e}^{\rm min}$. To perform this average experimentally, a set of measurements with different values of $T_{e}$ is required. While this can be technically challenging, the parameter $T_{e}$ can be adjusted by various methods depending on the type of entangled-photon source used. For example, in type-I SPDC (parallel-polarized photons), the width of the pump beam can be modified to change $T_{e}$ \cite{PhysRevA.50.3349}, whereas in type-II SPDC, the crystal length can be varied as $T_{e}$ is linearly proportional to it \cite{SHIH1994201}. As such, a set of wedge-shaped nonlinear crystals can be used to achieve the desired range of $T_{e}$.

Figure \ref{Fig:ETPA_1}b shows the Fourier transform of the normalized weighted-and-averaged ETPA transition probability for atomic hydrogen. Note the presence of different peaks, whose locations signal the frequencies of the intermediate $p$-transitions through which the 1s $\rightarrow $ 2s two-photon excitation takes place. These results demonstrate the potential of ETPA spectroscopy for retrieving relevant information about the electronic structure of arbitrary samples. Remarkably, even in the case when the SPDC crystal is pumped by a continuous-wave, monochromatic laser, i.e., $T_{p}\rightarrow\infty$, information regarding the various intermediate levels is still accessible. However, there are challenges associated with the original proposal for ETPA spectroscopy. First, it requires performing multiple experiments with two-photon states bearing different temporal correlations, which necessitates having hundreds of entangled-photon sources available. Second, as stated by the authors of Ref. \cite{saleh}, it requires \emph{a priori} knowledge of the lowest-lying intermediate electronic energy level of the system under study, which limits the applicability of ETPA spectroscopy for probing unknown samples.

In view of the aforementioned challenges, new approaches to ETPA spectroscopy have been proposed over the years. In the following sections, we will discuss some alternative schemes for obtaining information about the energy dynamics and electronic structure of unknown samples.

\subsection{ETPA Spectroscopy with Varying Entanglement Time and Pump Wavelengths}
\label{sec: ETPA_time_frequency}
Entangled photons possess unique nonclassical spectral and temporal features, which make them excellent probes for monitoring the excitation dynamics of molecular complexes. In striking contrast to classical light, entangled photons can circumvent the time-frequency uncertainty principle, meaning they exhibit independent temporal and spectral characteristics.

As described earlier, in SPDC, a photon from the pump pulse with frequency $\omega_{P}$ is down-converted in a birefringent crystal into a pair of entangled photons with frequencies $\omega_{S}$ and $\omega_{I}$. Energy conservation ($\omega_{P} = \omega_{S} + \omega_{I}$) induces a time-energy correlation between the pair of photons, where the bandwidth of the pump pulse limits the width of the sum of the down-converted photons' frequencies. However, different group velocities within the SPDC crystal can result in a broad bandwidth of the individual photons, which can exceed the bandwidth of their sum. This group velocity difference determines the entanglement time between the photons [see Eq. (\ref{Eq:entanglement_time})], and it sets a limit on the time delay between the absorption events of the correlated photons. Remarkably, this property has been successfully utilized by Schlawin \emph{et al.} \cite{schlawin2017-1} to reveal exciton dynamics in molecular complexes.

\begin{figure}[t!]
    \centering
    \includegraphics[width = \textwidth]{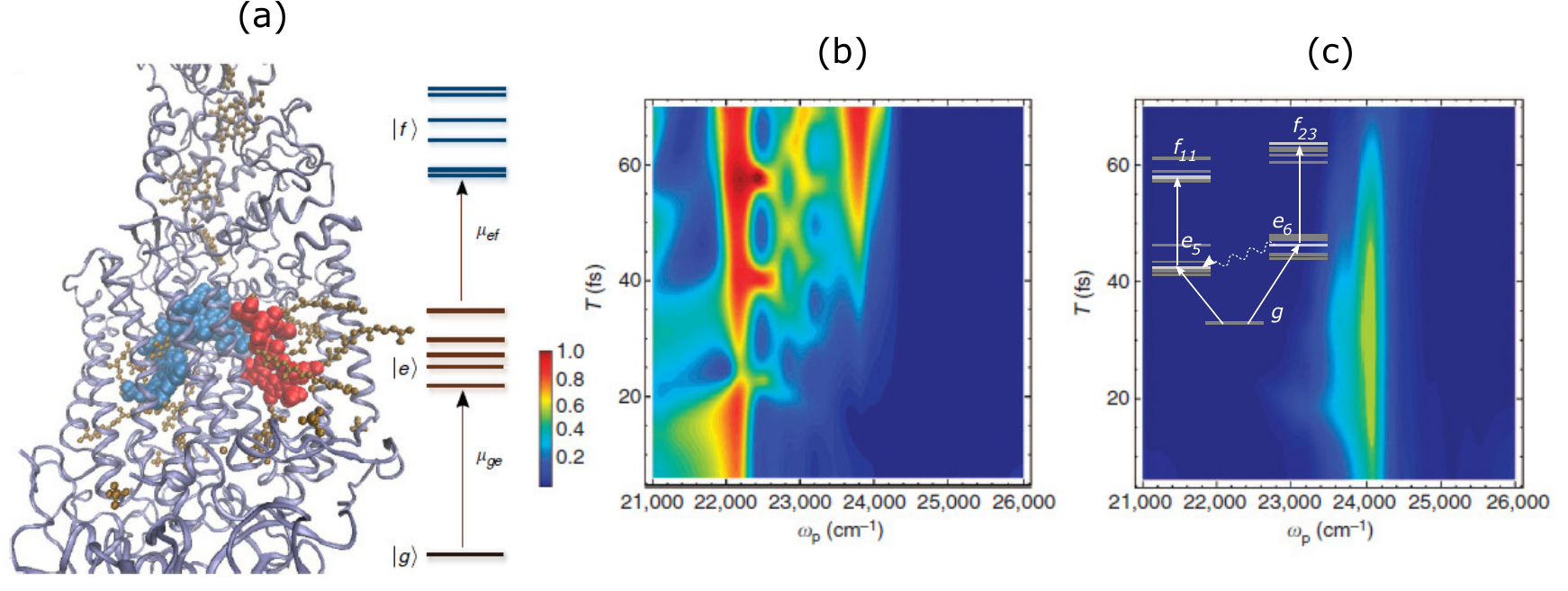}
    \caption{Control of exciton dynamics via ETPA. (a) Bacterial RC comprising 12 single-exciton and 41 two-exciton states, labeled as $\ket{e}$ and $\ket{f}$, respectively. (b) Two-photon absorption probability of state $f_{11}$ as a function of the entanglement time $T_{e}$ and the central pump frequency $\omega_{p}$. (c) The same for state $f_{23}$. Note that the entanglement time between photons defines whether two-photon absorption takes place via transport or direct excitation. Figures reprinted from Ref. \cite{schlawin2017-1}.}
    \label{Fig:ETPA_2}
\end{figure}

Consider the system shown in Fig.~\ref{Fig:ETPA_2}a, which depicts the bacterial reaction center (RC) of \emph{Blastochloris viridis}. The electronic structure of the RC can be modeled by including 12 single-exciton and 41 two-exciton states, labeled as $\ket{e}$ and $\ket{f}$, respectively. By varying the entanglement time and the pump frequency, one can reveal the influence of exciton dynamics in the single-exciton manifold of the RC.

If a two-photon absorption peak becomes stronger with increasing $T_{e}$, it indicates that the intermediate state is populated by transport rather than by direct excitation. This can be understood by monitoring the population (or, equivalently, the absorption probability) of the two-exciton state $f_{11}$, with frequency $\omega_{f_{11}} = 22,160\;\text{cm}^{-1}$. This state is mostly excited through the single-exciton state $e_{5}$, which can be populated by photon absorption or exciton transport.

As shown in Fig.~\ref{Fig:ETPA_2}b, when $T_{e}\rightarrow 0$, only direct excitation $g\rightarrow e_{5} \rightarrow f_{11}$ survives. However, as $T_{e}$ is increased, a number of resonances between $22,000$ and $24,000\;\text{cm}^{-1}$ appear. This is due to transport processes within the single-exciton manifold. It is particularly interesting to look at the $24,000\;\text{cm}^{-1}$ resonance, which corresponds to the excitation energy of the $f_{23}$ level. This state is solely excited through the short-lived single-exciton state $e_{6}$, which decays within $100$ fs to the state $e_{5}$. Figure \ref{Fig:ETPA_2}c shows that by increasing the entanglement time, direct excitation of the state $f_{23}$ is lost to the $f_{11}$ level due to the transport process between the $e_{6}$ and $e_{5}$ single-exciton levels, sketched in the inset in Fig.~\ref{Fig:ETPA_2}c.

The aforementioned results demonstrate the potential of entangled photon pulses for resolving specific single-exciton pathways through precise manipulation of two-photon states. By varying the entanglement time and the pump frequency, it becomes possible to selectively probe and characterize the dynamics of single-exciton states in complex systems. This level of control and resolution offered by entangled photon pulses opens up new avenues for studying and understanding the intricate processes and pathways involved in exciton dynamics. It provides a powerful tool for investigating and unraveling the detailed electronic structure and energy flow in various materials and molecular complexes.

\subsection{ETPA Spectroscopy with Temperature-Controlled SPDC}

Sec.~\ref{sec: ETPA_time_frequency} described how varying the entanglement time and the pump wavelength can bolster ETPA spectroscopy. There is, however, another feature of entangled photons that can be exploited to design realistic ETPA spectroscopy schemes, namely the non-degeneracy of entangled photons. Along this line, in 2019 Le\'on-Montiel \emph{et al.} \cite{RobertoTemperatureControlled}, proposed a rather simple technique for performing ETPA-based molecular spectroscopy with continuous-pumped SPDC . The main contribution of this proposed protocol is that it solves the two major drawbacks of the existing ETPA spectroscopy protocols, namely the need to perform multiple experiments with two-photon states bearing different
temporal correlations, which translates into the necessity to have at the experimenter’s disposal hundreds of entangled-photon sources, and the need to have a priori knowledge of the absorbing medium’s lowest-lying intermediate energy level.  

\begin{figure}[h!]
    \centering
    \includegraphics[width = 13.5cm]{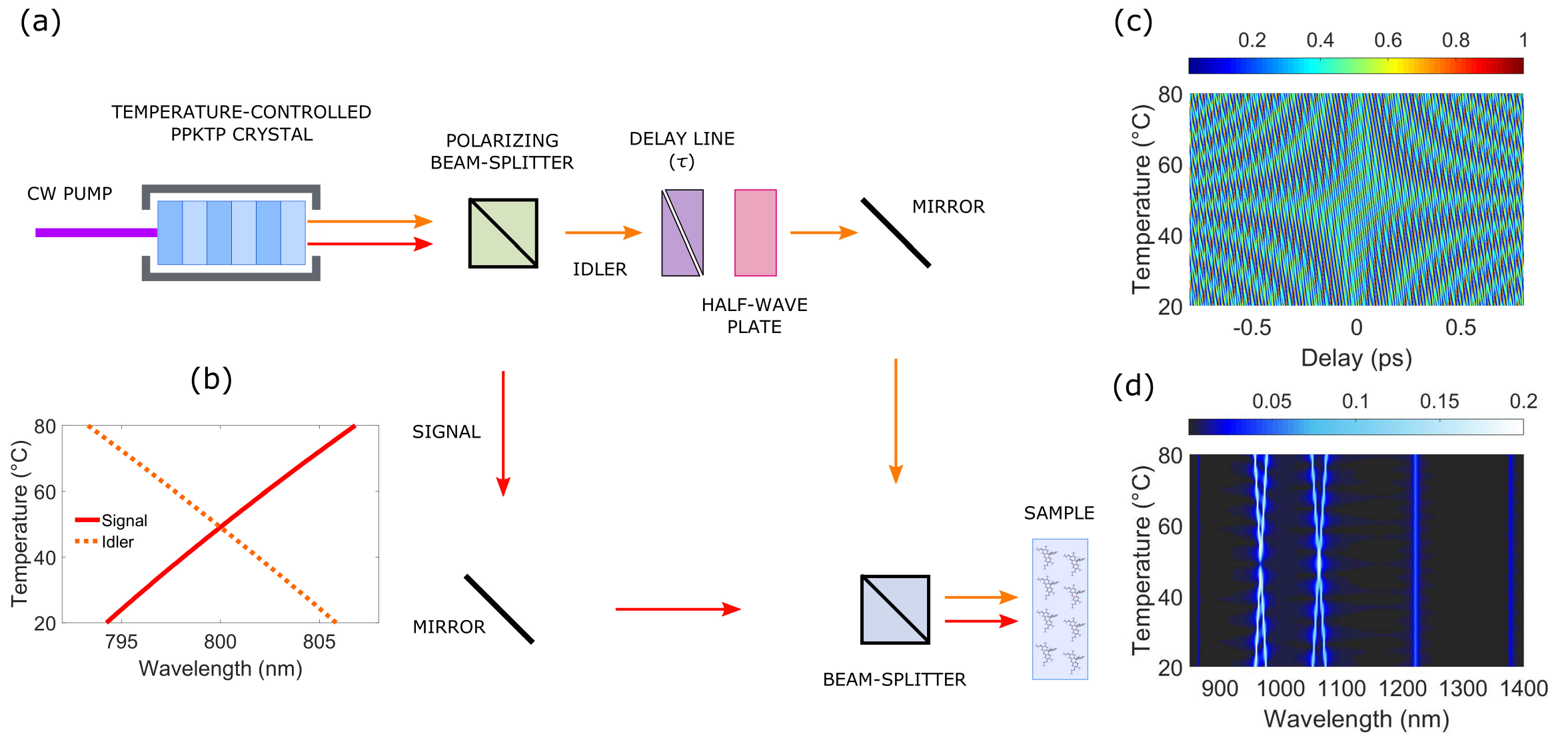}
    \caption{Temperature-controlled entangled two-photon absorption spectroscopy. (a) The proposed schematic. (b) Central wavelength of the SPDC photons as a function of the temperature of the nonlinear crystal, pumped by a continuous-wave laser at 400 nm. (c) Normalized ETPA signal as a function of the crystal temperature $T$ and the external delay between photons $\tau$. (d) Normalized Fourier transform of the ETPA signal with respect to the delay. Figures reprinted from Ref. \cite{RobertoTemperatureControlled}}
    \label{Fig:ETPA_3}
\end{figure}

The proposed protocol, depicted in Fig.~\ref{Fig:ETPA_3}a, makes use of a collinear, type-II SPDC source, where a PPKTP crystal is pumped by a continuous-wave laser centered at 400 nm. Interestingly, this configuration provides frequency anticorrelation of the down-converted photons, which guarantees the strongest ETPA signal \cite{roberto_spectral_shape}. The wavelengths of the paired photons can be tuned around the degenerate wavelength ($\omega_{0}=800$ nm) by controlling the crystal temperature \cite{fedrizzi2007wavelength, fedrizzi2}. The state produced by this temperature-dependent source is equivalent to the one depicted in Eq. (\ref{Eq:two-photon_1}), with $T_{p}\rightarrow\infty$ and $\omega_{s,i}^{0}(t) = \omega_{0} \pm \Delta\pare{T}$, where $\Delta\pare{T}$ stands for the temperature-dependent, linear frequency shift of the signal and idler photons, as drawn in Fig.~\ref{Fig:ETPA_3}b. Using this setup, we can thus monitor the ETPA signal as a function of the temperature of the crystal and the delay between photons. Figure~\ref{Fig:ETPA_3}c illustrates an example of a predicted ETPA signal for a model sample comprising two intermediate-state levels, whose wavelengths are arbitrarily chosen to be 967 and 1063 nm. Note the non-monotonic behavior of the signal results from the interference between different pathways through which two-photon excitation of the medium occurs. To reveal information about these pathways or equivalently, the intermediate-state energy locations, a Fourier transform is performed on the ETPA signal with respect to the delay. Figure~\ref{Fig:ETPA_3}d shows the normalized Fourier transform of the ETPA signal. Note that two characteristic patterns of $X$-shaped and straight lines appear. Remarkably, the $X$-shaped lines indicate $\varepsilon_{j}$, the energy location of intermediate states, whereas the straight lines appear at the combined frequencies $\pm\cor{\varepsilon_{j}\pm\varepsilon_{k}}$. This contrasting behavior between line-signals appear because the former contain frequency components that are temperature dependent, while the latter are constant with the temperature. Readers should refer to the Supplemental Material of Ref. \cite{RobertoTemperatureControlled} for details. 

In contrast to previous schemes for ETPA spectroscopy, the proposed protocol makes use of a single temperature-controlled SPDC crystal, a routinely-used technology in quantum optics laboratories. Moreover, the spectroscopic information about the probed samples can be obtained directly from the experimental data, without requiring complex post-processing. These features constituted a major simplification of the ETPA spectroscopy and established a new route toward its first experimental demonstration.

In summary, the investigation of entangled-light-matter interactions represents a current and captivating research domain at the forefront of physics and technology. It holds significant potential to revolutionize various fields of science and engineering, including material science, sensing, imaging, and spectroscopy. With the growing community dedicated to the ETPA spectroscopy, our objective through this concise review is to equip readers with the fundamental knowledge required to explore the ETPA research field. By advancing our comprehension of quantum light and its interactions with matter, we strive to develop innovative QIT that can positively transform our daily lives.

\section{Distributed Quantum Sensing}
\label{sec: distributed_quantum_sensing}

The quantum metrology, sensing, imaging, and spectroscopy protocols reviewed in the previous sections focus on measurement problems at a single sensor, while a multitude of real-world sensing tasks rely on a network or an array of sensors working collectively. Examples of sensor networks range from telescope arrays for astronomical observation to MEMS-based seismic networks\cite{walter2020distributed} for earthquake detection. 

In a sensor network, the impinging signals are recorded by independent sensors and extracted with joint signal processing \cite{krim1996two}. In general, a classical sensor network improves the signal-to-noise ratio by coherently adding the signals while the uncorrelated noise adds up incoherently. As a rule of thumb, collective measurements with $M$ independent sensors can improve the measurement sensitivity by $1/\sqrt{M}$, known as the standard quantum limit (SQL). Substantial progress has been made in the study of quantum multiparameter estimation \cite{proctor2018multiparameter,gessner2018sensitivity,goldberg2021intrinsic}, where the measurement sensitivity can potentially scale as 1/$M$, known as the Heisenberg limit, by utilizing entangled probes. Distributed quantum sensing (DQS) \cite{zhang2021dqs,ge2018distributed,zhuang2018distributed}, an emergent paradigm of quantum multiparameter estimation, aims to leverage entanglement to enhance the measurement sensitivity in estimating global properties shared by distributed sensors. 
\begin{figure}[htb]
    \centering
    \includegraphics[width=0.8\textwidth]{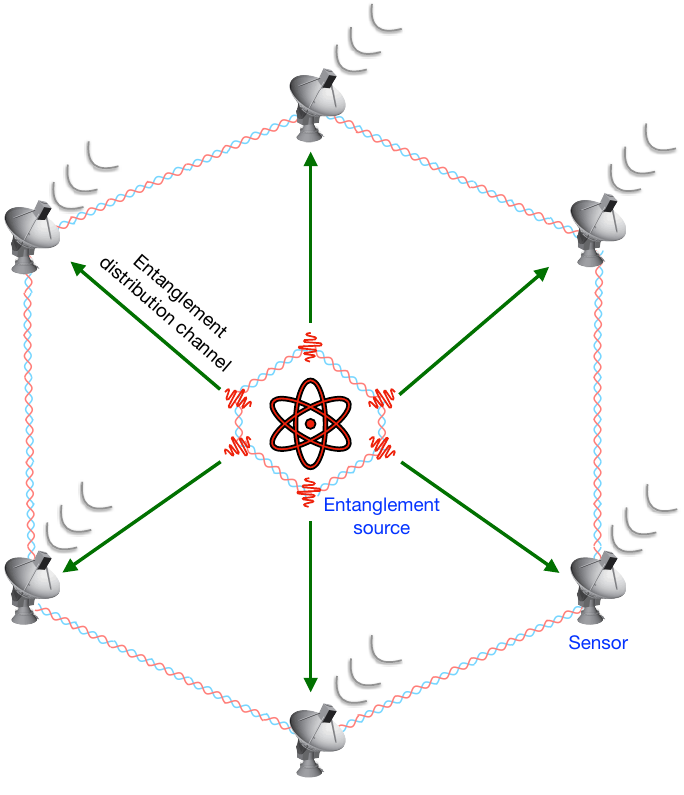}
    \caption{Concept of distributed quantum sensing. Multipartite entangled probes are prepared and distributed to spatially distant sensor nodes to tackle measurement problems in the sensor network.}
    \label{fig:DQS_concept}
\end{figure}

Figure~\ref{fig:DQS_concept} provides a schematic diagram of DQS. Multipartite entangled probes are prepared and delivered to distant sensors. Each sensor measures a local physical property of the object of interest. The measurement data from all sensors are then jointly post-processed to derive a global property of the sensor network, such as the average amplitude or a phase gradient seen by all sensors. To quantify the quantum advantage of a DQS protocol over its corresponding classical sensing (DCS) protocol based on separable probes, it is necessary define a resource constraint applicable to both protocols. Theoretical studies \cite{zhuang2018distributed,ge2018distributed, proctor2018multiparameter} have shown that a DQS protocol can approach a measurement sensitivity at the Heisenberg limit of $1/NM$, surpassing the optimal DCS protocol's $1/N\sqrt{M}$ and the SQL's $1/\sqrt{NM}$ scaling when the protocols are subject to a resource constraint of consuming on average $N$ photons in sensing. Intuitively, the quantum advantage stems from the correlated measurement noise arising from photon-number fluctuations among different sensors. Hence, the measurement noise from different sensors in DQS is canceled out during post-processing when combining the acquired data, in sharp contrast to the independent measurement noise in DCS. A detailed review of DQS protocols and their quantum advantage is given in Ref.\cite{zhang2021dqs}. We hereafter focus on the recent experimental advances of DQS in the optical domain. 

\subsection{Distributed Quantum Sensing for Optical Phase Estimation}
\begin{figure}
    \centering
    \includegraphics[width=\textwidth]{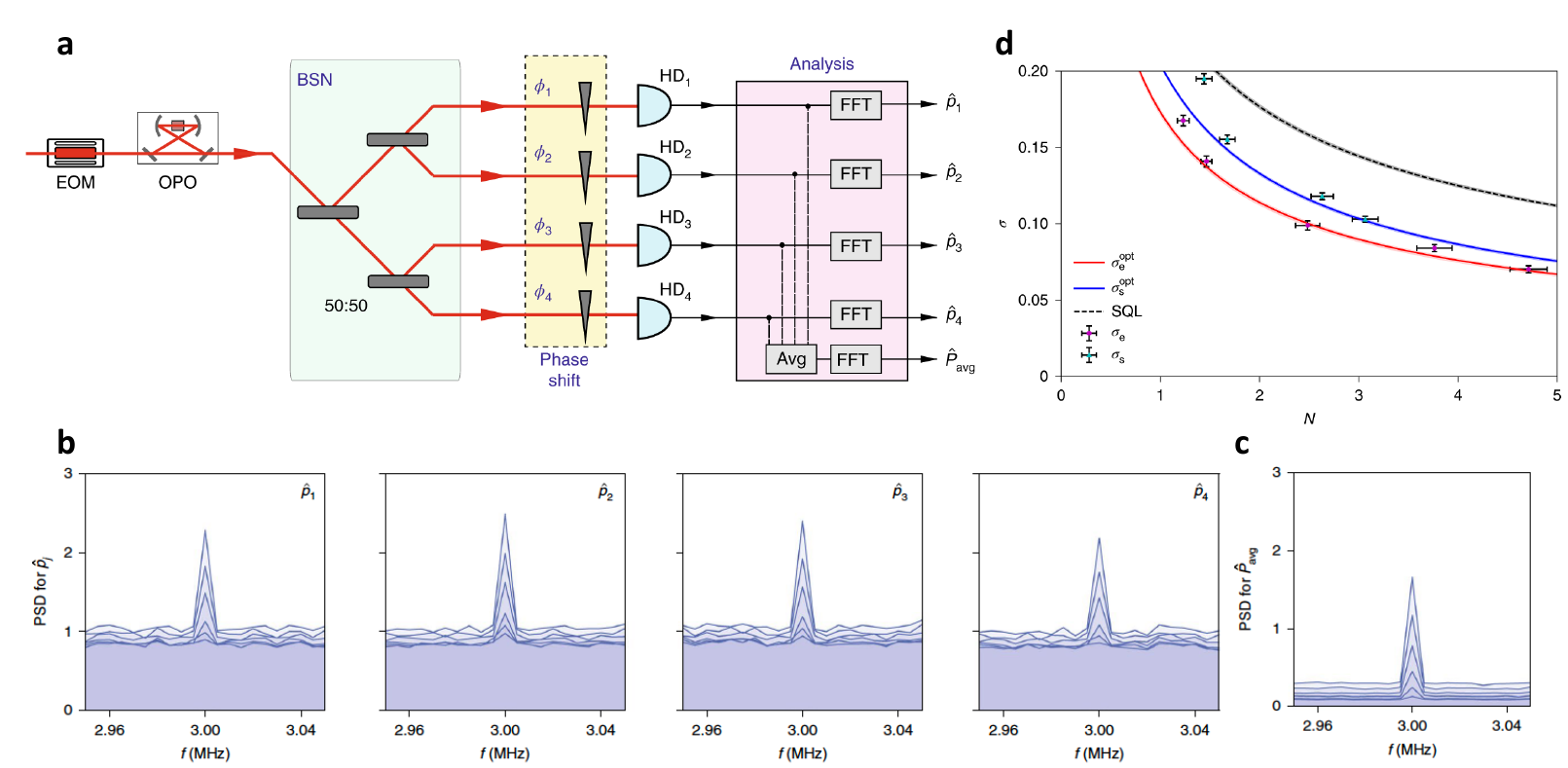}
    \caption{DQS for optical phase sensing with CV multipartite entangled states. (a) Experimental setup. Entangled states are prepared by splitting a displaced squeezed state with a beam splitter network. The phase quadratures of the four beams are measured by four homodyne detectors, followed by classical processing to derive the average phase. OPO: optical parametric oscillator; EOM: electro-optic modulator; FFT: fast Fourier transform. (b) The power spectral densities (PSDs) of the phase quadratures at four individual sensors for 6 different phase shifts. (c) Average PSDs from four detectors for average phase estimation show higher signal-to-noise ratios than that from a single sensor. (d) Comparison of measurement sensitivities $\sigma$ between DQS (red), DCS (blue), and SQL (dashed black) at different average number of photons $N$. Figures reprinted from Ref.~\cite{guo2020distributed}. }
    \label{fig:DQS_CV}
\end{figure}

Following the theoretical proposal in Ref.~\cite{zhuang2018distributed}, a proof-of-concept DQS experiment based on CV multipartite entanglement was reported by Guo \textit{et al.} \cite{guo2020distributed}. The schematic of the experiment is shown in Fig.~\ref{fig:DQS_CV}a. A displaced phase squeezed state is generated by injecting a phase-modulated seed laser to an OPO cavity pumped below the threshold.  It is then split by a balanced beam splitter network consisting of three 50:50 beam splitters into 4 modes to create entangled probes for DQS at 4 sensor nodes. At the $j$th sensor, a half-wave plate introduces a phase shift $\phi_j$. The objective of DQS is to estimate a global parameter, e.g., the averaged phase shift $P_{\rm ave}=(\sum_{j=1}^M \phi_j)/M$ across the sensor network. To this end, the phase quadratures at the 3 MHz sideband frequency of the entangled probes are measured by homodyne detectors at four sensor nodes followed by joint post-processing on the measurement data from all nodes. The power spectral densities at individual single sensors are depicted in Fig.~\ref{fig:DQS_CV}b, showing that signal peaks at 3 MHz increase with the optical phase shifts. The noise floor lies $0.8$ dB below the SNL due to the vacuum noise entered through the three spare ports of the beam splitter network. The power spectral densities of averaged signals are depicted in Fig.~\ref{fig:DQS_CV}c, showing that the noise floor is significantly lower than those of the individual sensors. The reduced noise floor corresponds to the single-mode squeezing level $5$ dB prior to the beam-splitter network \cite{zhuang2018distributed}. To carry out the same sensing task, DCS employs separable single-mode displaced phase squeezed states as probes while the SNL is defined as the performance of coherent-state probes. Figure~\ref{fig:DQS_CV}d compares the measurement sensitivities at various mean photon numbers between DQS, DCS, and the SNL. It should be noted that loss on the entangled probes diminishes the advantage of DQS. Quantum repeaters \cite{xia2019repeater} and quantum error correction \cite{zhuang2020distributed} will be necessary to reinstate DQS's advantage over DCS in the presence of loss. 

\begin{figure}
    \centering
    \includegraphics[width=\textwidth]{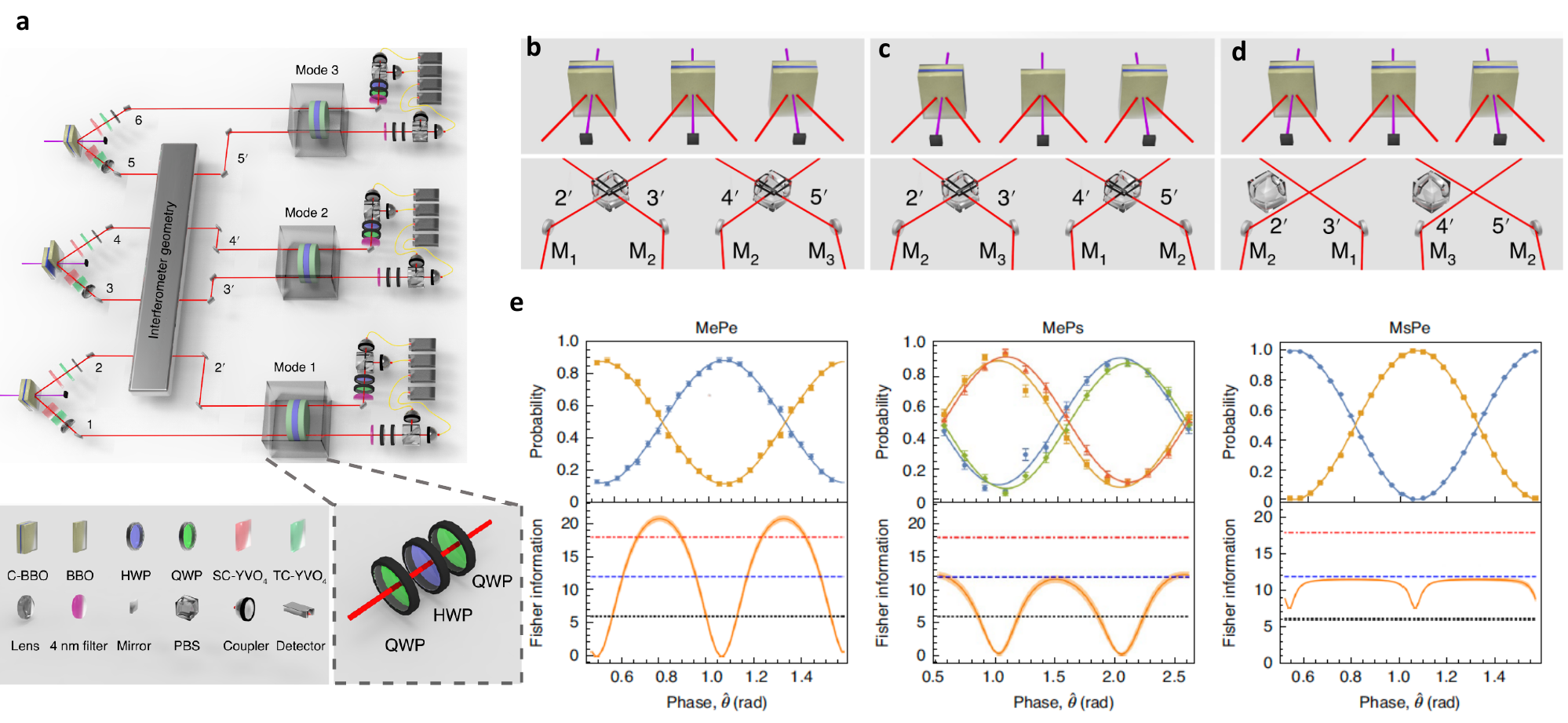}
    \caption{DQS for optical phase sensing with DV multipartite entangled states. (a) Experiment setup. SC-YVO4 and TC-YVO4: spatial compensation (SC) and temporal compensation (TC) yttrium orthovanadate crystals (YVO4); HWP: half-wave plate; QWP: quarter-wave plate; PBS: polarizing beam splitters. (b-d) The interferometer configurations to produce the mode-entangled particle-entangled (MePe), mode-entangled particle separable (MePs), and mode-separable particle-entangled (MsPe) quantum states. (e) (top) The average outcome probabilities in the measurement basis $\sigma_x^{\otimes 6}$, $\sigma_x^{\otimes 3}$, $\sigma_x^{\otimes 2}$ and (bottom) Fisher information (FI) for the MePe, MePs and MsPe states. The theoretical limit of FI is shown as the red dot-dashed line (MePs), blue dashed line (MsPe), and black dotted line (MsPs). Figures reprinted from Ref.~\cite{liu2021distributed}. }
    \label{fig:DQS_DV}
\end{figure}

Complementary to CV-DQS, Liu \textit{et al.} demonstrated a proof-of-concept DQS experiment based on DV entangled states \cite{liu2021distributed}. Figure~\ref{fig:DQS_DV}a shows their experimental setup: three BBO crystals are pumped by ultrafast ultraviolet lasers to produce three entangled photon pairs each in the state $\ket{\phi^+}=(\ket{HH}+\ket{VV})/\sqrt{2}$, where $H(V)$ denotes the horizontal (vertical) polarization. Mode-entangled and particle-entangled (MePe) probes are then prepared by passing three photon pairs through a tunable interferometer shown in Fig.~\ref{fig:DQS_DV}b, with the form of $\ket{\phi_{\rm MePe}}=(\ket{HH}_{M_1}\ket{HH}_{M_2}\ket{HH}_{M_3}+\ket{VV}_{M_1}\ket{VV}_{M_2}\ket{VV}_{M_3})/\sqrt{2}$. This three-mode entangled state consists of two entangled photons in each mode and is used to probe optical phases $\theta_{1,2,3}$ configured by a combination of a quarter-wave plate and a half-wave plate at three nodes. The goal is to estimate the average phase shifts $\hat{\theta}=\sum_{k=1}^3\theta_k/3$.
By performing projective measurements in $\sigma_x^{\otimes6}$ basis, an interference  fringe is obtained: $P_{\pm1}^{\rm MePe}=[1\pm V_{\pm}\cos{(NM\hat{\theta})}]/{2}$, where $V_{\pm}$ is the interference visibility, $N=2$ is the number of entangled photon in each mode and $M=3$ is the number of entangled modes. The Heisenberg limit of estimation sensitivity $\delta \hat{\theta}\sim1/NM$  is achieved when the visibility $V=1$. The quantum advantage of DQS based on the MePe state is benchmarked against the performance of DCS using mode-separable and particle-entangled (MsPe) probes in a local optimal state $\ket{\phi_{\rm MsPe}}=(\ket{HH}+\ket{VV})_{M_1}(\ket{HH}+\ket{VV})_{M_2}(\ket{HH}+\ket{VV})_{M_3}/2\sqrt{2}$, generated directly from three SPDC processes without passing through the interferometer as shown in Fig.~\ref{fig:DQS_DV}d. At each sensor node, the probability of projection outcome in the $\sigma_x^{\otimes2}$ basis is given by $P_{\pm1}^{\rm MsPe}=[1\pm V_{\pm}\cos{(N\hat\theta_i)}]/{2}$. The estimation sensitivity of individual phase shifts with unity visibility is $1/N$, resulting in an overall estimation sensitivity for averaged phase shifts $1/\sqrt{M}N$, similar to the separable squeezed states in the CV-DCS. Liu \textit{et al.} also prepared mode-entangled particle-separable (MePs) states $\ket{\phi_{\rm MePs}}=(\ket{H}_{M_1}\ket{H}_{M_2}\ket{H}_{M_3}+\ket{V}_{M_1}\ket{V}_{M_2}\ket{V}_{M_3})^{\otimes 2}/2$ by replacing one of the combinations of BBO crystals with a single piece of crystal shown in Fig.~\ref{fig:DQS_DV}c. This scheme is equivalent to repeating twice the measurements with mode entangled probes where only one photon is in each mode. The measurement outcome probability in $\sigma_x^{\otimes3}$ basis is given by $P_{\pm1}^{\rm MePs}=[1\pm V_{\pm}\cos{(M\hat{\theta})}]/{2}$, enabling a measurement sensitivity scaling $1/M$. The estimation uncertainty of $N$ measurements with one photon in each mode at a time follows $1/\sqrt{N}M$, also surpassing the standard quantum limit of $1/\sqrt{NM}$. The average probabilities for obtaining different phases in the measurement basis of $\sigma_x^{\otimes 6},\sigma_x^{\otimes 3},\sigma_x^{\otimes 2}$ for MePe, MePs, MsPe probe states are plotted in Fig.~\ref{fig:DQS_DV}e top. The corresponding Fisher information fitted from the measurement probabilities is shown in Fig.~\ref{fig:DQS_DV}e bottom. The blue (black) dashed line represents the theoretical limit of the FI for DCS (SQL). The red dot-dashed line represents the theoretical
limit of the FI for MePs. The DQS (MePe) and DCS (MsPe) achieve 2.7 dB and 1.43 dB reductions in the estimation variance as compared to the SQL. In addition, MePs probes outperform the performance of MsPe when the mode number is larger than the photon number in each mode, i.e., $M>N$.  Overall, the DQS protocol provides the best performance subject to a fixed total mean photon number. Combining the DQS protocol with the sequential scheme, where probes coherently interact with the sample for multiple times, could further improve the measurement sensitivity. A total number of photon passes at $n=21$ across 6 modes has been demonstrated. The joint detection probability for 6 modes is $P_{\pm1}^{\rm MePs}=[1\pm V_{\pm}\cos{(21\hat{\theta})}]/2$, and a 4.7 dB noise reduction compared to the SNL has been achieved, opening a new path to improving the measurement sensitivity of DQS.

Recently, Zhao \textit{et al.} demonstrate a field test of DV-DQS \cite{Zhao2021PRX} based on a loophole-free Bell test setup where entangled photon pairs are distributed to distant locations separated by 240 m. A global phase estimation precision of 0.916 dB below the SNL has been achieved unconditionally by virtue of a state-of-the-art average heralding efficiency of $73.88\%$.

\subsection{Distributed Quantum Sensing for Radio-Frequency Measurements}
\label{sec: DQS_RF}
DQS has been rapidly growing and has demonstrated great potential for practical quantum advantages in various sensing applications.
\begin{figure}
    \centering
    \includegraphics[width=\textwidth]{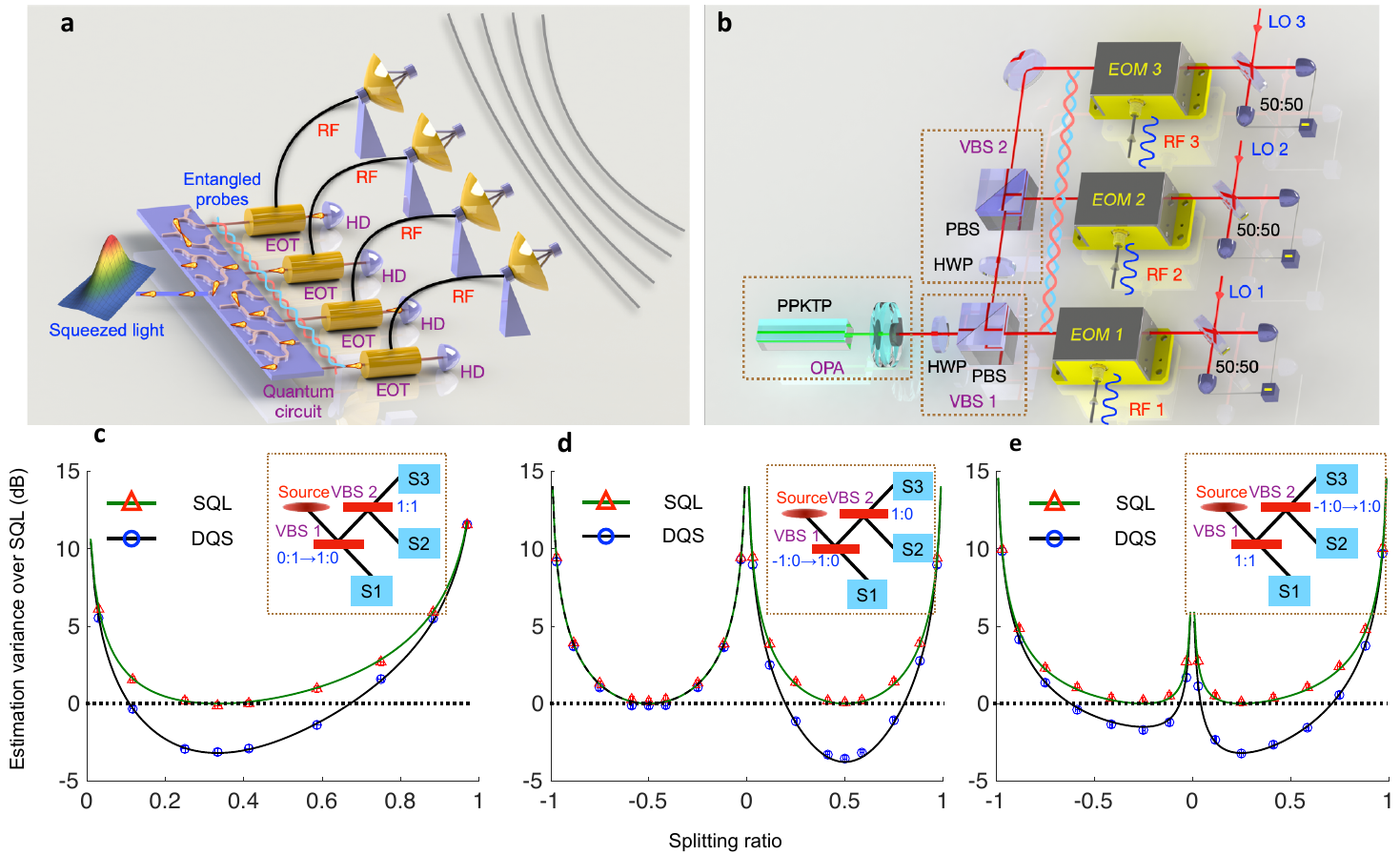}
    \caption{Entangled RF-photonic sensor network. (a) Concept of entangled RF-photonic sensor network. EOT: electro-optic transducer; HD: homodyne detector. (b) experiment diagram. HWP: half-wave plate; PBS: polarizing beam splitter; VBS: variable beam splitter; LO: local oscillator. (c-e) Optimization of multipartite entangled state for (c) RF-field average amplitude estimation, (d) phase-difference estimation at a central node, and (e) phase-difference estimation at an edge node. Insets show the tuning range of VBSs in preparing entangled states. (b-e) reprinted from Ref.~\cite{xia2020demonstration}.}
    \label{fig:DQS_RF}
\end{figure}
Beyond optical phase sensing, DQS can be harnessed to tackle measurement problems in a variety of physical domains by exploiting transducers or other types of quantum sensors, such as spin sensors~\cite{malia2022distributed}. Xia \textit{et al.} demonstrated CV-DQS of RF signals using RF-photonic transducers \cite{xia2020demonstration}, as sketched in Fig.~\ref{fig:DQS_RF} the concept and experimental setup. The entangled probes were generated by splitting squeezed light on two variable beam splitters each composed of a HWP and a PBS. The entangled probes were distributed to three distant sensor nodes. At each sensor node, an electro-optical modulator driven by an input RF signal created quadrature displacement at the sideband of the probe. Three balanced homodyne detectors characterized the quadrature displacement $\alpha_i\propto E_i\phi_i$, where $E_i (\phi_i)$ represents the amplitude (phase) of RF signal at $ i_{\rm th}$ sensor. A global property of the RF fields was derived $\alpha _g=\sum_i a_i\alpha _i$, for example, $a_i=1/3$ for mean amplitude estimation at a constant RF phase. An essential contribution of this work is the investigation of the optimal entanglement structure for different DQS tasks. Specifically, Xia \textit{et al.} optimized the entangled states for three different sensing cases shown in Fig.~\ref{fig:DQS_RF}: (c) averaged RF amplitude, (d) RF phase gradient at the center node, (e) RF phase gradient at the edge node. The estimation variance normalized to the SQL is plotted as a function of the beam-splitting ratios for different entangled states for the probes. The performance of the DQS is compared to that of DCS with laser light carrying the same optical power. It is important to note that the optimal performance is only achieved with the specific entangled state for a defined sensing task. The divergence of the estimation variance, as the splitting ratio approaches 0 or 1, indicates no information about the global property can be gained as no photon is sent to one of the sensors.
The minus sign of the splitting ratio in Fig.~\ref{fig:DQS_RF}d, e represents a $\pi$ phase delay to the signal $\phi_i \xrightarrow{} -\phi_i$ along with a sign flip $a_i \xrightarrow{} -a_i$ in post-processing to ensure an unbiased estimator. In these two cases, the performance curves of DQS display an asymmetric behavior in contrast to the symmetric DCS curves, which manifests the correlations in the measurement noise across different sensors.

\subsection{Distributed Quantum Sensing for Optomechanical Measurements}

\begin{figure}[!ht]
    \centering
    \includegraphics[width=\textwidth]{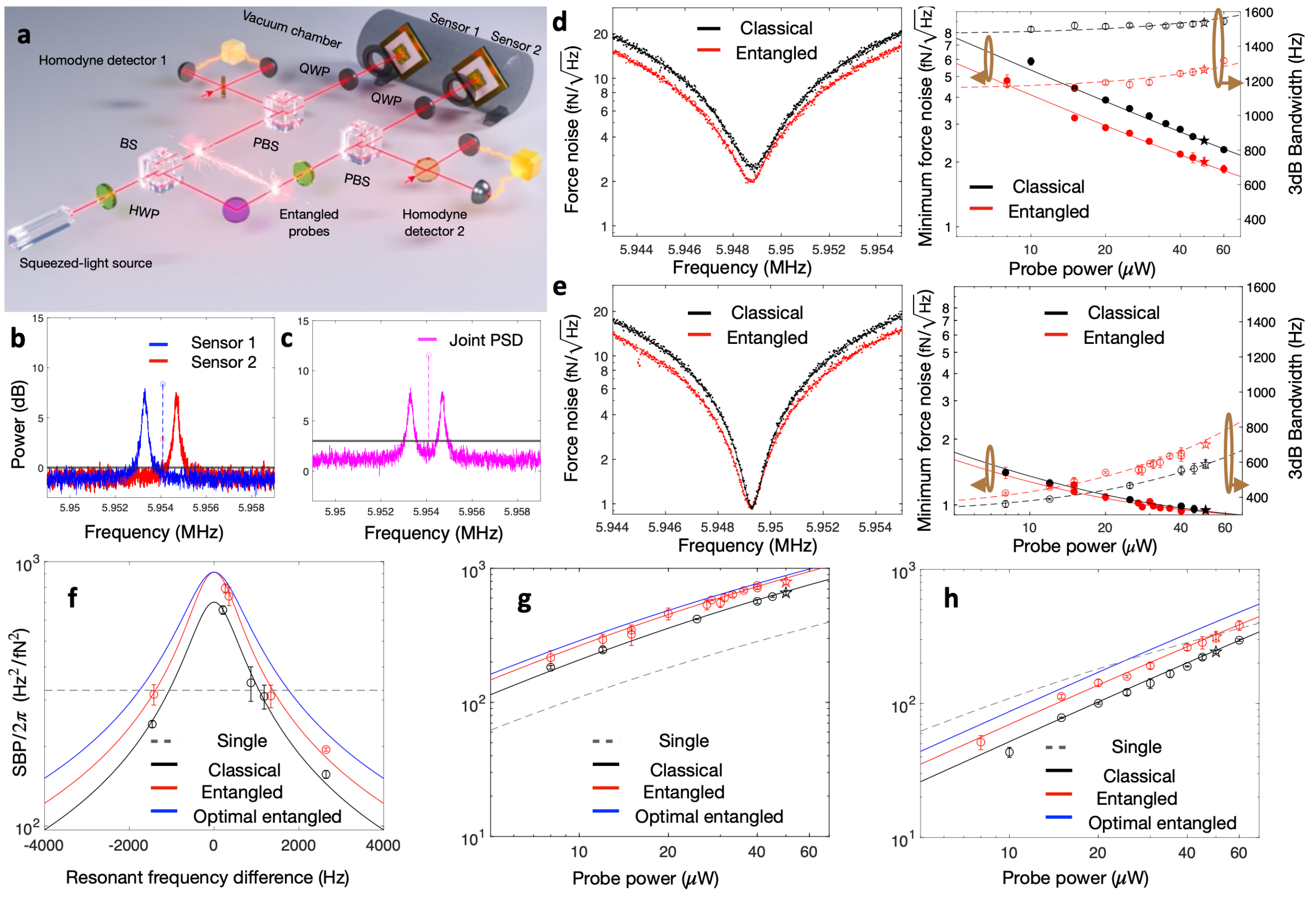}
    \caption{Entanglement-enhanced optomechanical sensing. (a) Experimental diagram. BS: beam splitter; HWP: half-wave plate; QWP: quarter-wave plate. PBS: polarizing beam splitter. (b) Normalized power-spectral density (PSD) for individual sensors. (c) Normalized joint PSD for both sensors. The circles mark the delta peaks yielded by test radiation pressure forces applied on the sensors. The SNL is shown as grey lines.  (d, e) Joint force noise at 50 $\mu$W probe power (left); minimum force noise and 3-dB bandwidth at different probe power (right). The resonant frequency difference is (d) 1422 Hz (e) and 262 Hz. (f) Sensitivity-bandwidth product (SBP) versus frequency difference at 50 $\mu$W probe power. SBP versus different probe power at resonant frequency difference of 262 Hz (g) and 1422 Hz (h). Figures reprinted from Ref.~\cite{xia2022entanglement}.}
    \label{fig:DQS_OM}
\end{figure}

Optomechanical sensors leverage the parametric coupling between a mechanical oscillator and a light field to achieve highly precise measurements of force, acceleration, mass, and magnetic fields \cite{li2021cavity}. Recently, Xia {\em et al.} demonstrated entanglement-enhanced joint force sensing using two optomechanical sensors, as depicted in Fig.~\ref{fig:DQS_OM}a \cite{xia2022entanglement}. The experiment employed entangled probe light at 1550 nm, similar to that used in continuous-variable distributed quantum sensing (CV-DQS) for RF measurements \cite{xia2020demonstration}. This entangled light was coupled into two optomechanical sensors, each comprising a mechanical membrane placed on top of a highly reflective mirror. The mechanical displacement at the $i$th sensor, encoded in the phase quadrature of the probe beam $Y^{(i)}_{\rm out}=Y^{(i)}_{\rm in}+\alpha_i\beta_i\chi_iF_i$, was measured using a balanced homodyne detector. Here, $|\alpha_i|^2$ represents the mean photon number of the coherent state at the carrier wavelength, $\beta_i$ denotes the optomechanical transduction efficiency, $\chi_i$ stands for the mechanical susceptibility, and $F_i$ encompasses both thermal force and signal forces. The entangled light exhibits quantum correlations between the input probe quadratures over a wide range around the mechanical resonant frequencies, thereby enhancing the performance of optomechanical sensors.

Figure~\ref{fig:DQS_OM}b presents the normalized power spectral densities (PSDs) of individual sensors, revealing two broad thermal peaks and two delta peaks originating from radiation pressure test forces. The noise floor is slightly below the SNL depicted by the gray line. In Fig.~\ref{fig:DQS_OM}c, the normalized joint PSD of the two sensors demonstrates a significantly improved signal-to-noise ratio (SNR), with the noise floor positioned 2 dB below the joint SNL. To assess the performance of averaged force sensing $(F_1+F_2)/2$ with entangled probes, two scenarios are compared: the imprecision-noise-dominated regime in Fig.~\ref{fig:DQS_OM}d and the thermal-noise-dominated regime in Fig.~\ref{fig:DQS_OM}e. When a substantial difference in mechanical resonant frequency exists, the minimum joint force noise $\sqrt{S_{FF}^{\rm min}}$ is achieved near the averaged mechanical resonant frequency, constrained by the laser shot noise. However, with entangled probes, the shot noise can be reduced by a factor of $\sqrt{V}$, as shown in Fig.~\ref{fig:DQS_OM}d (left), where $V=\left<(Y_{\rm in}^{(1)}+Y_{\rm in}^{(2)})^2\right>/2$ represents the squeezing factor offered by the entangled probes. The 3-dB bandwidth $\mathcal{B}_{3\text{dB}}$ with entangled probes is slightly inferior to that with classical probes due to the frequency-independent nature of the entangled states in the experiment. To fully exploit the quantum correlations, frequency-dependent entanglement akin to frequency-dependent squeezing\cite{kimble2001conversion} is required. The minimum force noise and 3-dB bandwidth at different probe powers are depicted in Fig.~\ref{fig:DQS_OM}d (right). Increasing the mean photon number $\alpha_c^2$ at the carrier wavelength leads to a reduction in force noise at a rate of $1/\alpha_c$. The bandwidth is primarily determined by the resonant frequency difference and exhibits a slight increase with the probe power.

When the mechanical resonant frequencies are nearly identical, as illustrated in Fig.~\ref{fig:DQS_OM}e, both classical and entangled probes experience a minimum force noise limited by thermal noise. However, entangled probes offer a larger bandwidth by a factor of $1/\sqrt{V}$, as demonstrated by the force noise curves in Fig.~\ref{fig:DQS_OM}e (left). The minimum force noise and 3-dB bandwidth at different probe powers are plotted in Fig.~\ref{fig:DQS_OM}e (right). As the probe power increases, the minimum force noise converges to the thermal noise limit, while the bandwidth scales with $\alpha_c$. 

The performance of the sensor array for broadband sensing tasks is characterized by the sensitivity-bandwidth product (SBP)\cite{brady2022entanglement}, given by $\mathcal{S}\times \mathcal{B}_{3 \text{dB}} \propto 1/\sqrt{VS_{FF}^{\rm min}}$, where $\mathcal{S} = 1/\sqrt{S_{FF}^{\rm min}}$ and is similar to integrated sensitivity. Fig.~\ref{fig:DQS_OM}f illustrates the SBP of classical, entangled, and optimal entangled probes at different resonant frequency differences, with the probe power fixed at 50 $\mu$W for each sensor. In the experiment, the SBP of entangled probes was suboptimal, but it approached the performance of optimal entangled probes at small resonant frequencies, surpassing the SBP of the classical probe by approximately $1/\sqrt{V} \approx 1.25$. The SBPs increase with respect to the square root of the probe power in the thermal noise dominant regime, as shown in Fig.~\ref{fig:DQS_OM}g, and with respect to the probe power in the imprecision noise dominant regime, as shown in Fig.~\ref{fig:DQS_OM}h.

Before concluding this section, we would like to make a few remarks. It is worth noting that DQS is not limited to the photonic platform and has rapidly expanded to encompass various physical systems, including atomic spins\cite{malia2022distributed} and microwave cavities\cite{brady2022entangled}. These advancements have opened up a wide range of applications, from inertial navigation to dark matter search\cite{brady2022entanglement,carney2021mechanical}. Furthermore, we would like to highlight a recent milestone in the field, namely the successful distribution of squeezed light over a 40-km fiber link\cite{suleiman202240}. In this experiment, the transmitted squeezed light was detected using a real local oscillator, demonstrating a promising pathway for implementing CV-DQS in large-scale quantum networks. These developments underscore the growing significance and potential of DQS in various physical systems, as well as its applicability in enabling advanced QIT applications.

\section{Quantum Machine Learning}
\label{sec: quantum_machine_learning}

\begin{figure}[!ht]
    \centering
    \includegraphics[width=1\textwidth]{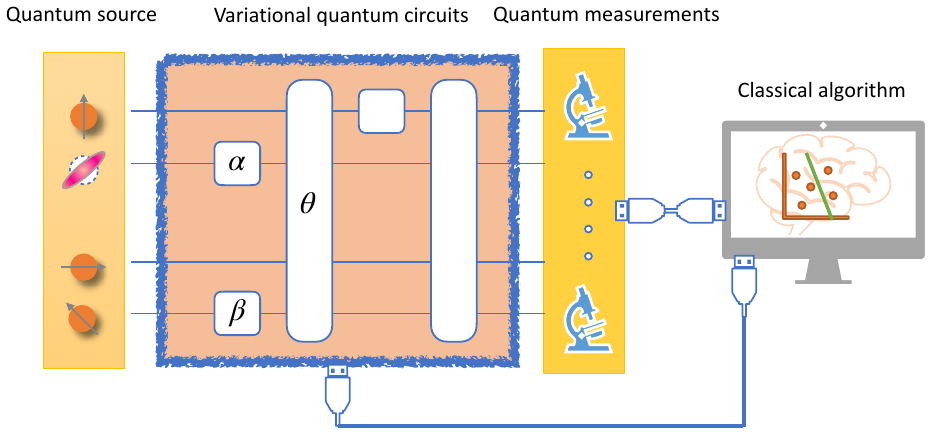}
    \caption{Schematic for quantum machine learning with photons. The quantum state of light from a source is processed by a variational quantum circuit configured by a classical algorithm to generate the quantum state for a defined classical or quantum processing task. Measurements on the quantum state produce data used by a classical algorithm to optimize the parameters of the variational quantum circuit and seek solutions of the data-processing task. }
    \label{fig:QML_concept}
\end{figure}

Machine learning has rapidly grown over the last two decades and continues to have a significant impact on everyday life, from speech recognition to autonomous vehicles. In the meantime, quantum hardware is advancing at a great pace, leading to quantum computation advantage in solving certain computation problems. It is thus expected that quantum computers can also significantly enhance the performance of machine learning tasks in several aspects, including quantum speedups in optimization, improved classification accuracy, and sampling of classically intractable systems \cite{biamonte2017quantum}. Many existing quantum machine-learning algorithms require a large-scale fault-tolerant quantum computer, which is not yet available in the current era of noisy intermediate-scale quantum (NISQ) technology. Instead, hybrid approaches combining variational quantum circuits and classical processing, as schematically illustrated in Fig.~\ref{fig:QML_concept}, have emerged as a leading strategy for quantum machine learning tasks\cite{cerezo2021variational} such as variation quantum eigensolver\cite{kandala2017hardware,wang2019accelerated,parrish2019quantum} and quantum kernel method\cite{schuld2019quantum,havlivcek2019supervised}. Along this line of research, researchers have been exploring quantum photonic platforms in conjunction with classical processing to improve classical processing~\cite{saggio2021experimental,xia2021quantum} as well as quantum processing~\cite{carolan2020variational}. In the same vein as quantum machine learning in photonic platforms, a more recent experiment in the trapped-ion platform exploited a classical algorithm in tandem with variational quantum circuits~\cite{marciniak2022optimal} to produce entangled spin states for a Ramsey interferometer to approach the ultimate measurement precision at the Heisenberg limit. We will next review the recent advances in quantum machine learning for both classical and quantum processing.

\subsection{Quantum Machine Learning for Classical Processing}
\label{subsec: quantum_machine_learning_classical_processing}

\begin{figure}[!ht]
    \centering
    \includegraphics[width=1\textwidth]{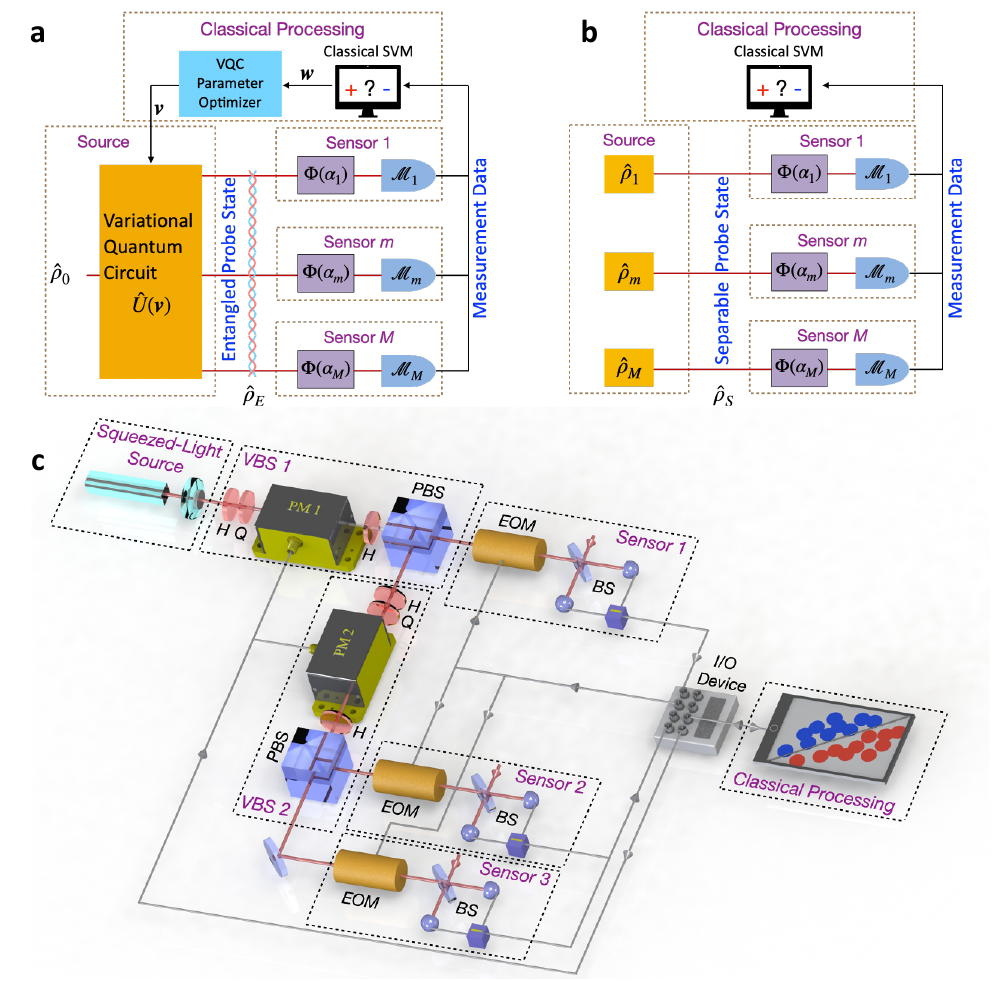}
    \caption{Entanglement enhanced data classification. (a) Schematic of supervised learning assisted by an entangled sensor network (SLAEN). (b) Classical classifiers built on separable states and pure classical processing. (c) Experiment diagram for RF signals classification using SLAEN. Figure reprinted from Ref.~\cite{xia2021quantum}. }
    \label{fig:QML_RF_illustration}
\end{figure}

Quantum machine learning algorithms, such as quantum support vector machine \cite{rebentrost2014quantum} and quantum principal analysis \cite{lloyd2014quantum}, often require quantum memories to load classical data, which could hinder their near-term quantum advantage for big-data problems. Supervised learning assisted by an entangled sensor network (SLAEN) \cite{zhuang2019physical} is a different paradigm that takes advantage of quantum metrology to acquire data at the physical layer instead of using classical data given a priori. SALEN can improve the accuracy of classical data classification and compressing. The schematics of SLAEN and classical classifiers are compared in Fig.~\ref{fig:QML_RF_illustration} a, b. SLAEN consists of a variational quantum circuit (VQC) parameterized by $\boldsymbol{v}$ to generate an entangled probe state shared by $M$ sensors.  The objects, modeled by quantum channels $\Phi(\alpha_m)$, are probed by the entangled states, which are subsequently measured to produce results $\tilde{\alpha}_m$. The measurement data are processed in a classical computer running a support vector machine (SVM) algorithm. The goal of the training stage of SLAEN is to find the optimum entangled probes and the hyperplane used to generate labels for the data. This is done by using training data to optimize the hyperplane $\{\boldsymbol{w},b\}$ stored in the classical computer with the measurement data $\tilde{\alpha}_m$ and known labels $y$. The VQC is reconfigured in real-time to generate entangled probes that minimize measurement noise by mapping $\boldsymbol{w} \rightarrow \boldsymbol{v}$ at each optimization step. At the end of the training stage, the obtained optimal entangled state and hyperplane would facilitate the data classification with reduced error probability. In contrast, the classical classifier acquires measurement data with separable probes and conducts training in post-processing in the classical computing without invoking physical hardware.

\begin{figure}[!ht]
    \centering
    \includegraphics[width=1\textwidth]{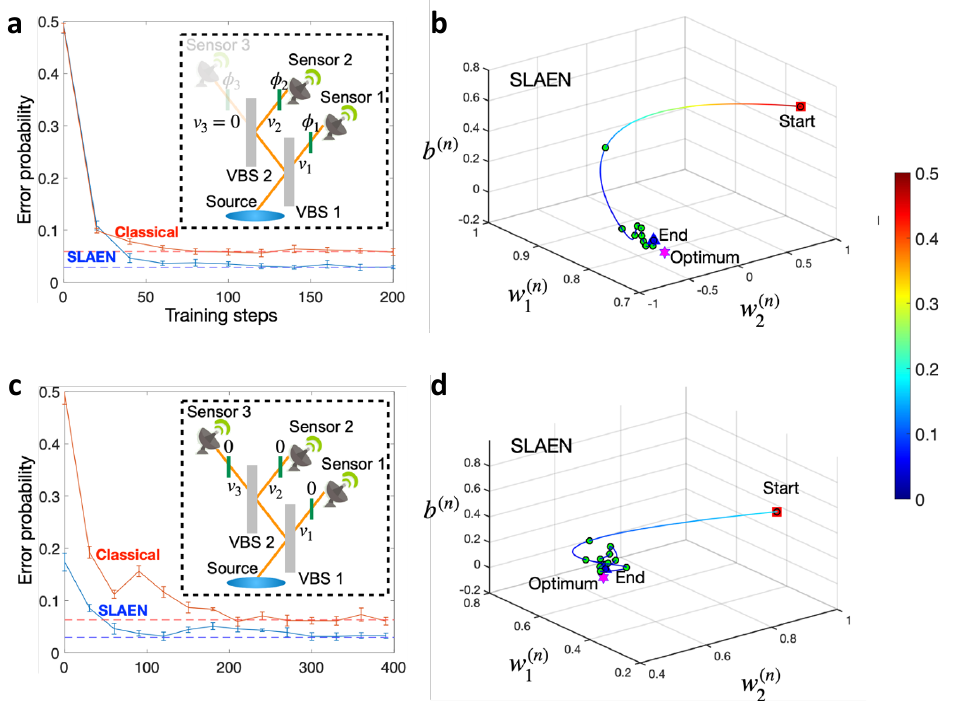}
    \caption{Convergence of error probability during training for (a) RF phase gradient classification with two sensors and (c) RF mean amplitude classification with three sensors. (b, d) Evolution of the hyperplane parameters during training, corresponding to (a, c). Colorbar: error probability. Figure reprinted from Ref.~\cite{xia2021quantum}. }
    \label{fig:QML_RF_exp}
\end{figure}

Xia \textit{et al.} demonstrated improved data classification accuracy of RF signals using a variational entangled sensor network on a continuous-variable quantum optics platform\cite{xia2021quantum}. The experiment setup is shown in Fig.~\ref{fig:QML_RF_illustration}c. A source generated squeezed light, which was then processed by a variational quantum circuit composing two variable beam splitters (VBSs) and three phase shifters. The splitting ratio of each VBS was controlled by an external DC voltage. The output entangled probes were delivered to three sensors equipped with electro-optic modulators that converted RF signals into displacements on the phase quadrature of the $m_{th}$ mode. The optical phase quadratures were subsequently measured by three homodyne detectors and the measurement results $\Tilde{\alpha}_m$ were fed into a classical computer for training, classification, and VQC optimization. Fig.~\ref{fig:QML_RF_exp} shows the experimental training results of SLEAN and classical classifiers using laser light aiming to classify (a,b) RF-phase gradient with two sensors and (c,d) RF mean amplitude with three sensors. SLAEN was provided with $N$ sets of training data and labels for two training tasks and the hyperplane was optimized iteratively by minimizing a cost function. The setting for the VQC was updated at each step. Fig.~\ref{fig:QML_RF_exp}b, d show the trajectories of the hyperplanes $\{\boldsymbol{w},b\}$ evolving toward the optimum hyperplanes (Hexagrams) during the two training processes. The error probabilities measured at different training steps of SLAEN and classical classifier are plotted in Fig.~\ref{fig:QML_RF_exp}a, c, which demonstrate two advantages of SLAEN over the classical classifier. First, SLAEN takes less step to converge to the optimal configuration. Second, once trained SLAEN enjoys reduced error probabilities as compared to the classical classifier. The reduction in classification error probability with SLAEN  stems from the quantum correlation between measurement noise at different sensors, akin to the estimation of global properties in DQS tasks. The experiment Xia \textit{et al.} verified SLAEN in RF sensing. As a versatile architecture, SLAEN can also be readily adapted to cope with other types of quantum sensors\cite{degen2017quantum} such as NV centers, nanomechanical oscillators, etc. 

\begin{figure}[!ht]
    \centering
    \includegraphics[width=\textwidth]{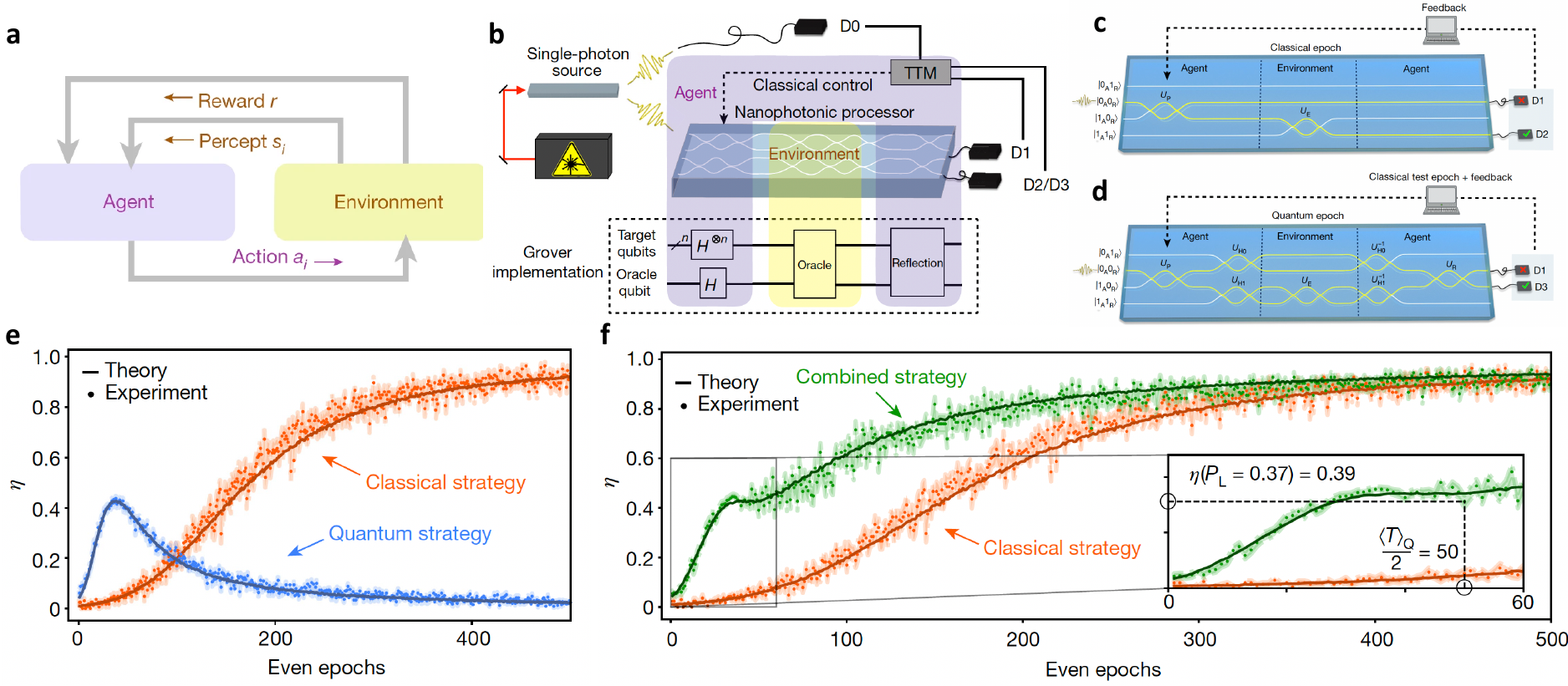}
    \caption{Quantum speedup in reinforcement learning. (a) Schematic of reinforcement learning. (b) Experimental diagram. Configurations of the nanophotonic chip for (c) classical and (d) quantum epoch. (e) Averaged reward for quantum (blue) and classical (orange) strategy. (f) Average reward for the combined strategy (green), where the agent switches to a classical strategy after reaching the optimal point of the quantum strategy, as compared to the classical-only strategy (orange). Figures reprinted from Ref.~\cite{saggio2021experimental}.} 
    \label{fig:QML_reinforcement}
\end{figure}

Reinforcement learning (RL)\cite{kaelbling1996reinforcement}, a powerful paradigm in artificial intelligence with a wide range of applications including robotics and autonomous driving, has recently been incorporated into quantum photonic platforms. The RL process involves iterative interactions between an agent and the environment through a classical channel, as shown in Fig.~\ref{fig:QML_reinforcement}a, referred to as an epoch. In general, the agent's learning process can be divided into two critical steps within each epoch and could potentially benefit from quantum speedups through multiple pathways. First, the agent takes actions based on received rewards from the environment. Quantum computers could potentially reduce the running time of agent's internal program for selecting actions that lead to higher rewards. Second, quantum communication could enable a quadratic reduction in environment queries, similar to Grover's search algorithm. Along the second pathway, Saggio \textit{et al.} experimentally demonstrated quantum speed-up in reinforcement learning \cite{saggio2021experimental} on an integrated nanophotonics platform. They achieved this by amplifying the amplitudes of states corresponding to high reward actions, as shown in Fig.~\ref{fig:QML_reinforcement}b. The configurations of the nanophotonic processor for classical RL and quantum-enhanced RL are shown in Fig.~\ref{fig:QML_reinforcement}c, d respectively. In their experiment, the wining (losing) action state $\ket{w}_A$ ($\ket{l}_A$) of the agent was represented by a single photon qubit via $\ket{1}_A$ ($\ket{0}_A$), and the reward state by the environment was denoted by another qubit state $\ket{1}_R$ or $\ket{0}_R$. This four-level Hilbert space was encoded in the path degree of freedom of a single photon passing through four waveguides. The behavior of the environment was modeled by a controlled-NOT gate $U_E$ such that the reward state was flipped only in the case of the winning state $\ket{w}_A$.
A single photon was initially coupled into the mode $\ket{0_A0_R}$ and then interfered at a Mach-Zehnder interferometer (MZI), resulting in the state $\ket{\psi}_A\ket{\psi}_R=(\cos(\xi)\ket{0}_A+\sin(\xi)\ket{1}_A)\ket{0}_R$ where $\varepsilon=\sin^2(\xi)$ is the winning probability of winning action state $\ket{1}_A$.  During the classical RL, the environment applied a controlled-NOT gate on the two qubits, by implementing another MZI between the third and fourth waveguide. The photon was then coupled out and detected by two single-photon detectors D1 and D2 at the output of two waveguides, corresponding to states $\ket{0_A0_R}$ and $\ket{1_A1_R}$ with success probability $\cos^2(\xi)$ and $\sin^2(\xi)$. The winning probability $\varepsilon_j$ after $j_{th}$ photon arriving at D2 was updated according to $\varepsilon_j={(1+2j)}/{(100+2j)}$, where the initial winning probability was set to $\varepsilon_0=1/100$. In quantum-enhanced RL, the initial state $\ket{\psi}_A\ket{\psi}_R$ was first transformed to $(\cos(\xi)\ket{0}_A+\sin(\xi)\ket{1}_A)\ket{-}_R$ via two unitary operators $U_{H0}$ and $U_{H1}$, where $\ket{-}_R=(\ket{0}_R-\ket{1}_R)/\sqrt{2}$. Then, the environment applied $U_E$, which flipped the sign of the winning state, $(\cos(\xi)\ket{0}_A-\sin(\xi)\ket{1}_A)\ket{-}_R$. After interaction with the environment, the quantum agent reversed operations $U_{H0}$ and $U_{H1}$, and performed a reflection $U_R=2\ket{\psi}\bra{\psi}_A-1_A$, amplifying winning probability $\varepsilon=\sin^2(3\xi)$ at detector D3.  This quantum-enhanced winning probability, in contrast to classical winning probability $\sin^2(\xi)$, enabled a speed-up in RL. Since the reward was not revealed in each quantum epoch, an additional classical test epoch was required and the update rule of winning probability remained the same as in the classical case. Fig.~\ref{fig:QML_reinforcement}e shows the average reward of classical (orange) and quantum (blue) RL strategies. To ensure a fair comparison with classical RL, the reward after each classical test epoch in quantum RL was distributed (averaged) over both the quantum and classical test epochs. The quantum strategy reached its optimum point when the averaged reward $\eta_Q=\sin^2(3\xi)/2$ equaled that of classical strategy $\eta_C=\sin^2(\xi)$ with $\eta_Q=\eta_C=0.396$, then started to decrease, which is common to any Grover-like algorithm. After reaching the optimum point, the agent switched back to the classical strategy while maintaining the quantum advantage as shown in Fig.~\ref{fig:QML_reinforcement}f, where a combined strategy (green) is compared to the classical case.

\begin{figure}
    \centering
    \includegraphics[width=\textwidth]{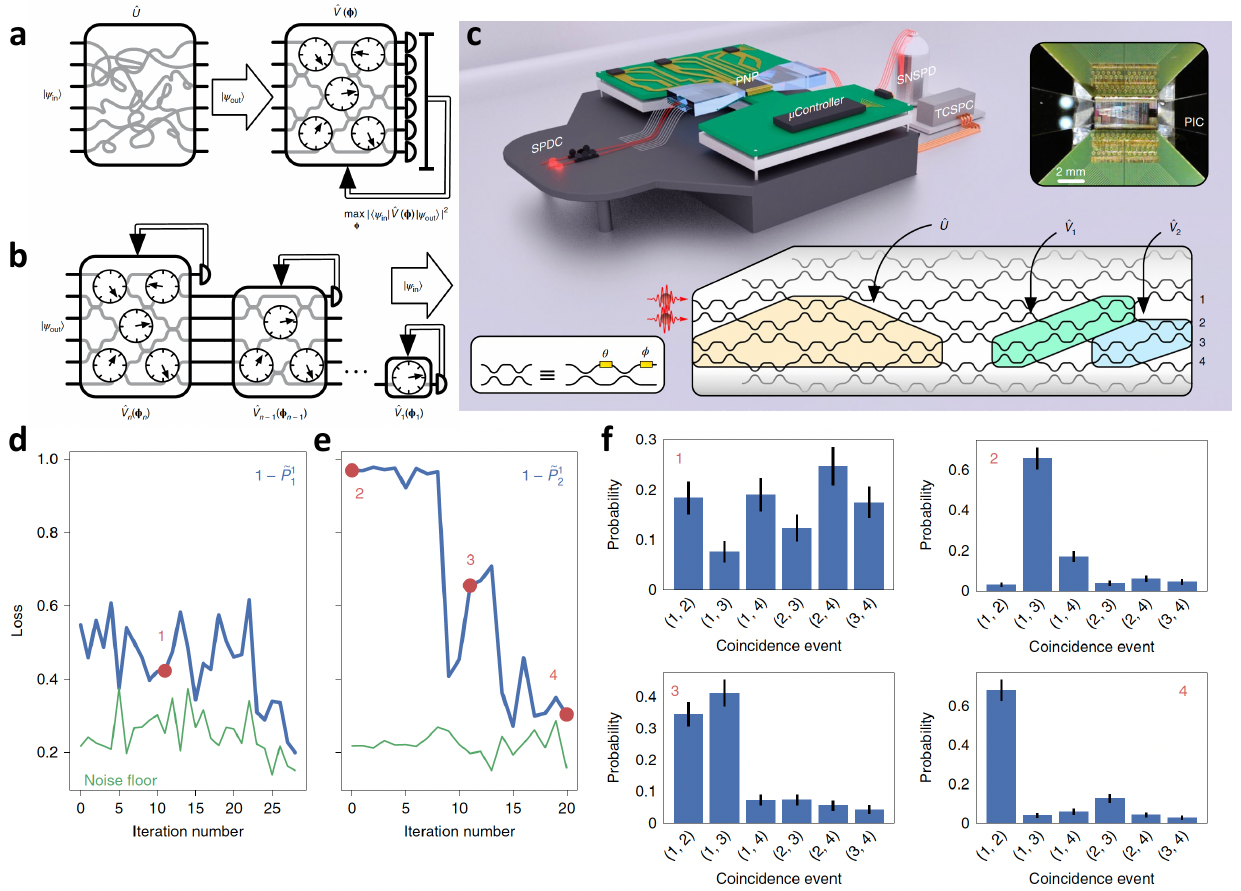}
    \caption{Quantum unsampling with a photonic processor. (a) Illustration of quantum unsampling. (b) Layer-wised approach for quantum unsampling. (c) Experiment diagram and layout of the programmable nanophotonic processor. Minimizing the loss function of the (d) first layer to find a photon in Mode 1 and (e) second layer to find a photon in Mode 2. The four red points show the probabilities for all six twofold coincidence events in (f). Figures reprinted from Ref.~\cite{carolan2020variational}.}
    \label{fig:QML_unsampling}
\end{figure}

\subsection{Quantum Machine Learning for Quantum Processing}
\label{subsec: quantum_machine_learning_quantum_processing}
In addition to improving the performance of classical processing, one of the most promising applications of quantum machine learning is to validate and calibrate quantum states and quantum dynamics\cite{gebhart2023learning}, which are of great importance to NISQ devices. Carolan \textit{et al.} introduced a novel variational quantum unsampling  protocol to unravel quantum dynamics on the known input states via a layer-wised approach and demonstrated it on a quantum photonic processor\cite{carolan2020variational}. Quantum unsampling is a process aimed at learning the inverse of the quantum dynamics of a black box. Given the known input state $\ket{\psi_{\rm in}}$ and unknown quantum operation $\hat{U}$, a quantum unsampling protocol seeks an appropriately parametrized circuit $\hat{V}(\pmb{\phi})$ such that $\hat{V}(\pmb{\phi})\approx\hat{U}^\dagger$ by minimizing a loss function: $L(\pmb{\phi})=1-|\langle \psi_{\rm in}|\hat{V}(\pmb{\phi})|\psi_{\rm out}\rangle|^2$, as illustrated in Fig.~\ref{fig:QML_unsampling}a. In general, the probability of an individual event is exponentially unlikely and the gradient-based optimization becomes exponentially inefficient\cite{mcclean2018barren}. To address this challenge, researchers have developed a layer-by-layer training approach shown in Fig.~\ref{fig:QML_unsampling}b, where each layer optimizes over only a polynomially sized subset of the full Hilbert space by disentangling one qubit at a time. Consider an $n$-qubit input state $\ket{\psi_{\rm in}}=\ket{\alpha_1,\alpha_2,...\alpha_n}$, the output state $\hat{U}\ket{\psi_{\rm in}}$ is fed into the first layer $\hat{V}_n(\pmb{\phi}_n)$ acting on all $n$ qubits $\rho_1=\hat{V}_n(\pmb{\phi}_n)\ket{\psi_{\rm out}}\bra{\psi_{\rm out}}\hat{V}_n^\dagger(\pmb{\phi}_n)$ and an optimization algorithm varies the circuit parameter $\pmb{\phi}_n$ to minimize the loss function $L_1(\pmb{\phi}_n)=1-|\langle \alpha_1|\text{tr}_{2...n}(\rho_1)|\alpha_1\rangle|^2$, aiming to find the first qubit in state $\ket{\alpha_1}$. The reminder $n-1$ qubit state is then fed to the next circuit layer $\hat{V}_{n-1}(\pmb{\phi}_{n-1})$, which maximizes the overlap between the second qubit state $\ket{\alpha_2}$. The optimization continues until all qubits are processed. The overall unsampling quantum operation is thus $\hat{V}=\Pi_{i=1}^n \hat{I}_{n-i}\oplus\hat{V}_i(\pmb{\phi}_i)$, and the probability at each stage is exponentially enhanced. The unsampling quantum circuits are over-parameterized and only partially characterize the unknown quantum operations over certain input states. Additional information of $\hat{U}$ can be obtained by training with more input basis states. 

A proof of concept variational quantum unsampling experiment of boson sampling process was carried out on a state-of-the-art programmable nanophotonic processor (PNP) drawn in Fig.~\ref{fig:QML_unsampling}c. Pairs of photons generated via SPDC from a PPKTP crystal were coupled into the two optical modes of PNP which comprised reconfigurable MZIs arranged in a mesh. Two photons initially passed through the four-mode sampling circuit (orange block) in Fig.~\ref{fig:QML_unsampling}c. The output state was then fed into a two-layer upsampling circuit (green and blue blocks) and detected by an array of single photon detectors. During the training of first unsampling layer (green block), a classical optimizer running the local gradient-free BOBYQA algorithm \cite{powell2009bobyqa} varied the phase shifters of MZIs to minimize the loss function $L_1(\pmb{\phi}_4)=1-\tilde{P}_1^1$, which maximized the probability of a single photon in optical mode 1  $\tilde{P}_1^1$. The loss function  $L_1(\pmb{\phi}_4)$ was initialized to 0.55 and converged to around 0.2 after 28 iterations to the noise floor, as shown in Fig.~\ref{fig:QML_unsampling}d. Then the output state was fed into the second unsampling layer (blue block) while leaving first unsampling layer untouched. The classical optimizer now aimed to find a single photon in mode 2 by minimizing the loss function $L_2(\pmb{\phi}_2)$. The loss function was initialized to 0.97 and converged to 0.31 after 20 steps, as plotted in Fig.~\ref{fig:QML_unsampling}. The probabilities of six twofold coincidence events at different training steps (red dots in Fig.~\ref{fig:QML_unsampling}d, e) are depicted in Fig.~\ref{fig:QML_unsampling}f. At the end of the training, the coincidence count probability in optical mode (1,2) was found to be $P=0.695\pm0.053$, equivalent to the overlap between the initialization state and the output unsampling state $|\bra{\psi_{\rm in}}\hat{V}(\pmb{\phi})\ket{\psi_{\rm out}}|^2$.

Quantum machine learning is still in its early stages and is rapidly evolving with the potential to solve complex problems and advance the capabilities of quantum computing. PIC platforms hold the advantage of providing scalable architectures for quantum machine learning tasks where quantum-light sources, beam splitters, and single-photon detectors can be packed onto small chips, facilitating the development of compact and efficient quantum photonic circuits\cite{bao2023very,wang2020integrated,elshaari2020hybrid,pelucchi2022potential}.

\section{Entanglement-Assisted Communication}
\label{sec: EACOMM}

Entanglement is a critical ingredient to enable a multitude of communication capabilities beyond the reach of classical communication. This section introduce the fundamentals and recent advances in entanglement-assisted communication (EACOMM), a paradigm that harnesses the pre-shared entanglement between communicating parties to increase the rate of transmitting classical information. A representative EACOMM regime is the well-known quantum superdense coding protocol~\cite{bennett1992communication}. This section will review the recent advancements of EACOMM, with a particular focus on experimental demonstrations. For a comprehensive review of quantum superdense coding, readers can refer to a recent review \cite{guo2019advances}. This review article will not cover entanglement-based quantum key distribution. Readers interested in this topic may refer to recent reviews~\cite{pirandola2020advances,xu2020secure}.

\begin{figure}[t!]
    \centering
    \includegraphics[width = 0.8\textwidth]{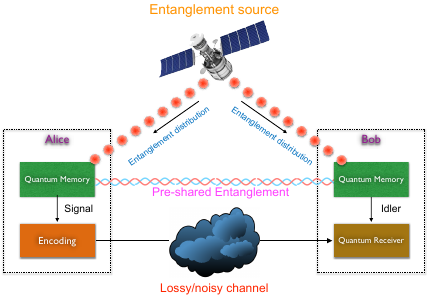}
    \caption{Schematic for entanglement-assisted communication. Entanglement is distributed to two communicating users, Alice and Bob, prior to classical communication. To communicate, Alice retrieves her share of the entangled state, signal, from a quantum memory and encodes on the state. The encoded signal is transmitted to Bob through a lossy and noisy quantum channel. At Bob's terminal, he exploits a quantum receiver to take a joint measurement on the received signal from the channel and the idler. In doing so, Bob infers Alice's encoded classical information.}
    \label{Fig:EACOMM_concept}
\end{figure}

EACOMM enables the transmission of classical information at rates surpassing those achievable by classical communication alone, given the same channel parameters and power constraints. The configuration of EACOMM is illustrated in Fig.~\ref{Fig:EACOMM_concept}: prior to transmitting classical information, two communicating parties, Alice and Bob, establish shared entangled states, the signal and the reference, through entanglement distribution channels and store this pre-shared entanglement in their quantum memories. To initiate communication, Alice retrieves her portion of the entangled state from her quantum memory and encodes the classical information through modulations. The encoded signal is then transmitted to Bob over a noisy and lossy channel. Upon receiving the signal, Bob retrieves his share of the entangled state, i.e., the reference, from his quantum memory. Both the signal and reference are forwarded to a quantum receiver, which performs a joint measurement on them to decode the classical information originally encoded by Alice.

The concept of EACOMM originated from the development of a protocol known as quantum superdense coding \cite{bennett1992communication}, where two classical bits of information can be transmitted using one qubit of an entangled pair. The discovery of the enhanced classical communication rates achievable with pre-shared entanglement led to the exploration of the entanglement-assisted (EA) classical capacity. Early seminal results from the early 2000s demonstrated that the ratio between the EA classical capacity and the ordinary classical capacity could become infinite under certain circumstances \cite{bennett2002entanglement}. However, the derived EA capacity did not provide specific information regarding the encoding format and quantum measurements required to fully exploit the advantage over classical communication. In recent years, structured designs and implementations of encoders and quantum receivers for EACOMM have emerged, utilizing both DV and CV states. 

\subsection{Superdense Coding with DV Entanglement}
\label{subsec: EACOMM_DV}

The principles of superdense coding, also known as dense coding, were initially introduced by Bennett and Wiesner in the context of classical communication using EPR states \cite{bennett1992communication}. In this scenario, Alice and Bob each possess one particle from an entangled pair, prepared in a known entangled state. For simplicity, let us consider the particles as qubits, and the prepared state as the Bell state
\begin{equation}
    \ket{\Phi^{+}} = \frac{1}{\sqrt{2}}\left(|0\rangle_A|0\rangle_B + |1\rangle_A|1\rangle_B\right).
\end{equation}
Alice chooses one of the operations $\{I, Z, X, Y\}$ from the set of Pauli matrices to apply to her particle $A$. Each operation maps one of the four possible bit pairs $\{00, 01, 10, 11\}$ to one of the four orthogonal Bell states: $|\Phi^\pm\rangle = \left(|0\rangle_A|0\rangle_B \pm |1\rangle_A|1\rangle_B\right)/\sqrt{2}$ and $|\Psi^\pm\rangle = \left(|0\rangle_A|1\rangle_B \pm |1\rangle_A|0\rangle_B\right)/\sqrt{2}$. After applying the chosen operation, Alice sends her particle to Bob. Bob, who now controls both particles, performs a complete Bell-state measurement to determine the identity of the joint state. Bob then extracts the two bits representing the state created by Alice, allowing her to transmit two bits of information to Bob using only a single qubit and shared entanglement. It is important to note that without entanglement, Alice's transmission of a qubit can convey only one classical bit of information. This can be proven by the communication channel's capacity, defined as the supremum of the mutual information between the input and output of the channel, which quantifies the maximum rate of information transmission per qubit.

Several significant experimental demonstrations of superdense coding have showcased the achievable capacity using different transmission and measurement techniques. In 1996, Mattel {\em et al.} demonstrated a capacity of 1.58 bits per photon using polarization-encoded states generated at 702 nm with a type-II degenerate SPDC source \cite{PhysRevLett.76.4656}. Their experiment employed a joint measurement device that interfered the two photons to reveal the encoded Bell state through polarization detection. However, this setup based on linear optics allowed only partial Bell-state measurement, limiting the observed communication capacity below the quantum limit of 2 bits per channel use. Barriero {\em et al.} later demonstrated a method to surpass the restriction of detecting only three out of the four Bell states by employing hyperentanglement \cite{barreiro2008beating}. By utilizing the additional orbital angular degree of freedom in the generated photons, they achieved a channel capacity of $1.630 \pm 0.006$ bits per channel. By first distributing a separable joint state entangled in polarization and orbital angular momentum, the resulting measurement could directly detect any of the four polarization-entangled Bell states.

\begin{figure}[t!]
    \centering
    \includegraphics[width = 1\textwidth]{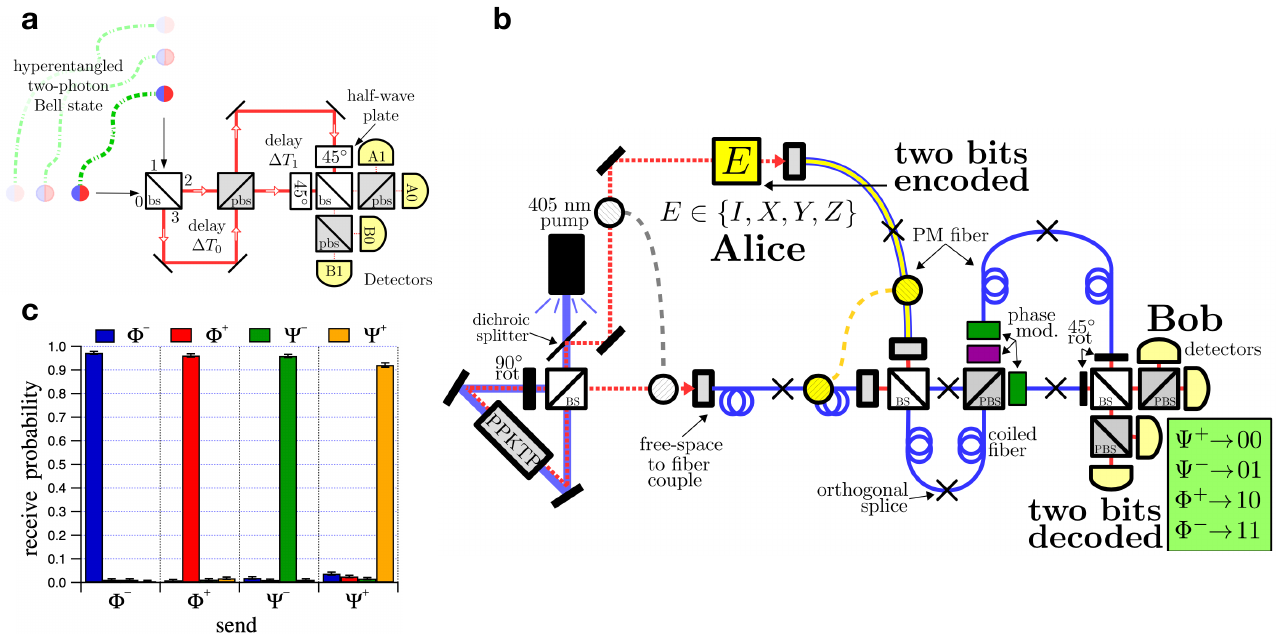}
    \caption{Experiment of superdense coding with complete Bell-state measurements. (a) Schematic for complete Bell-state measurement based on 
    linear optics and time-polarization hyper-entanglement. (b) Experimental diagram. (c) Experimental data for Bell-state discrimination. Figures reprinted from Ref.~\cite{PhysRevLett.118.050501}.}
    \label{fig: superdense_Travis}
\end{figure}

In a recent experiment, Williams {\em et al.} used similar methods to demonstrate superdense coding from complete Bell-state measurements on time-polarization hyper-entanglement that takes advantage of an extra degree of freedom in the time domain~\cite{PhysRevLett.118.050501}. The quantum states of the hyper-entangled photon pairs can be described as
\begin{align}
    |\Phi^{\pm}(t)\rangle &= \frac{1}{\sqrt 2}\left(|H\rangle_A|H\rangle_B \pm |V\rangle_A|V\rangle_B\right) \otimes |\phi(t)\rangle\\
    |\Psi^{\pm}(t)\rangle &= \frac{1}{\sqrt 2}\left(|H\rangle_A|V\rangle_B \pm |V\rangle_A|H\rangle_B\right) \otimes |\phi(t)\rangle,
\end{align}
where $|H\rangle$ and $|V\rangle$ are, respectively, the horizontal and vertical polarization states, and $\phi(t)$ is the temporal quantum state that exhibits two-photon coherence within the timescale relevant to the measurements. The complete linear-optics Bell-state measurement apparatus comprised 50:50 beam splitters, polarizing beam splitters, delay lines, half-wave plates for polarization rotation, and single-photon detectors, as sketched in Fig.~\ref{fig: superdense_Travis}a. 
The temporal and polarization coherence of the hyper-entangled photon pairs allowed for the following transformations on the input Bell states:
\begin{align}
    |\Phi^+\rangle & \rightarrow \frac{1}{\sqrt 2}\left(|H\rangle_A|V\rangle_A + |V\rangle_B|H\rangle_B\right) \\
    |\Phi^-\rangle & \rightarrow \frac{1}{\sqrt 2}\left(|H\rangle_A|H\rangle_B - |V\rangle_A|V\rangle_B\right) \\
    |\Psi^+\rangle & \rightarrow \frac{1}{\sqrt 8}(|H\rangle_A|H'\rangle_B + |H'\rangle_A|H\rangle_B + |V\rangle_A|V'\rangle_B + |V'\rangle_A|V\rangle_B \notag\\
    &+|H\rangle_A|V'\rangle_B - |H'\rangle_A|V\rangle_B + |V\rangle_A|H'\rangle_B + |V'\rangle_A|H\rangle_B) \\
    |\Psi^-\rangle & \rightarrow \frac{1}{\sqrt 8}(|H\rangle_A|V''\rangle_A + |H''\rangle_A|V\rangle_A - |H\rangle_B|V''\rangle_B - |H''\rangle_B|V\rangle_B \notag\\
    &+i|H\rangle_A|V''\rangle_B - i|H''\rangle_A|V\rangle_B + i|V\rangle_A|H''\rangle_B - i|V''\rangle_A|H\rangle_B),
\end{align}
where the subscripts $A$ and $B$ denote the two output ports of the second 50:50 beam splitter, and the prime (double prime) suggests that the photon carries a $\Delta T_0$ ($\Delta T_1$) delay introduced by the first (second) delay line. As such, a time and polarization resolving measurement on both photons discriminates the four Bell states. The complete Bell-state measurement apparatus was integrated in the experimental setup depicted in Fig.~\ref{fig: superdense_Travis}b. The hyper-entanglement source entailed a PPKTP crystal embedded in a Sagnac interferometer introduced in Sec.~\ref{sec: polarization_entanglement}. The crystal was pumped by a continuous-wave 405 nm diode laser to produce more than 200,000 hyper-entangled photon pairs per second. Alice executed the encoding operations to map the photon pairs to one of the four Bell states while Bob leveraged the complete Bell-state measurement apparatus to infer Alice's encoded classical bits. The measurement data plotted in Fig.~\ref{fig: superdense_Travis}c show that the success probability in discriminating all four Bell states exceeded 90\%. This quantum superdense experiment demonstrated a capacity of $1.665\pm0.018$ per detected photon pair, approaching the limit of 2 bits derived from qubit transmission.

\begin{figure}[t!]
    \centering
    \includegraphics[width = 1\textwidth]{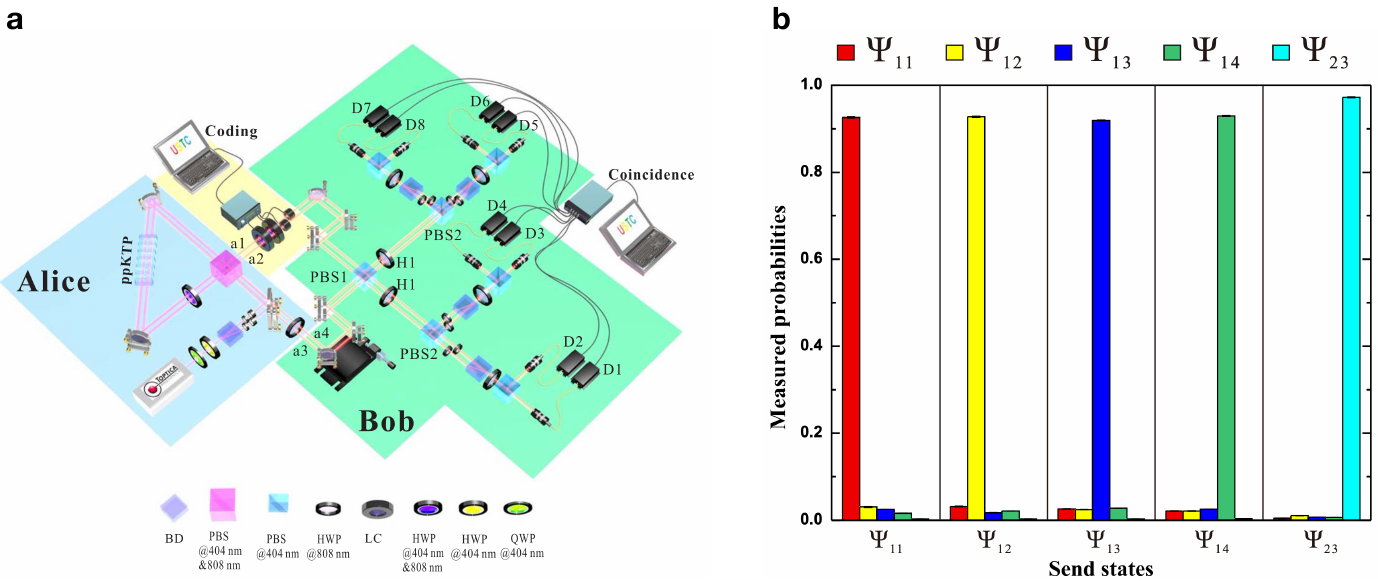}
    \caption{Experiment of superdense coding with quqarts. (a) Experiment setup comprising a source for entangled quqarts, encoding module based on liquid crystal variable retarders, and a measurement apparatus for quantum-state discrimination. (b) Experimental data for discrimination between the five entangled quqart states, showing high success probabilities. Figures reprinted from Ref.~\cite{hu2018beating}.}
    \label{fig: superdense_quqarts}
\end{figure}

In 2018, Hu {\em et al.} demonstrated the idea of superdense coding using photon pairs that were encoded in path and polarization \cite{hu2018beating} degrees of freedom. As illustrated in Fig.~\ref{fig: superdense_quqarts}a, the experiment hinged on a polarization-path entanglement source comprising a PPKTP crystal in a Sagnac interferometer. The source was pumped bidirectionally by two beams to create entangled photon pairs encoded in both polarization and path degrees of freedom. This joint path-polarization encoding was used to address four-dimensional subsystems called ``ququarts'' that express an extension of qubits. Each output arm of the Sagnac interferometer consisted of two paths, i.e., a1, a2 in the upper arm and a3, a4 in the lower arm. The quantum state of the upper-arm photon was specified as follows: (H, a1) $\rightarrow |0\rangle$, (V, a1) $\rightarrow |1\rangle$, (H, a2) $\rightarrow |2\rangle$, and (V, a2) $\rightarrow |3\rangle$, while the quantum state of the lower-arm photon was specified as: (H, a3) $\rightarrow |0\rangle$, (V, a3) $\rightarrow |1\rangle$, (H, a4) $\rightarrow |2\rangle$, and (V, a4) $\rightarrow |3\rangle$. Alice exploited four computer-controlled liquid crystal variable retarders to produce five entangled ququart states that could be discriminated by Bob's measurement apparatus:
\begin{align}
    |\Psi\rangle_{11} &= \frac{1}{2} \left(|00\rangle + |11\rangle + |22\rangle + |33\rangle \right)\\
    |\Psi\rangle_{12} &= \frac{1}{2} \left(|00\rangle - |11\rangle + |22\rangle - |33\rangle \right)\\
    |\Psi\rangle_{13} &= \frac{1}{2} \left(|00\rangle + |11\rangle - |22\rangle - |33\rangle \right)\\
    |\Psi\rangle_{14} &= \frac{1}{2} \left(|00\rangle - |11\rangle - |22\rangle + |33\rangle \right)\\
    |\Psi\rangle_{23} &= \frac{1}{2} \left(|01\rangle + |10\rangle - |23\rangle - |32\rangle \right).
\end{align}
The theoretical channel capacity subject to the encoding scheme was $\log_{2} 5 = 2.32$ bits per channel use. The measurement apparatus consisted of a polarizing beam splitter to interfere the two photons followed by additional polarizing beam splitters and beam displacers to extract the polarization and path information. The experimental data for quantum-state discrimination are plotted in Fig.~\ref{fig: superdense_quqarts}b, showing success probabilities in excess of 90\% for all five entangled quqart states. This led to a measured $2.09 \pm 0.01$ bits per channel use, exceeding the 2 bit per channel use limit for qubit-based quantum superdense coding protocol.

\subsection{Communication Assisted by CV Entanglement}
\label{subsec: EACOMM_CV}

\begin{figure}[t!]
    \centering
    \includegraphics[width = 0.8\textwidth]{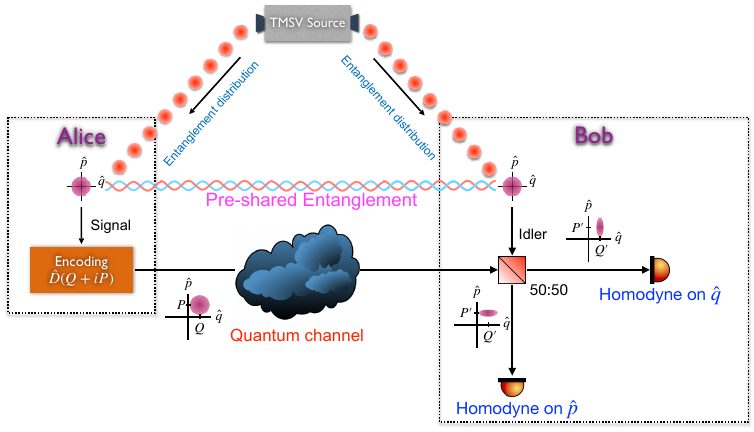}
    \caption{Schematic for quantum superdense coding with continuous variables. Alice and Bob pre-share a TMSV state prior to communication. Alice's encoding entails a displacement operation that shifts the signal in the phase space. The encoded signal is transmitted to Bob through a quantum channel. Bob interfers the received signal and the idler on a 50:50 beam splitter to produce two single-mode displaced squeezed states. Two homodyne measurements are followed to detect the quadratures and infer Alice's encoding.}
    \label{fig:CV_superdense}
\end{figure}

The CV version of quantum superdense coding schematically shown in Fig.~\ref{fig:CV_superdense} was proposed by Braunstein and Kimble~\cite{braunstein2000dense} and first demonstrated by Li {\em et al.}~\cite{li2002quantum}. In the protocol, Alice and Bob establish pre-shared entanglement in a TMSV state via an entanglement-distribution channel. To encode, Alice performs a displacement operation $\hat D(Q+iP)$ to shift her share of the TMSV state, the signal, in the phase space. The displacement on the $\hat q$ ($\hat p$) quadrature is parametrized by $Q$ ($P$). The encoded signal is sent to Bob through a quantum channel. Bob's quantum receiver comprises a 50:50 beamp splitter and two homodyne detectors that measure, respectively, the $\hat q$ and $\hat p$ quadrature. The beam splitter mixes Bob's share of the TMSV state, the idler, with the received signal, generating two single-mode state squeezed in the $\hat q$ and $\hat p$ quadrature. The mean of the the two single-mode squeezed state is $Q'+iP'$, where $Q'=Q/\sqrt{2}$ and $P'=P/\sqrt{2}$ assuming an ideal channel. As such, multiplying the measurement outcomes from both homodyne detectors by $\sqrt{2}$ yields an unbiased estimator for Alice's encoded message. 

Braunstein and Kimble showed that the CV quantum superdense coding protocol outperformed a classical communication scheme based on transmitting displaced coherent states with the same power in tandem with a heterodyne detector~\cite{braunstein2000dense}. However, one would need 6.78 dB of ``break-even squeezing'' to beat the ultimate classical capacity, and in the limit of infinite squeezing CV quantum superdense coding enables a two-fold enhancement in the communication rate over the optimal classical scheme. Such a squeezing level would be quite challenging to achieve in the presence of many imperfect components in a system-level demonstration. Worse, the amount of break-even squeezing is derived based on a perfect quantum channel with no loss and additive noise. Channel loss and noise is known to quickly diminish the squeezing at the quantum receiver, thereby questioning the practical benefit of the CV quantum superdense coding protocol. 

\begin{figure}[t!]
    \centering
    \includegraphics[width = 1\textwidth]{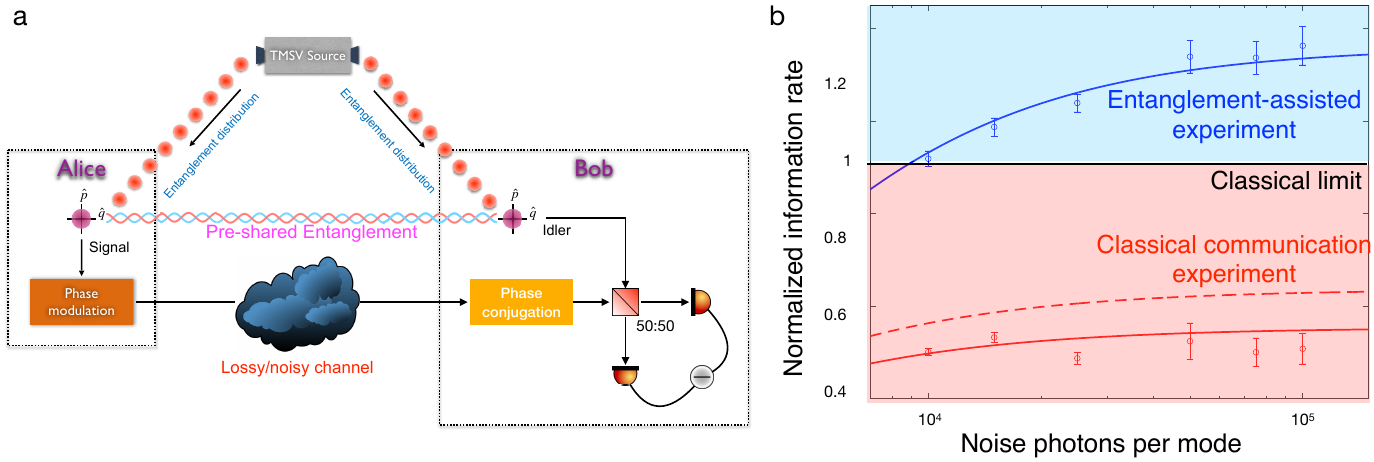}
    \caption{Entanglement-assisted communication experiment. (a) The EACOMM experimental setup comprising a TMSV source, a phase-modulation encoder operated by Alice, and a PCR at Bob's terminal. (b) Normalized information rate for EACOMM (blue) and classical communication (red) at different channel background noise levels. Black solid line is the classical limit specified by the HSW capacity. Solid line: theory; dot: experimental data; dashed line: theory with ideal experimental parameters. (b) reprinted from Ref.~\cite{hao2021entanglement}.}
    \label{fig:EACOMM_exp}
\end{figure}

The challenge faced by CV quantum superdense coding over practical channels behooves researchers to quantify the advantage of EACOMM over the ultimate classical capacity. To this end, let us compare the EA classical capacity and the classical capacity without EA, i.e., the Holevo-Schumacher-Westmoreland (HSW) capacity. The EA classical capacity over a lossy channel plagued by a large noise background is found to be
\begin{equation}
    C_E(\mathcal{L}^{\kappa, N_B})= \frac{1}{N_B}\kappa N_S (1+N_S)\log_2\left(1+\frac{1}{N_S}\right),
\end{equation}
where $N_B \gg 1$ is the mean photon number of the background mode, $\kappa$ is the transmissivity of the channel, and $N_S$ is the mean photon number signal or idler mode of the TMSV state. By comparison, the classical capacity without EA, i.e., the Holevo-Schumacher-Westmoreland (HSW) capacity reads
\begin{equation}
    C(\mathcal{L}^{\kappa, N_B}) = g(\kappa N_S + N_B)-g(N_B),
\end{equation}
where $g(N) = (N+1)\log_2(N+1)-N\log_2(N)$ is the von Neumann entropy of a thermal state with mean photon number $N$. In the limit of a low-brightness entanglement transmitter, 
\begin{equation}
    \lim_{N_S \rightarrow 0}\frac{C_E(\mathcal{L}^{\kappa, N_B})}{C(\mathcal{L}^{\kappa, N_B})} = \infty,
\end{equation}
showing that entanglement can, in principle, enable an infinite-fold advantage in the communication rate over a very noisy channel. To reap the quantum advantage, a recent theoretical work devised the entanglement sources, encoding formats, and quantum receivers~\cite{shi2020practical}. Among the investigated quantum receivers, the phase-conjugate receiver (PCR) presents an experimentally viable route toward surpassing the HSW capacity to achieve a quantum advantage in communication. Ref.~\cite{hao2021entanglement} reported an EACOMM experiment based on low-brightness non-degenerate TMSV source and a PCR, as sketched in Fig.~\ref{fig:EACOMM_exp}a. The entanglement transmitter entailed a PPLN crystal that generated the signal and idler with $N_S \ll 1$. The signal and idler were distributed to Alice and Bob through low-loss optical fibers. Alice's encoded her classical information by performing binary phase-shift keying on the signal and sent the encoded signal to Bob through a very lossy and noisy channel. Akin to the optical QI experiment, the channel noise was emulated by mixing thermal light with the signal. Bob's PCR first exploited a second PPLN crystal to produce the phase conjugate of the received signal. A 50:50 beam splitter then interfered the idler and the phase conjugate and transmited the light on its two output ports to a pair of high-efficiency photodiodes in a balanced setting. The difference photocurrent of the two photodiodes was used to infer Alice's encoded bit. The experimental data plotted in Fig.~\ref{fig:EACOMM_exp}b show that the EACOMM rate surpassed the classical limit set by the HSW bound. Moreover, the achieved EACOMM rate substantially outperformed the rate attained by a practical classical communication setup also built in the experiment.

\section{Outlook}
\label{sec: outlook}
\begin{figure}[t!]
    \centering
    \includegraphics[width = \textwidth]{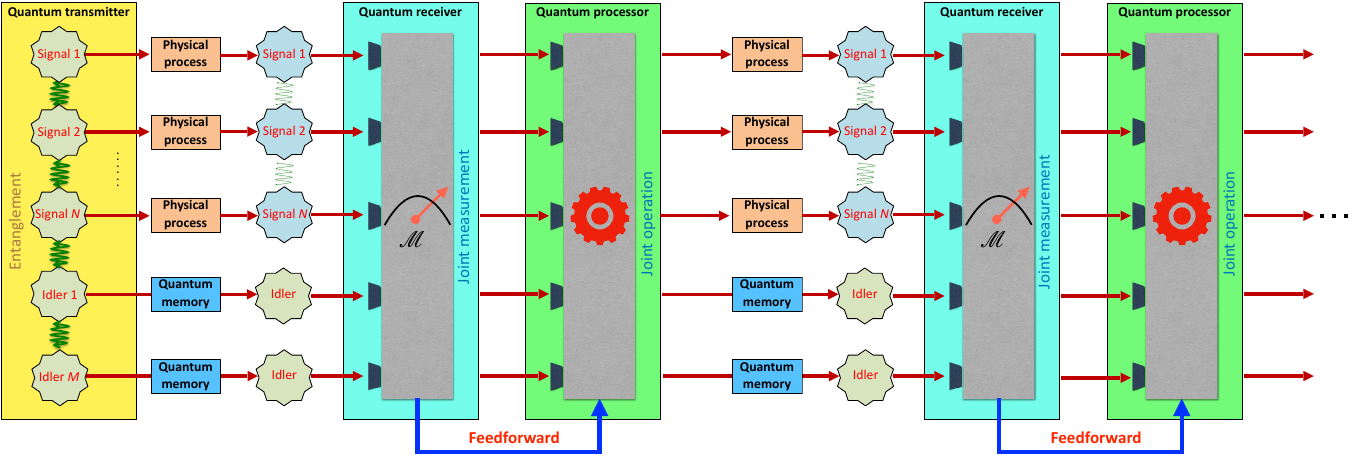}
    \caption{Outlook for entanglement-based QIT architecture. Future QIT protocols would build on multiple stages of entanglement preparation, processing, and detection, in tandem with an adaptive feedforward mechanism to unlock the full power of quantum information processing.}
    \label{fig:outlook_architecture}
\end{figure}

The next-generation entanglement-based QIT beyond the three classes of protocols and their associated applications as discussed in this article is on the verge, thanks to the remarkable strides across the spectrum of functional quantum modules including large-scale entanglement sources~\cite{reimer2016generation,kues2019quantum,yang2021squeezed,schwartz2016deterministic,larsen2019deterministic,chen2014experimental}, efficient quantum transducers~\cite{brubaker2022optomechanical,mirhosseini2020superconducting,han2021microwave,forsch2020microwave,jiang2020efficient}, long-lived quantum memories~\cite{sukachev2017silicon,sun2018single,wallucks2020quantum,wang2019efficient,liu2021heralded,heshami2016quantum}, high-fidelity quantum gates~\cite{sung2021realization,xu2020experimental,zhang2020error,egan2021fault}, near-unity-efficiency~\cite{reddy2020superconducting} and photon-number-resolving~\cite{cheng2022100,eaton2022resolution} detectors, and novel quantum receivers~\cite{cui2022quantum,hao2021entanglement,izumi2020tomography,tsujino2011quantum,chen2012optical,becerra2013experimental,becerra2015photon,ferdinand2017multi,burenkov2018quantum,izumi2020experimental,rengaswamy2021belief,dimario2020phase}, in conjunction with auxiliary classical real-time processing capabilities. An envisaged architecture for QIT protocols based on the emergent enabling quantum modules is illustrated in Fig.~\ref{fig:outlook_architecture}. The quantum transmitter entails sources of encoded entangled states to address distributed sensing, data processing, and decision problems. The encoded entangled states, in tandem with the quantum receiver and processor, are tailored to combat imperfections such as environmental loss and noise~\cite{noh2020encoding,Gottesman2001,zhuang2020distributed,wu2022optimal,wu2022continuous,xu2022qubit,rozpkedek2021quantum}. Over recent years, large-scale deterministic quantum sources have been extensively investigated to serve future scalable QIT~\cite{yang2021squeezed,larsen2019deterministic,schwartz2016deterministic,asavanant2019generation,kues2019quantum,kues2017chip,reimer2016generation}. The signals from the quantum transmitter interface with physical processes, which are operations realized by quantum transduction to interconnect various quantum information processing platforms~\cite{awschalom2021development}. The idlers emitted from the quantum transmitter are loaded into quantum memories, awaiting on-demand retrieval by a next-generation quantum receiver that features the following characteristics. First, it allows for more general joint measurements on the signals and idlers, which are proven an essential ingredient to approach the ultimate quantum-limited performance~\cite{guha2011structured}. Second, the measurements are non-destructive, producing signals and idlers for the next stage of processing~\cite{raha2020optical,nakajima2019quantum,bowden2020improving,xue2020repetitive,nakajima2019quantum}. Notably, new materials on the horizon exhibit extraordinary nonlinearities~\cite{yoshioka2021strongly,wang2021fully}, rendering the quest for quantum non-demolition measurements more practical. Third, with state-of-the-art electronics, the quantum receiver is capable of generating real-time feedforward to configure a quantum processor comprising high-fidelity quantum gates to conduct a joint operation on the signals and idlers. The quantum processor enhances, as needed, the entanglement shared by the signals and idlers for the next cycle of information processing. Such an adaptive architecture based on real-time measurements, feedforward, and reconfiguration of quantum operations has been demonstrated in quantum receivers for semi-classical state discrimination. As a recent advance in this subject, machine-learning techniques were utilized to assist in the design and implementation of quantum receivers to tackle environmental noise~\cite{cui2022quantum}. One can envision that the adaptive architecture, in conjunction with classical artificial intelligence, will empower a wealth of entanglement-based QIT protocols. 

The prospective QIT architecture based on the emergent functional quantum modules will give rise to unprecedented communication, sensing, and data processing capabilities. For instance, with the quantum transducers quantum information can be transferred between stationary qubits and flying photonic qubits to facilitate long-distance quantum communications~\cite{mirhosseini2020superconducting,bhaskar2020experimental,yu2020entanglement}. The quantum error correction components can ensure the fidelity of such conversion~\cite{noh2020encoding,rozpkedek2021quantum}. Moreover, the joint encoding at the quantum transmitter and joint measurement at the quantum receiver will unlock quantum-communication regimes with no classical analogue, such as superadditive quantum capacities~\cite{hastings2009superadditivity,zhu2017superadditivity,zhu2018superadditivity} and superactivation over quantum channels~\cite{duan2009super,shor2003superactivation,cubitt2012extreme}. On the sensing front, recent experiments have demonstrated that entanglement in the optical domain together with transducers can be leveraged to enhance the performance of arrayed RF~\cite{xia2020demonstration} or optomechanical sensors~\cite{xia2022entanglement}. One may imagine that large-scale optical entanglement in tandem with efficient quantum transducers, quantum error correction, and quantum-limited detector arrays would constitute new powerful metrological tools for a variety of realms such as the search for new physics~\cite{backes2021quantum,dixit2021searching,wurtz2021cavity,estrada2021searching,brady2022entangled,You2023npj}, inertial navigation~\cite{grace2020quantum,rico2015quantum,jiao2023noisy}, and remote sensing based on microwave photonics~\cite{ghelfi2014fully,zou2016photonics,yao2020microwave,hervas2017microwave} enhanced by QIT. With respect to data processing, recent proof-of-concept experiments have leveraged quantum resources to tackle physical simulation and machine-learning problems~\cite{arrazola2021quantum,havlivcek2019supervised,xia2021quantum,carolan2020variational,saggio2021experimental}. The potential of quantum-enhanced data processing will be further unleashed with the new QIT architecture by virtue of the capability of generating a wide variety of multipartite entangled state, general joint measurements, real-time feedforward, and quantum error correction. 

Over the past three decades, the humankind has witnessed entanglement's transition from a verifiable physical phenomenon to a profound resource underpinning a plethora of new technologies. With the highly interdisciplinary, synergistic, and holistic efforts devoted by the scientific and technological communities, entanglement-enhanced QIT is set to boom in the years to come, bringing excitements that have yet to imagine.

\begin{backmatter}
\bmsection{Funding}
National Science Foundation Award No.~1920742, 2134830, 2144057, and 2304118, Defense Advanced Research Projects Agency (DARPA) under Young Faculty Award (YFA) Grant No. N660012014029, and University of Michigan. C. Y. and O. S. M. L. acknowledge support from the Army Research Office (ARO), through the Early Career Program (ECP) under the grant no. W911NF-22-1-0088 and the U. S. Department of Energy for supporting this research through the Office of Basic Energy, Office of Basic Energy Sciences, Division of Materials Sciences and Engineering under the Award DE-SC0021069.

\bmsection{Acknowledgments}
We thank Sophia Economou for the feedback on the manuscript.

\bmsection{Disclosures}
The authors declare no conflicts of interest.

\bmsection{Data Availability Statement}
No data were generated or analyzed in the presented research.

\end{backmatter}


\end{document}